\pdfoutput=1
\RequirePackage{ifpdf}
\ifpdf 
\documentclass[pdftex]{sigma}
\else
\documentclass{sigma}
\fi

\usepackage{subfig}
\usepackage[labelsep=period,labelfont=bf]{caption}
\usepackage{tikz-cd}
\usepackage{extarrows}
\usepackage{mathtools}
\mathtoolsset{showonlyrefs=true}
\usepackage[customcolors,norndcorners]{hf-tikz}
\usetikzlibrary{decorations.pathmorphing}

\numberwithin{equation}{section}

\newtheorem{Theorem}{Theorem}[section]

\newtheorem{Proposition}[Theorem]{Proposition}
\newtheorem{Conjecture}[Theorem]{Conjecture}

{\theoremstyle{definition}
\newtheorem{Example}[Theorem]{Example}
\newtheorem{Remark}[Theorem]{Remark}

\newtheorem{Definition}[Theorem]{Definition}}

\definecolor{ared}{RGB}{128,0,0}
\definecolor{agreen}{RGB}{99,160,99}
\definecolor{ablue}{RGB}{0,66,170}
\definecolor{aorange}{RGB}{255,100,0}

\begin{document}

\newcommand{\arXivNumber}{2009.00426}

\renewcommand{\thefootnote}{}

\renewcommand{\PaperNumber}{032}

\FirstPageHeading

\ShortArticleName{An Introduction to Motivic Feynman Integrals}

\ArticleName{An Introduction to Motivic Feynman Integrals\footnote{This paper is a~contribution to the Special Issue on Representation Theory and Integrable Systems in honor of Vitaly Tarasov on the 60th birthday and Alexander Varchenko on the 70th birthday. The full collection is available at \href{https://www.emis.de/journals/SIGMA/Tarasov-Varchenko.html}{https://www.emis.de/journals/SIGMA/Tarasov-Varchenko.html}}}

\Author{Claudia RELLA}

\AuthorNameForHeading{C.~Rella}

\Address{Section de Math\'ematiques, Universit\'e de Gen\`eve, Gen\`eve, CH-1211 Switzerland}
\Email{\href{mailto:claudia.rella@unige.ch}{claudia.rella@unige.ch}}

\ArticleDates{Received August 30, 2020, in final form March 03, 2021; Published online March 26, 2021}

\Abstract{This article gives a short step-by-step introduction to the representation of~para\-metric Feynman integrals in scalar perturbative quantum field theory as periods of motives. The application of~moti\-vic Galois theory to the algebro-geometric and categorical structures underlying Feynman graphs is reviewed up to the current state of research. The~example of primitive log-divergent Feynman graphs in scalar massless $\phi^4$ quantum field theory is analysed in detail.}


\Classification{81-02; 14-02; 81Q30; 81T18; 81T15; 14C15; 14C30; 14F40; 11R32}

\tableofcontents

\renewcommand{\thefootnote}{\arabic{footnote}}
\setcounter{footnote}{0}

\section{Introduction}

In Section~\ref{sec: graphs}, we describe the graph-theoretic framework for the investigation of the algebraic information contained in the topology of scalar Feynman diagrams. Perturbative quantum field theories possess an inherent algebraic structure, which underlies the combinatorics of recursion governing renormalisation theory, and are thus deeply connected to the theory of graphs.

\looseness=1 In Section~\ref{sec: geometry}, we broadly review preliminary notions in algebraic geometry and algebraic topology.
An algebraic variety over $\mathbb{Q}$ gives rise to two distinct rational structures via algebraic de Rham cohomology and Betti cohomology, which are compatible with each other only after complexification.
The coexistence of these two cohomologies and their peculiar compatibility are linked to a specific class of complex numbers, known as periods. The cohomology of an~alge\-braic variety is equipped with two filtrations, and the mixed Hodge structure arising from their interaction constitutes the bridge between the theory of periods and the theory of motives.

In Section~\ref{sec: periods}, we introduce the set of periods, lying between $\bar{\mathbb{Q}}$ and $\mathbb{C}$, among which are the numbers that come from evaluating parametric Feynman integrals, and we briefly review their remarkable properties. Suitable cohomological structures are exploited to derive non-trivial information about these numbers.

In Section~\ref{sec: motives}, we describe how Feynman integrals are promoted to periods of motives. Technical issues arising from the presence of singularities are tackled by blow up. We~adopt the category-theoretic Tannakian formalism where motivic periods, and motivic Feynman integrals in particular, reveal their most intriguing properties. We~present an overview of the current progress of research towards the general understanding of the structure of scattering amplitudes via the theory of motivic periods, giving particular attention to recent results in massless scalar~$\phi^4$ quantum field theory.

\section{Scalar Feynman graphs} \label{sec: graphs}

\subsection{Perturbative quantum field theory}
A quantum field theory encodes in its Lagrangian every admissible interaction among particles, but it does it in a way that makes decoding this information a difficult task.
The probability amplitude associated to the interaction process between given initial and final states, called its \textit{Feynman amplitude}, is determined by the set of kinematic and interaction terms in the Lag\-ran\-gian. However, individual Lagrangian terms correspond to propagators and interaction vertices which can be linked together in infinitely many distinct ways to connect the same pair of initial and final states.
Each of these admissible realisations of the same interaction process has to be accounted for in an infinite sum of contributions to the probability amplitude. We~associate to each of these possibilities a graphical representation, called its \textit{Feynman diagram}, whose individual contribution to the probability amplitude is explicitly written in the form of a \textit{Feynman integral} by applying the formal correspondence between Lagrangian terms and graphical components, which is established by convention through the set of Feynman rules of~the theory. It~is only the sum of the contributing Feynman integrals to a given process that has a physical meaning and not the individual integrals, which are themselves interrelated by~the gauge symmetry of the Lagrangian.

\begin{figure}[htb!]
\centering
\subfloat[One-particle irreducible]{\includegraphics[scale=.5]{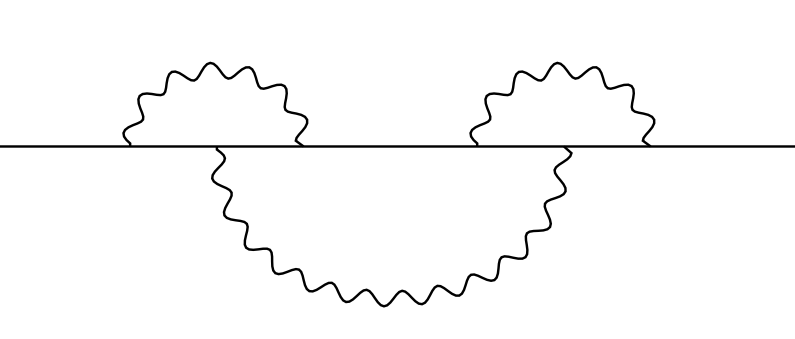}} \quad
\subfloat[One-particle reducible]{\includegraphics[scale=.5]{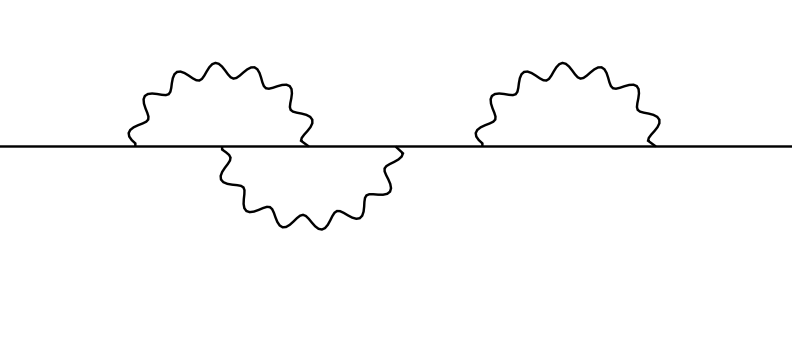}}
\caption{Examples of 1PI and non-1PI diagrams.}
\label{fig:1PI}
\end{figure}

In perturbative quantum field theory, the sum of individual Feynman integrals is a \textit{perturbative expansion} in some small parameter of the theory, typically a suitable coupling constant. Thus, the Feynman amplitude can be expanded in a formal power series, which has been shown to be divergent\footnote{Serone et al~\cite{SSV17} characterised the conditions under which some class of asymptotic perturbative series are Borel resummable, leading to exact results without introducing non-perturbative effects in the form of trans-series.} by Dyson~\cite{Dys52}. The divergence does not, however, undermine the accuracy of~predictions that can be made with the theory.
Indeed, although a Feynman amplitude recei\-ves contributions to any order in perturbation theory, practical calculations are made by truncating the sum at a certain order and directly evaluating only the remaining finitely many terms.
Moreover, the explicit calculation of a Feynman amplitude only includes those diagrams which are one-particle irreducible, or 1PI, that is, diagrams which cannot be divided in two by~cutting through a single propagator. See Fig.~\ref{fig:1PI}.
The contribution from a non-1PI diagram at some given order can be expressed as a combination of lower-order 1PI contributions, which have already been accounted for in the formal series.

\looseness=1
The leading order terms in the perturbative expansion of a Feynman amplitude are called tree-level contributions. Higher order diagrams are obtained from tree-level diagrams by adding internal loops. Each independent loop in a diagram is associated to an unconstrained momentum and integrals over unconstrained loop momenta are the origin of \textit{singularities} in Feynman integrals. We~distinguish two classes of singularities.
The \textit{ultraviolet} (UV) divergences arise in the limit of infinite loop momentum, a regime that is far beyond the energy scale that we have currently experimental access to and where we expect new physical phenomena to become relevant and corresponding new terms to enter the Lagrangian. Sensitivity to the high loop momentum region is treated by means of \textit{renormalisation theory}. For a renormalizable theory, a~suitable adjustment of the Lagrangian parameters allows to systematically re-express the predictions of the theory in terms of renormalized physical couplings, so that they decouple from UV physics. Thus, the theory gives a finite and well-defined relation between physical observables.
The \textit{infrared} (IR) divergences only arise in theories with massless particles as they originate in the limit of infinitesimal loop momentum. They cannot be removed by renormalisation and introduce numerous subtleties in the evaluation of Feynman integrals which we are not dealing with in the present text. For a detailed and comprehensive presentation of perturbative quantum field theory we refer to Zee~\cite{Zee03} and Srednicki~\cite{Sre10}.

\looseness=1 Evaluating Feynman integrals over loop momenta has been of fundamental concern since the early days of perturbative quantum field theory.
Since the first insights into the problem of UV divergences in a quantum field theory presented by Dyson~\cite{Dys49,Dys52}, Salam~\cite{Sal51_2,Sal51_1} and Weinberg~\cite{Wei60} in the 1950s and 60s, our understanding has vastly improved.
In 2004, Smirnov~\cite{Smi06} summarised more than fifty years of advancements in the field, providing an overview of the most powerful, successful and well-established methods available at the time for evaluating Feynman integrals in a systematic way, showing how the problem of evaluation had become more and more critical. What could be easily evaluated had already been evaluated years ago.
Nowadays, new approaches, based on the symmetry properties of the loop integrands\footnote{We refer to Elvang and Huang~\cite{EH13} for a review of the subject, including unitarity methods, BCFW recursion relations, and the methods of leading singularities and maximal cuts.} and the complementary perspective of differential equations,\footnote{Henn~\cite{Henn15} gives an overview of the method of differential equations, using tools such as Chen iterated integrals, multiple polylogarithms, and the Drinfeld associator.} are available and vastly studied.
Despite progress, the mathematical understanding and the computation of Feynman integrals are still far from being complete.
Overlapping divergences can be treated iteratively, thus revealing in the first place the recursive nature of renormalisation theory. However, this combinatorics of subdivergences is only the first hint to a more fundamental algebraic structure inherent in all renormalizable quantum field theories and deeply connected to the theory of graphs.\footnote{A first discussion about the appearance of transcendental numbers in Feynman integrals and its relation to the topology of Feynman graphs is presented by Kreimer~\cite{Kre97} in the framework of knot theory and link diagrams. A~recent review on the theory of numbers and single-valued functions on the complex plane which arise in quantum field theory is presented by Schnetz~\cite{Sch16} in the modern context of the theory of motivic periods.}

\subsection{Feynman parametrisation}
\looseness=1
We consider a scalar quantum field theory in an even number $D$ of space-time dimensions with Euclidean metric\footnote{It is common practice to compute amplitudes in Euclidean space. Moving to Minkowski space involves performing an extension by analytic continuation known as Wick rotation. See for example~\cite{Smi06} and~\cite{Sre10}.} and allow different propagators to have different masses. A~Feynman diagram is a \textit{connected directed graph} where each edge represents a propagator and is assigned a~momentum and a mass and each vertex stands for a tree-level interaction. Exter\-nal half-edges, also known as external legs, represent incoming or outgoing particles, while internal edges are the internal propagators of the diagram. We~define the \textit{loop number} to be the number of~inde\-pen\-dent cycles of the graph. We~adopt the convention for which all external legs have arrows pointing inwards, and consequently distinguish incoming and outgoing particles depending on~the momentum being positive or negative, respectively. Momentum is positive when it points in~the same direction of the arrow of the corresponding directed edge, and it is negative otherwise. We~fix momenta on external lines and for each internal loop we choose an arbitrary orientation of the edges which is consistent with momentum conservation at each vertex of the graph and globally. Momentum conservation leaves one unconstrained free momentum variable for each loop. Thus, the loop number is equal to the number of independent loop momentum vectors. We~only consider graphs that are one-particle irreducible.

{\samepage
Let $G$ be such a Feynman graph with $m$ external legs, $n$ internal edges, and $l$ independent loops. Its Feynman integral, up to numerical prefactors, is
\begin{gather} \label{eq: I_G_old}
I_G= \big(\mu^2\big)^{n-lD/2} \int \prod_{r=1}^l \frac{{\rm d}^Dk_r}{{\rm i} \pi^{D/2}} \; \prod_{j=1}^n \frac{1}{-q_j^2+m_j^2-{\rm i} \varepsilon},
\end{gather}
where $\varepsilon$ is a small positive parameter,\footnote{The $-{\rm i} \varepsilon$ term is required by the choice of Feynman pole prescription for the computation of the propagators and it allows to perform a Wick rotation to Minkowski space. In what follows, however, the $-{\rm i} \varepsilon$ term does not play a role. We~set $\varepsilon = 0$ for simplicity of notation.} $\mu$ is a scale introduced to make the expression dimensionless,\footnote{In what follows, the scale $\mu$ remains factored out. We~set $\mu^2 = 1$ for simplicity of notation.} $k_1,\dots,k_l$ are the independent loop momenta, $m_1,\dots,m_n$ are the masses of the internal lines, and $q_1,\dots,q_n$ are the momenta flowing through them, which can be expressed as
\begin{gather}
q_j=\sum_{i=1}^l \lambda_{ji} k_i + \sum_{i=1}^m \sigma_{ji} p_i,
\end{gather}
where $p_1,\dots,p_m$ are the external momenta and $\lambda_{ji}$, $\sigma_{ji} \in \{ -1,0,1 \}$ are constants depending on~the particular graph structure.

}

Feynman~\cite{Fey49} introduced the well-known manipulation consisting of defining a set of parameters $x_1,\dots,x_n$, called \textit{Feynman parameters}, one for each internal edge of the graph, and applying the formula
\begin{gather}
\prod_{j=1}^n \frac{1}{P_j} = \Gamma(n) \int_{\{x_j \ge 0\}} {\rm d}^nx \,\delta\bigg(1- \sum_{j=1}^n x_j \bigg) \frac{1}{\left(\sum_{j=1}^n x_j P_j \right)^n}
\end{gather}
with the choice $P_j=-q_j^2+m_j^2$ for $j=1,\dots,n$. Here, $\Gamma$ is the Euler gamma function and $\delta$ is the Dirac delta distribution.
Indeed, we can write
\begin{gather}
\sum_{j=1}^n x_j \big({-}q_j^2+m_j^2\big) = - \sum_{r=1}^l \sum_{s=1}^l k_r \cdot (M_{rs} k_s) + \sum_{r=1}^l 2 k_r \cdot Q_r + J,
\end{gather}
where $M$ is an $l \times l$-matrix with scalar entries, $Q$ is an $l$-vector with momentum vectors as entries and $J$ is a scalar. $M$, $Q$ and $J$ can be suitably expressed in terms of the graph parameters $\{x_j, q_j, m_j\}_{j=1}^n$.
Applying Feynman parametrisation to~\eqref{eq: I_G_old}, the $l$-dimensional integral over the loop momenta becomes an $(n-1)$-dimensional integral over the Feynman parameters
\begin{gather} \label{eq: I_G_new}
I_G = \Gamma\bigg(n-\frac{lD}{2}\bigg) \int_{\{ x_j \ge 0 \}} {\rm d}^nx \, \delta\bigg(1- \sum_{j=1}^n x_j \bigg) \frac{\mathcal{U}^{n-(l+1)D/2}}{\mathcal{F}^{n-lD/2}},
\end{gather}
which is characterised by the polynomials $\mathcal{U}= \det (M)$ and $\mathcal{F}= \det (M) \big(J+ Q M^{-1} Q \big)$, called \textit{first} and \textit{second Symanzik polynomials} of the Feynman graph, respectively.
Notice that the dimension~$D$ of space-time, entering the exponents in the integrand
of~\eqref{eq: I_G_new}, acts as regularisation. We~use dimensional regularisation\footnote{The dimensional regularisation procedure has been first introduced by 't~Hooft and Veltman~\cite{tHooftVeltman}.} with $D=4-2\epsilon$ and $\epsilon$ a small parameter. A~detailed description of~Feynman parametrisation can be found in Srednicki~\cite{Sre10}.

\begin{Example}
Consider a one-loop diagram with $m=n$ external legs, as the one shown in~Fig.~\ref{fig: nbox}.
\begin{figure}[htb!]
\centering
\includegraphics[scale=.5]{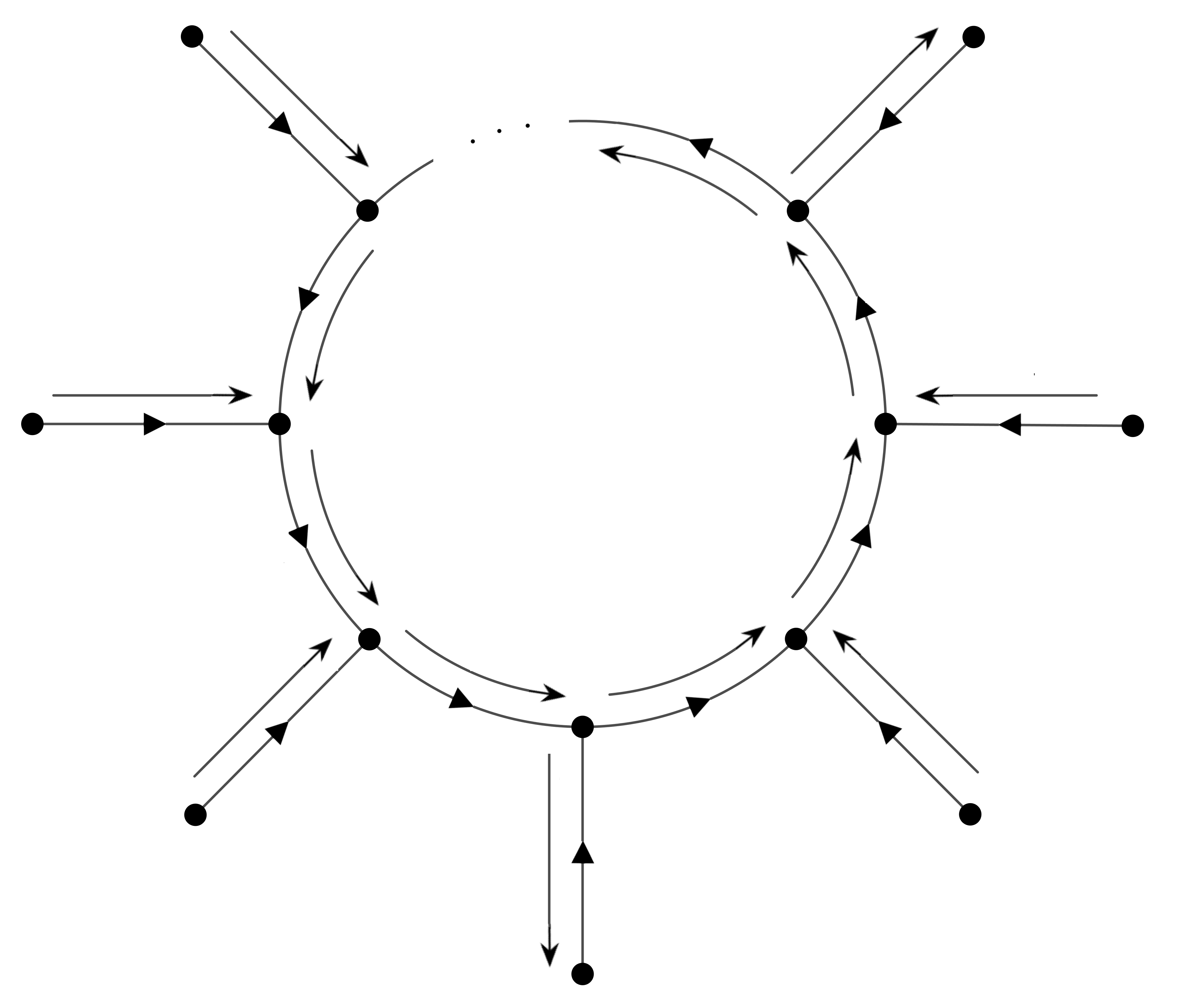}
\put(-182,130){\makebox(0,0)[lb]{\small $m_n$}}
\put(-95,166){\makebox(0,0)[lb]{\small $m_6$}}
\put(-55,130){\makebox(0,0)[lb]{\small $m_5$}}
\put(-55,85){\makebox(0,0)[lb]{\small $m_4$}}
\put(-95,45){\makebox(0,0)[lb]{\small $m_3$}}
\put(-141,45){\makebox(0,0)[lb]{\small $m_2$}}
\put(-182,85){\makebox(0,0)[lb]{\small $m_1$}}
\put(-155,125){\makebox(0,0)[lb]{\small $q_n$}}
\put(-98,146){\makebox(0,0)[lb]{\small $q_6$}}
\put(-77,125){\makebox(0,0)[lb]{\small $q_5$}}
\put(-77,90){\makebox(0,0)[lb]{\small $q_4$}}
\put(-98,66){\makebox(0,0)[lb]{\small $q_3$}}
\put(-136,68){\makebox(0,0)[lb]{\small $q_2$}}
\put(-155,90){\makebox(0,0)[lb]{\small $q_1$}}
\put(-198,120){\makebox(0,0)[lb]{\small $p_n$}}
\put(-156,170){\makebox(0,0)[lb]{\small $p_{n-1}$}}
\put(-75,170){\makebox(0,0)[lb]{\small $p_5$}}
\put(-35,120){\makebox(0,0)[lb]{\small $p_4$}}
\put(-44,55){\makebox(0,0)[lb]{\small $p_3$}}
\put(-132,20){\makebox(0,0)[lb]{\small $p_2$}}
\put(-184,55){\makebox(0,0)[lb]{\small $p_1$}}
\caption{Example of a one-loop Feynman diagram with $m=n$ external legs.}
\label{fig: nbox}
\end{figure}
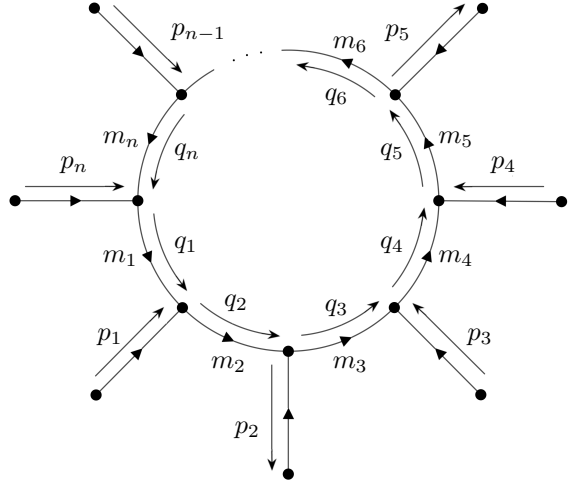

\noindent
Its Symanzik polynomials are
\begin{gather}\label{eq: 1Sym}
\mathcal{U}_{\text{1-loop}} = \sum_{j=1}^n x_j,
\\
\mathcal{F}_{\text{1-loop}} = \mathcal{U}_{\text{1-loop}} \sum_{j=1}^n m_j^2 x_j
+ \sum_{\substack{i,j=1 \\ i < j}}^n (q_i-q_j)^2 x_i x_j,
\end{gather}
where the internal momenta are given by $q_1=k$, $q_i=k+p_1+\dots+p_{i-1}$ for $1 < i \leq n$. Here,~$k$~is the unique loop momentum of the graph, and $p_1+\dots+p_m=0$ by global momentum conservation.
\end{Example}

\subsection{Graph polynomials}
Re-expression of Feynman integrals in parametric form shows that the correspondence bet\-ween scalar Feynman diagrams and Feynman integrals can be reformulated in different terms.
The~infor\-mation contained in a Feynman graph is shared out among its multiple components, which can be identified as the underlying graph structure, the directionality of edges and the various edge labels. If~we destructure a Feynman graph in these layers and momentarily neglect the extra information apart from the graph structure, we observe that its integral is insensitive to changes of the graph which leave its topology unaltered.
Focusing on the underlying graph topology, the Symanzik polynomials can be suitably re-interpreted and they are commonly called \textit{graph polynomials} in this context.

Let $G$ be a finite graph without isolated vertices. $G$ is specified by the pair $(V_G, E_G)$, where $V_G$ is the collection of vertices and $E_G$ is the collection of edges. We~choose an arbitrary orientation of its edges and define the map
\begin{align}
\mathbb{Z}^{E_G} &\longrightarrow \mathbb{Z}^{V_G},
\\
e &\longmapsto t(e)-s(e),
\end{align}
where $e \in E_G$ is an edge and $s(e),t(e) \in V_G$ are its \textit{source} and \textit{target} endpoints with respect to the edge orientation. Let~us extend this map to the exact sequence
\begin{gather}
0 \rightarrow H_1(G, \mathbb{Z}) \rightarrow \mathbb{Z}^{E_G} \rightarrow \mathbb{Z}^{V_G} \rightarrow H_0(G, \mathbb{Z}) \rightarrow 0,
\end{gather}
where $H_0(G, \mathbb{Z})$ and $H_1(G, \mathbb{Z})$ are the zeroth and first homology groups of the graph.
As~a~con\-sequence, the graph loop number $l_G$ is related to the number of edges $n_G$, the number of verti\-ces~$v_G$, and the number of connected components $c_G$ by\footnote{The loop number is equivalently defined as the rank of the first homology group of the graph, while the number of connected components corresponds to the rank of the zeroth homology group of the graph.}
\begin{gather}
l_G = \text{rank}(H_1(G, \mathbb{Z}))
 = |E_G |- |V_G| + \text{rank}(H_0(G, \mathbb{Z}))
 = n_G - v_G + c_G.
\end{gather}

Assume $G$ is a graph of Feynman type, that is, finite, connected and one-particle irreducible. Let~the \textit{valence} of a vertex be the number of edges attached to it.
Since they do not contribute to~the braid pattern of Feynman graphs, both vertices of valence one, corresponding to the source endpoints of external legs, and vertices of valence two, corresponding to mass insertions, do not play a role here.
To such a graph $G$ we wish to assign an integral $I_G$ which corresponds to the one previously defined in~\eqref{eq: I_G_new} when the neglected extra information is re-inserted. We~start by~associating a variable $x_e$ to every internal edge $e \in E_G$ of the graph. These variables are known as \textit{Schwinger parameters} and they are the graph-theoretic analogues of Feynman parameters. Let~$\mathcal{T}_1$ be the set of \textit{spanning trees}\footnote{A graph of zero loop number with $k$ connected components is called a $k$-forest. When $k=1$, the forest is called a tree.
Given an arbitrary connected graph $G$, a spanning $k$-forest of $G$ is a subgraph $T \subseteq G$ such that $V_T=V_G$ and $T$ is a $k$-forest. A~spanning $k$-forest of $G$ is usually denoted by the collection of its trees. A~connected graph has always at least one spanning tree.} of $G$.
The \textit{first graph polynomial} of $G$ is defined as
\begin{gather} \label{eq: Psi_G}
\Psi_G = \sum_{\substack{T \in \mathcal{T}_1}} \prod_{e \notin E_T} x_e.
\end{gather}
It is a homogeneous polynomial of degree $l_G$ in the Schwinger parameters. Note that each monomial of $\Psi_G$ has coefficient one, and $\Psi_G$ is linear in each Schwinger parameter.

\begin{Example}
The first graph polynomial of the Feynman graph shown in Fig.~\ref{fig: loops} is $\Psi_G = x_1 \cdots x_n \big( \frac{1}{x_1}+\dots+ \frac{1}{x_n}\big)$.
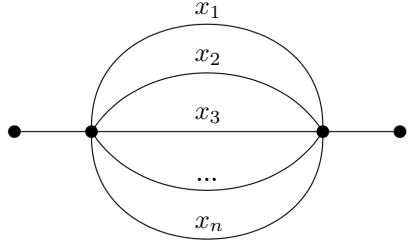
\begin{figure}[htb!]
\centering
\includegraphics[scale=.37]{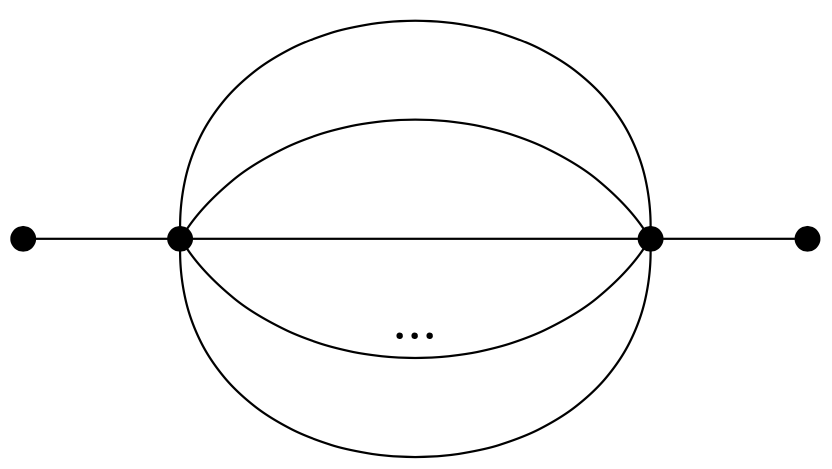}
\put(-82,86){\makebox(0,0)[lb]{\small $x_1$}}
\put(-82,68){\makebox(0,0)[lb]{\small $x_2$}}
\put(-82,46){\makebox(0,0)[lb]{\small $x_3$}}
\put(-82,5){\makebox(0,0)[lb]{\small $x_n$}}
\caption{Example of a scalar Feynman graph with $n$ internal propagators.}
\label{fig: loops}
\end{figure}
\end{Example}

By construction, the first Symanzik polynomial $\mathcal{U}$ of a Feynman graph $G$ does not depend on momenta and masses involved in the diagram, but is only dependent on the graph topology. Indeed, it explicitly identifies with the first graph polynomial $\Psi_G$ of the corresponding pure graph structure. The same is not true for the second Symanzik polynomial $\mathcal{F}$, which is a~function of~external momenta and internal masses. However, we can re-express $\mathcal{F}$ in a way that clearly separates the contribution to $\mathcal{F}$ coming from the graph topology from its other
dependences. To this end, momenta and masses edge labels must re-enter our
discussion. Let~$\mathcal{T}_2$ be the set of \textit{spanning $2$-forests} of $G$ and $P_{T_i}$ be the set of external momenta of $G$ attached to its tree $T_i$.
The \textit{second graph polynomial} of $G$ is defined as
\begin{gather}
\Xi_G (\{p_j, m_e\}) = \bigg( \sum_{e \in E_G} m_e^2 x_e \bigg) \Psi_G \; - \sum_{(T_1,T_2) \in \mathcal{T}_2} \bigg( \prod_{e \notin E_{T_1} \cup E_{T_2}} x_e \bigg) \Bigg(\sum_{\substack{p_j \in P_{T_1} \\ p_k \in P_{T_2}}} p_j \cdot p_k \Bigg).
\end{gather}
It is a homogeneous polynomial of degree $l_G+1$ in the Schwinger parameters. Note that, if all internal masses are zero, then $\Xi_G$ is linear in each Schwinger parameter. It~follows from their definitions that the second Symanzik polynomial and the second graph polynomial of a~Feynman graph are, indeed, the same. Moreover, having fixed the momenta of external particles and the masses of internal propagators, we are left with the explicit dependence of $\mathcal{F}$ on the graph structure given in terms of spanning $2$-forests.

\begin{figure}[htb!]
\centering
\subfloat[Full Feynman diagram]{\includegraphics[scale=.35]{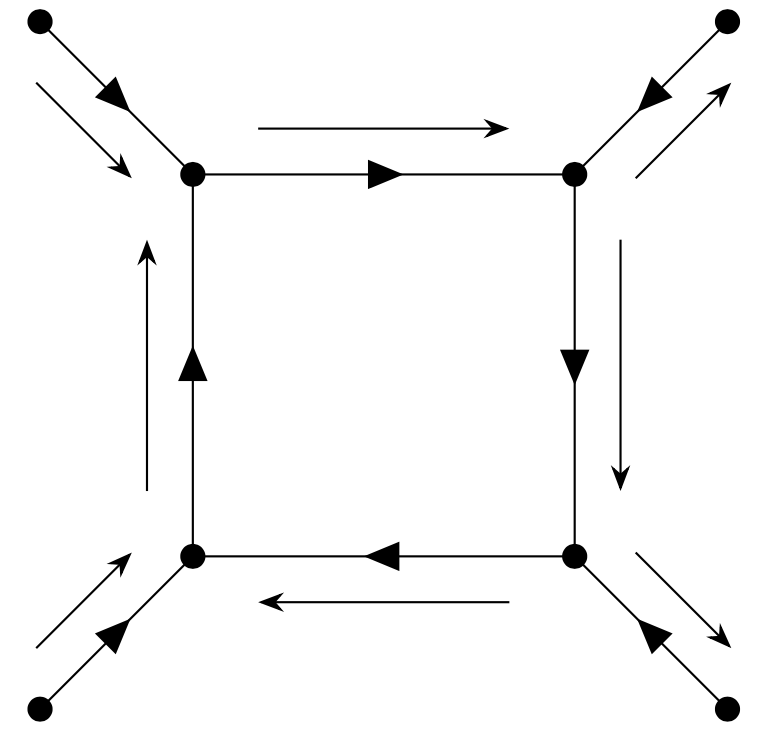}
\put(-70,111){\makebox(0,0)[lb]{\small $q_3$}}
\put(-120,62){\makebox(0,0)[lb]{\small $q_2$}}
\put(-22,62){\makebox(0,0)[lb]{\small $q_4$}}
\put(-70,14){\makebox(0,0)[lb]{\small $q_1$}}
\put(-71,88){\makebox(0,0)[lb]{\small $m_3$}}
\put(-96,62){\makebox(0,0)[lb]{\small $m_2$}}
\put(-50,62){\makebox(0,0)[lb]{\small $m_4$}}
\put(-71,36){\makebox(0,0)[lb]{\small $m_1$}}
\put(-14,98){\makebox(0,0)[lb]{\small $p_3$}}
\put(-128,98){\makebox(0,0)[lb]{\small $p_2$}}
\put(-14,26){\makebox(0,0)[lb]{\small $p_4$}}
\put(-128,26){\makebox(0,0)[lb]{\small $p_1$}}}
\qquad\qquad
\subfloat[Underlying graph structure]{\includegraphics[scale=.35]{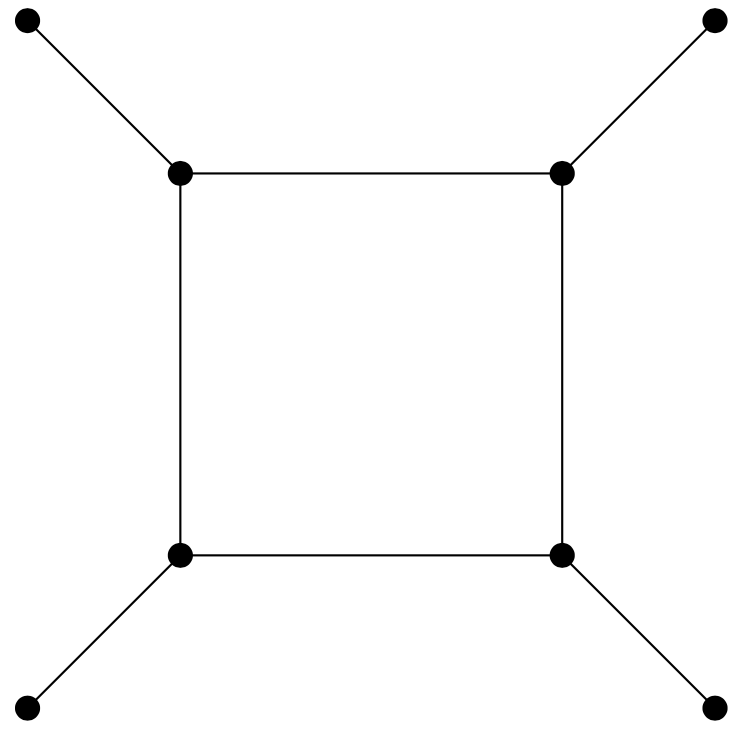}
\put(-70,102){\makebox(0,0)[lb]{\small $x_3$}}
\put(-110,62){\makebox(0,0)[lb]{\small $x_2$}}
\put(-28,62){\makebox(0,0)[lb]{\small $x_4$}}
\put(-70,22){\makebox(0,0)[lb]{\small $x_1$}}}
\caption{Box diagram with four legs.}
\label{fig: box}
\end{figure}
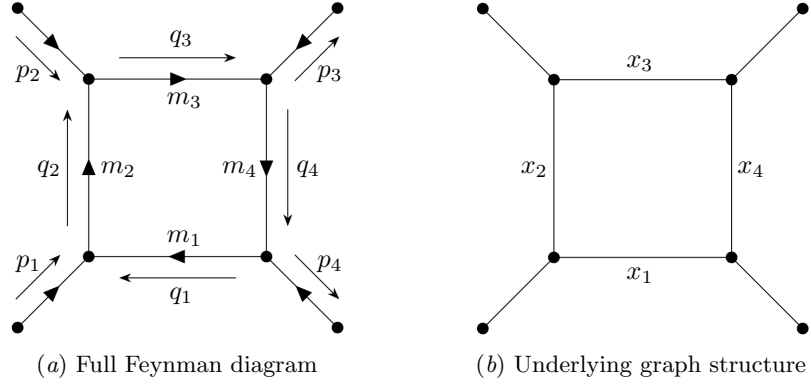

\begin{Example} \label{ex: box}
To explicitly see how the individual terms in the graph polynomials arise from the knot structure of the diagram, we look closer at the one-loop Feynman graph with $m=4$ external legs, also called \textit{box diagram},\footnote{This gives a next-to-leading order contribution to the two-to-two particle scattering process. Srednicki~\cite{Sre10} gives a detailed discussion of two particles elastic scattering at one-loop using standard methods in perturbative quantum field theory.} which is shown in Fig.~\ref{fig: box}.
Its Symanzik polynomials are
\begin{gather}\label{eq: box}
\mathcal{U}_{\text{box}} = x_1+x_2+x_3+x_4,
\\
\mathcal{F}_{\text{box}} = \big[(x_1+x_2+x_3+x_4)\big(m_1^2x_1+m_2^2x_2+m_3^2x_3+m_4^2x_4\big)+ x_1x_2 p_1^2+ x_2x_3 p_2^2 + x_3x_4 p_3^2
\\ \hphantom{\mathcal{F}_{\text{box}} =}
{} + x_4x_1 p_4^2+ x_1x_3 (p_1+p_2)^2 + x_2x_4 (p_2+p_3)^2\big].
\end{gather}
Neglecting mass terms, the remaining monomials correspond to the spanning forests shown in~Figs.~\ref{fig: trees} and~\ref{fig: forests}.
\begin{figure}[htb!]
\centering
\subfloat[$+x_1$]{\includegraphics[width=0.19\textwidth]{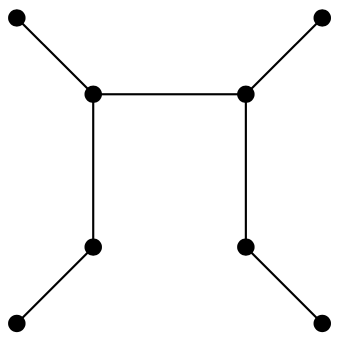}
\put(-48,65){\makebox(0,0)[lb]{\small $x_3$}}
\put(-75,39){\makebox(0,0)[lb]{\small $x_2$}}
\put(-20,39){\makebox(0,0)[lb]{\small $x_4$}}
} \qquad
\subfloat[$+x_2$]{\includegraphics[width=0.19\textwidth]{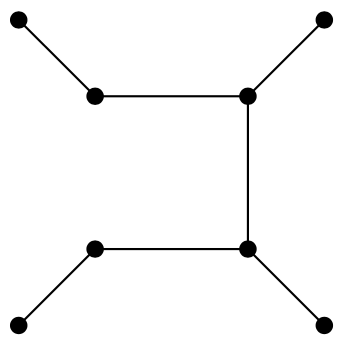}
\put(-48,65){\makebox(0,0)[lb]{\small $x_3$}}
\put(-20,39){\makebox(0,0)[lb]{\small $x_4$}}
\put(-48,12){\makebox(0,0)[lb]{\small $x_1$}}} \qquad
\subfloat[$+x_3$]{\includegraphics[width=0.19\textwidth]{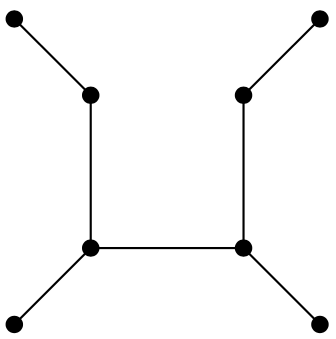}
\put(-75,39){\makebox(0,0)[lb]{\small $x_2$}}
\put(-20,39){\makebox(0,0)[lb]{\small $x_4$}}
\put(-48,12){\makebox(0,0)[lb]{\small $x_1$}}} \qquad
\subfloat[$+x_4$]{\includegraphics[width=0.19\textwidth]{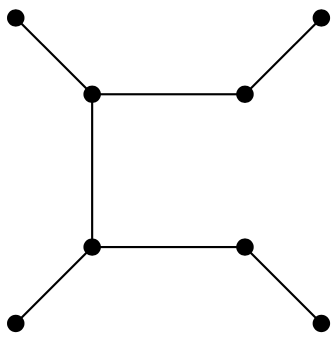}
\put(-48,65){\makebox(0,0)[lb]{\small $x_3$}}
\put(-75,39){\makebox(0,0)[lb]{\small $x_2$}}
\put(-48,12){\makebox(0,0)[lb]{\small $x_1$}}}
\caption{Spanning trees in the box diagram with four legs and corresponding terms in $\mathcal{U}_{\text{box}}$.}
\label{fig: trees}
\end{figure}
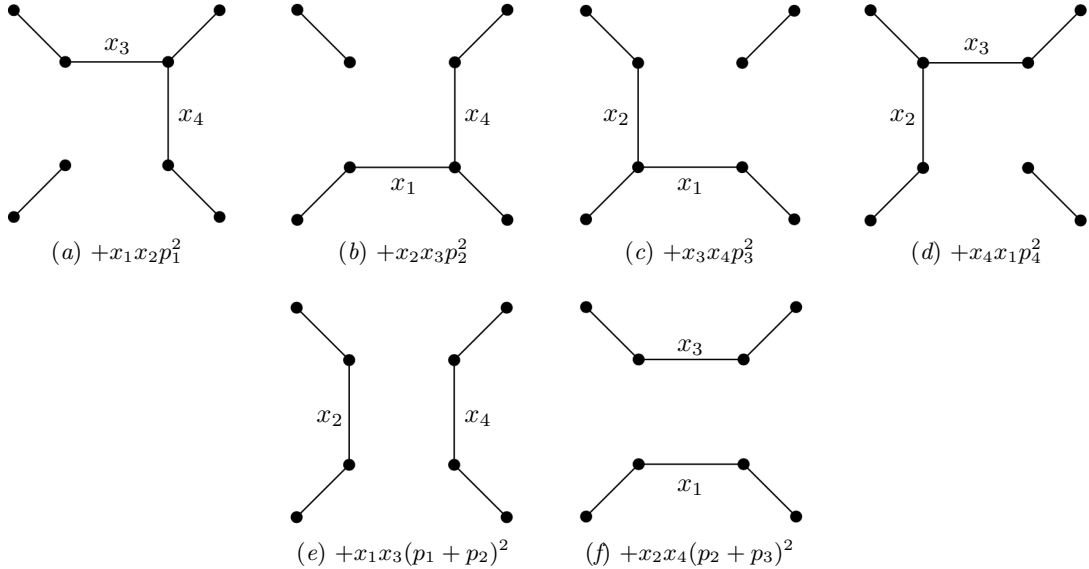
\begin{figure}[htb!]
\centering
\subfloat[$+x_1 x_2 p_1^2$]{\includegraphics[width=0.19\textwidth]{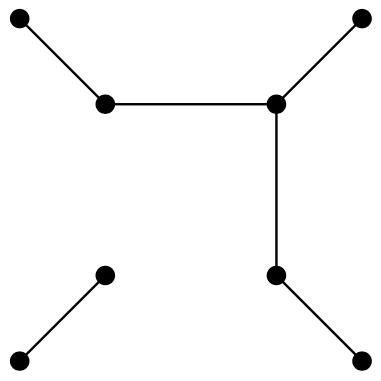}
\put(-20,39){\makebox(0,0)[lb]{\small $x_4$}}
\put(-48,66){\makebox(0,0)[lb]{\small $x_3$}}} \qquad
\subfloat[$+x_2 x_3 p_2^2$]{\includegraphics[width=0.19\textwidth]{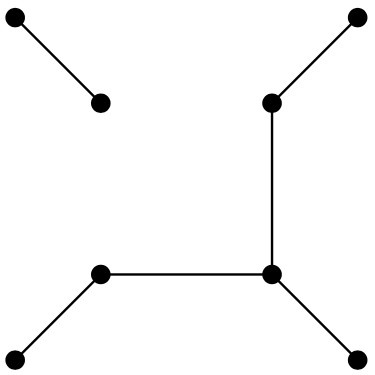}
\put(-20,39){\makebox(0,0)[lb]{\small $x_4$}}
\put(-48,13){\makebox(0,0)[lb]{\small $x_1$}}} \qquad
\subfloat[$+x_3 x_4 p_3^2$]{\includegraphics[width=0.19\textwidth]{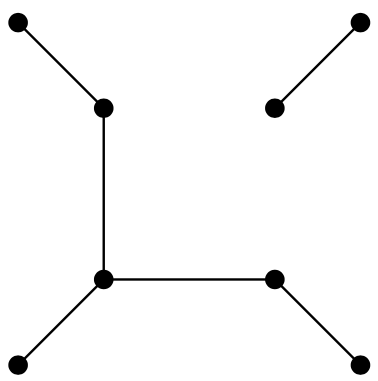}
\put(-76,39){\makebox(0,0)[lb]{\small $x_2$}}
\put(-48,13){\makebox(0,0)[lb]{\small $x_1$}}} \qquad
\subfloat[$+x_4 x_1 p_4^2$]{\includegraphics[width=0.19\textwidth]{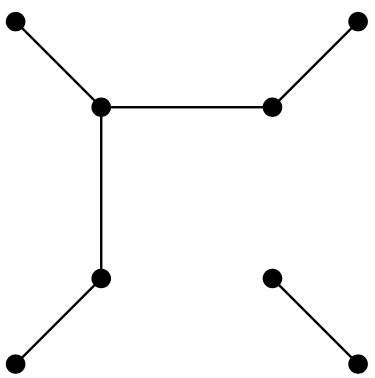}
\put(-48,66){\makebox(0,0)[lb]{\small $x_3$}}
\put(-76,39){\makebox(0,0)[lb]{\small $x_2$}}} \\
\subfloat[$+x_1 x_3 (p_1+p_2)^2$]{\includegraphics[width=0.19\textwidth]{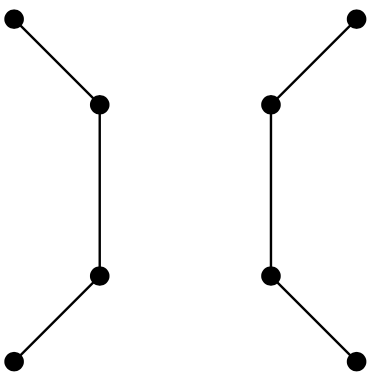}
\put(-20,39){\makebox(0,0)[lb]{\small $x_4$}}
\put(-76,39){\makebox(0,0)[lb]{\small $x_2$}}} \qquad
\subfloat[$+x_2 x_4 (p_2+p_3)^2$]{\includegraphics[width=0.19\textwidth]{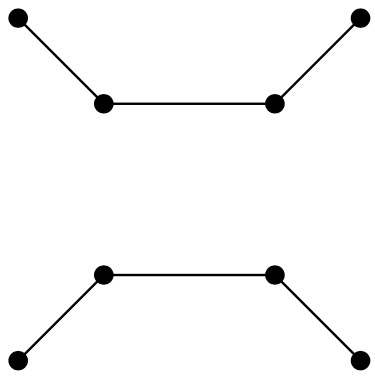}
\put(-48,66){\makebox(0,0)[lb]{\small $x_3$}}
\put(-48,13){\makebox(0,0)[lb]{\small $x_1$}}}
\caption{Spanning $2$-forests in the box diagram with four legs and corresponding terms in~$\mathcal{F}_{\text{box}}$.}
\label{fig: forests}
\end{figure}
\end{Example}
Thus, the Symanzik or graph polynomials capture the algebraic information contained in the topology of a Feynman diagram and they prove to be the first tool to be used in the tentative investigation of renormalisation theory via the algebraic manipulation of concatenated one-loop integrals. For a more detailed overview of the properties of Feynman graph polynomials we refer to Bogner and Weinzierl~\cite{BW10}.

\subsection{Primitive log-divergent graphs} \label{sec: primitive}
The parametric Feynman integral in~\eqref{eq: I_G_new} can be written in a slightly different notation, which turns out to be particularly useful henceforth.
Neglecting prefactors and assuming $D=4$, it is equivalent to the \textit{projective integral}
\begin{gather}\label{eq: I_G}
I_G(\{p_j,m_e\}) = \int_{\sigma} \frac{\Omega}{\Psi_G^2} \bigg(\frac{\Psi_G}{\Xi_G(\{p_j,m_e\})} \bigg)^{n_G-2l_G},
\end{gather}
where $\sigma$ is the real projective simplex given by
\begin{gather}
\sigma = \big\{ [x_1:\dots:x_{n_G}] \in \mathbb{P}^{n_G-1}(\mathbb{R}) \,|\, x_e \ge 0,\, e=1,\dots,n_G \big\}
\end{gather}
and $\Omega$ is the top-degree differential form on $\mathbb{P}^{n_G-1}$ expressed in local coordinates as
\begin{gather}
\Omega = \sum_{e=1}^{n_G}(-1)^e x_e \, {\rm d}x_1 \wedge \dots \wedge \widehat{{\rm d}x_e} \wedge \dots \wedge {\rm d}x_{n_G}.
\end{gather}
One can check that the integrand is homogeneous of degree zero, so that the integral in projective space is well-defined and equivalent, under the affine constraint $x_{n_G}=1$, to the previous parametric integral in affine space.
Integral~\eqref{eq: I_G} is in general divergent, as singularities may arise if the zero sets of the graph polynomials $\Psi_G$ and $\Xi_G$ intersect the domain of integration.

Graphs satisfying the condition $n_G=2l_G$ are called \textit{logarithmically divergent} and constitute a particularly interesting class of graphs. In fact, their Feynman integral simplifies to
\begin{gather}\label{eq: I_G_last}
I_G= \int_{\sigma} \frac{\Omega}{\Psi_G^2},
\end{gather}
where the dependence on the second Symanzik polynomial, and consequently on momenta and masses, has vanished. Being uniquely sensitive to the graph topology, such a Feynman graph describes a so-called \textit{single-scale process}.\footnote{Among other contexts, the feature of no-scaling also occurs in the evaluation of Feynman diagrams concerning the anomalous magnetic moment of the electron, as presented by Laporta and Remiddi~\cite{LR96}.}
For a logarithmically divergent graph $G$, we define the \textit{graph hypersurface} as the zero set of its first Symanzik polynomial
\begin{gather} \label{eq: X_G}
X_G = \big\{[x_1:\dots:x_{n_G}] \in \mathbb{P}^{n_G-1} \,|\, \Psi_G(x_1,\dots,x_{n_G})=0 \big\}
\end{gather}
which describes the singularities of its Feynman integral $I_G$.
The following theorem on the convergence of logarithmically divergent graphs is proven by Bloch, Esnault and Kreimer~\cite{BEK06}.

\begin{figure}[htb!]
\centering
\subfloat[][\emph{$I_G=6 \zeta(3)$}]
 {\includegraphics[width=0.18\textwidth]{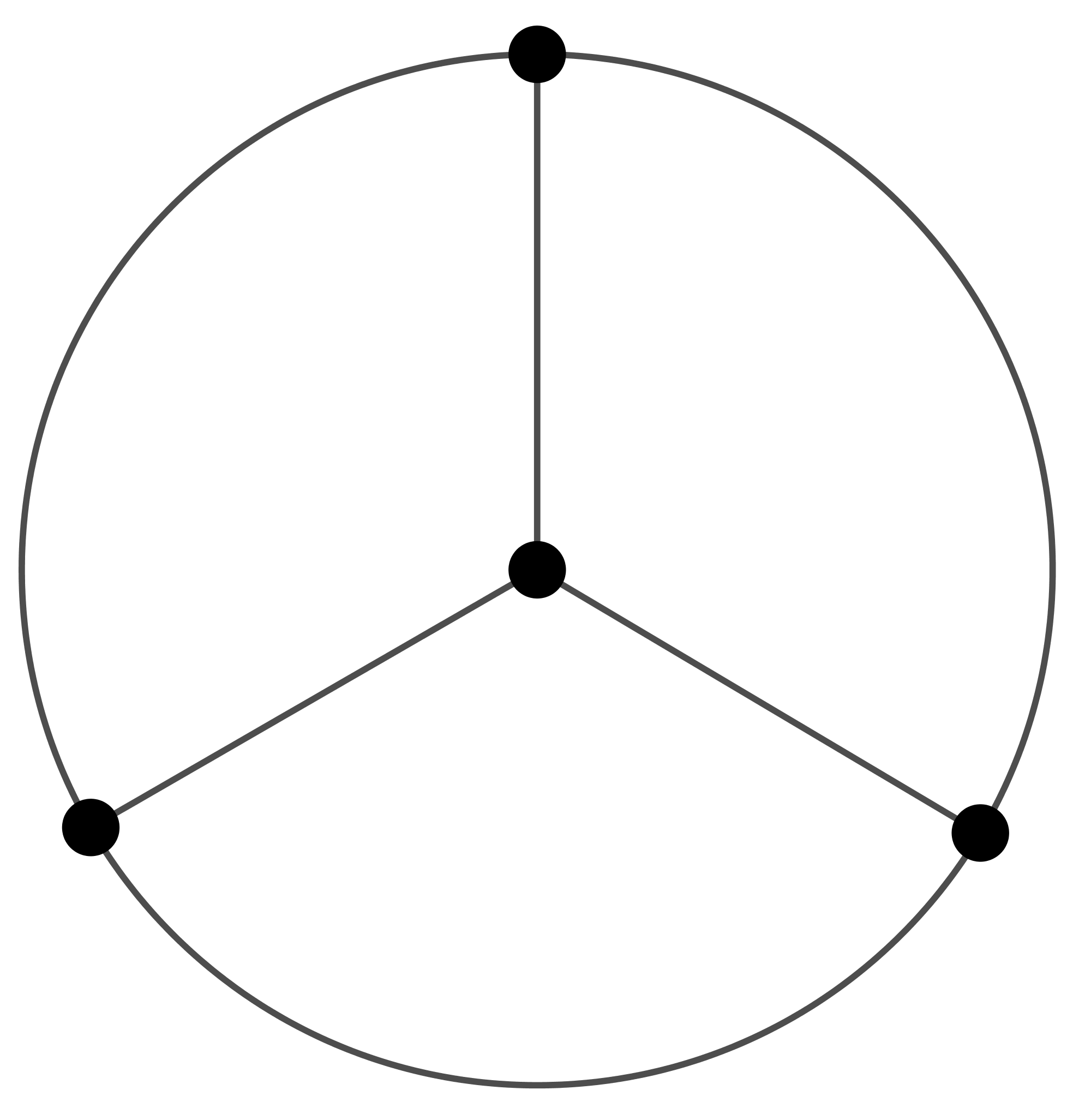}} \qquad
\subfloat[][\emph{$I_G=20 \zeta(5)$}]
 {\includegraphics[width=0.18\textwidth]{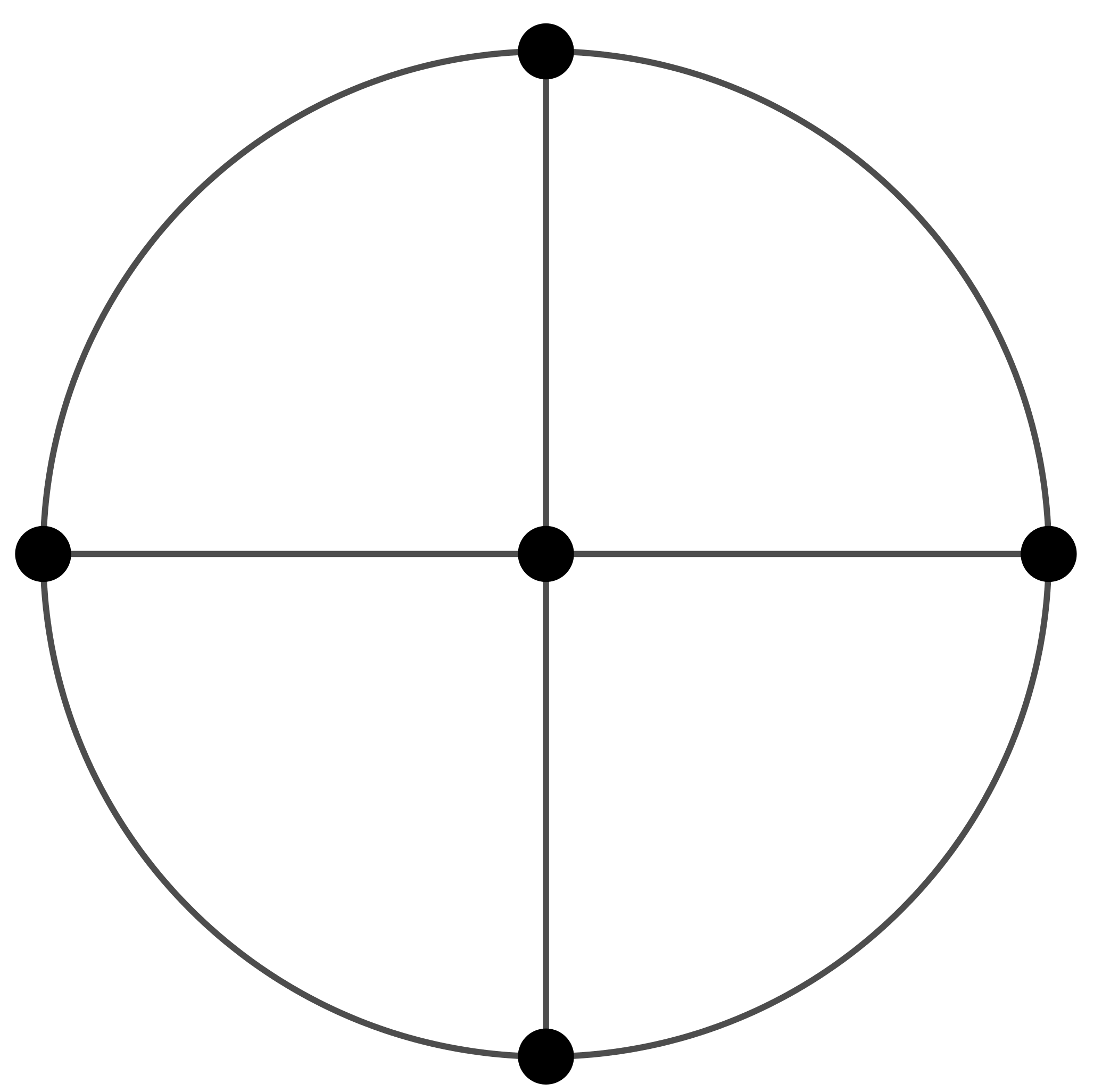}} \qquad
\subfloat[][\emph{$I_G=36 \zeta(3)^2$}]
 {\includegraphics[width=0.18\textwidth]{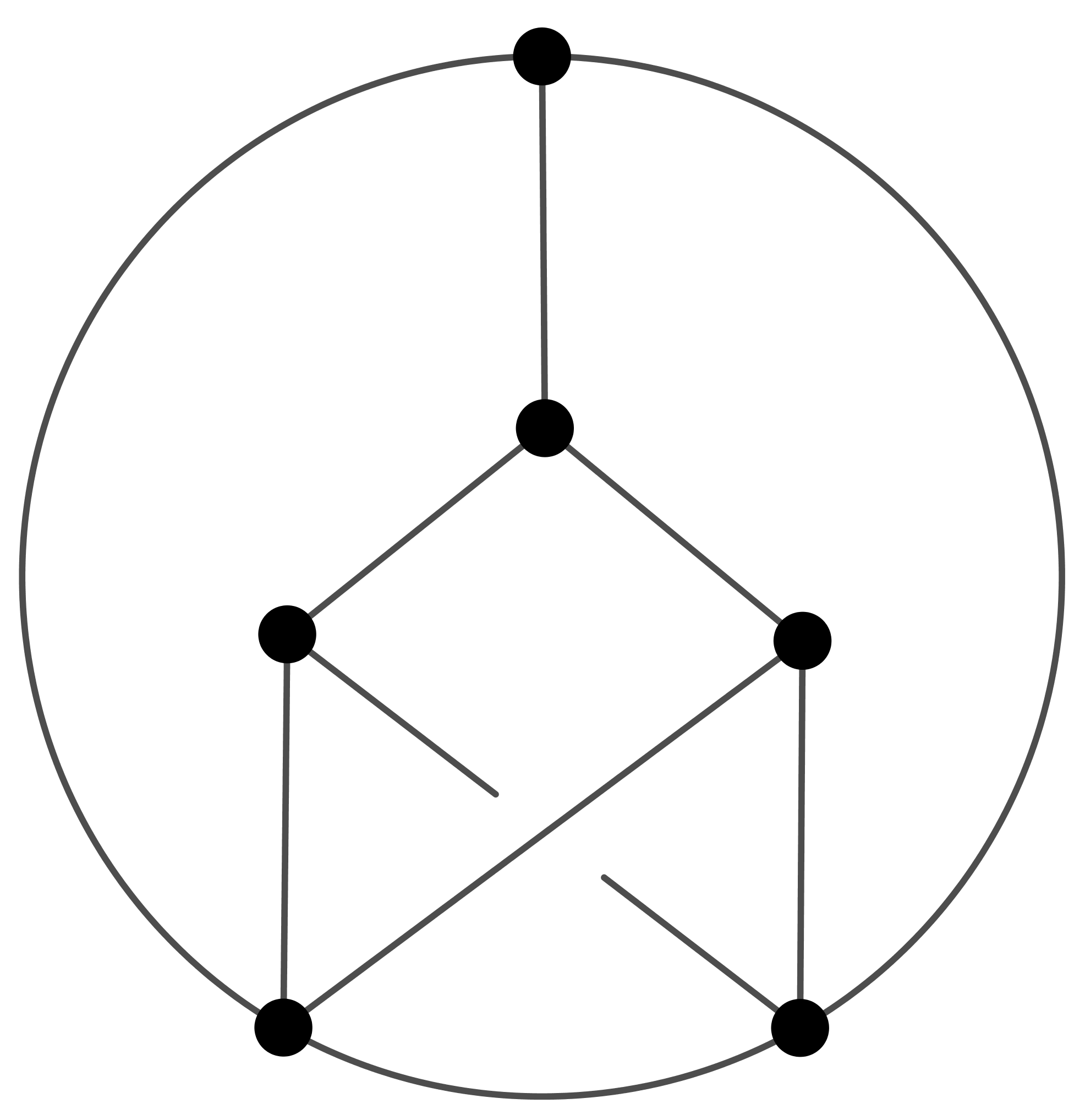}} \qquad
\subfloat[][\emph{$I_G=32 P_{3,5}$}]
 {\includegraphics[width=0.18\textwidth]{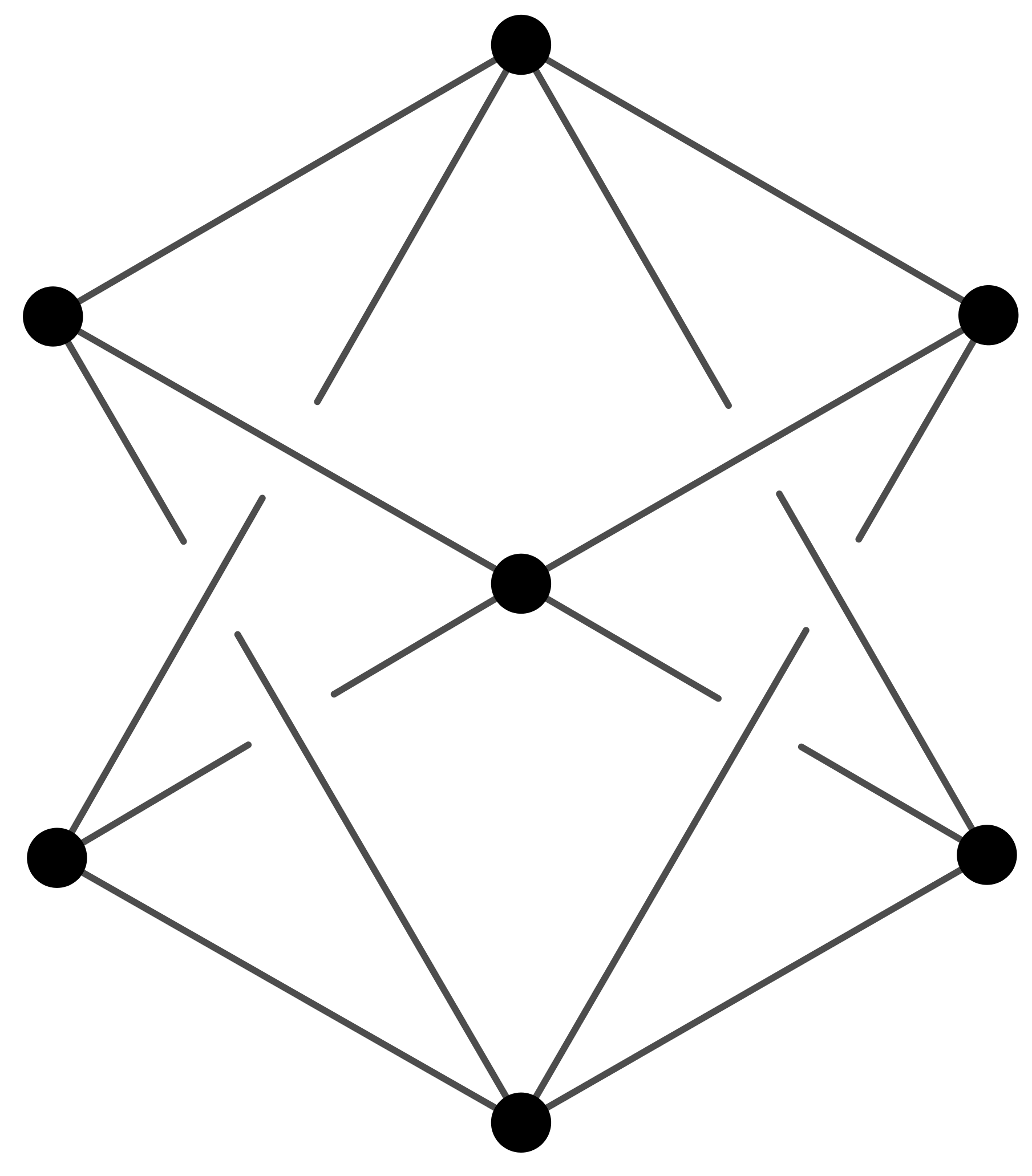}}
\caption{Examples of $\phi^4$-graphs with 3, 4, 5 and 6 loops.}
\label{fig: phi4}
\end{figure}

\begin{Theorem}\label{th: log}
Let $G$ be logarithmically divergent. The integral $I_G$ converges if and only if every proper subgraph $\varnothing \ne \gamma \subset G$ satisfies the condition $n_{\gamma} > 2 l_{\gamma}$.
\end{Theorem}
A logarithmically divergent graph $G$ such that $I_G$ is convergent is called \textit{primitive log-divergent}, or simply \textit{primitive}. We~give particular attention to the class of primitive log-divergent graphs in scalar massless $\phi^4$ quantum field theory. They are called \textit{$\phi^4$-graphs}, and have vertices with valence at most four.
Feynman amplitudes in $\phi^4$ theory have been computed to much higher loop orders than most other quantum field theories thanks to the work of Broadhurst and Kreimer~\cite{BK95,BK97}, and Schnetz~\cite{Sch10}. Some of the simplest $\phi^4$-graphs are shown in Fig.~\ref{fig: phi4} along with the values of the associated Feynman integrals. Here, $\zeta$ is the Riemann zeta function, and $P_{3,5}=-\frac{216}{5} \zeta(3,5) - 81 \zeta(5)\zeta(3) + \frac{522}{5} \zeta(8)$.

\subsection{Multiple zeta values} \label{sec: mzvs}
The Riemann zeta function is defined on the half-plane of complex numbers $s \in \mathbb{C}$ with $\mathop{\rm Re}(s) > 1$ by the absolutely convergent series
\begin{gather} \label{eq: zeta}
\zeta(s) = \sum_{n=1}^{\infty} \frac{1}{n^s}
\end{gather}
and extended to a meromorphic function on the whole complex plane with a single pole at $s = 1$.
The first tentative attempts to find polynomial relations among zeta values by multiplying terms of the form~\eqref{eq: zeta} have led to a generalisation of the notion of Riemann zeta value.
Multiple zeta values, or MZVs, are the real numbers
\begin{gather} \label{eq: zsum}
\zeta(s_1,\dots,s_l) = \sum_{n_1 > n_2 > \dots > n_l \ge 1} \frac{1}{n_1^{s_1} \cdots n_l^{s_l}}
\end{gather}
associated to tuples of integers $\mathbf{s}=(s_1,\dots,s_l)$, called \textit{multi-indices}. To guarantee the convergence of the infinite series, only multi-indices such that $s_i \ge 1$ for $i=1,\dots,l$ and $s_1 \ge 2$ are considered. They are called \textit{admissible} multi-indices. The integers $\mathop{\rm wt}(\mathbf{s})=s_1+\dots+s_l$ and $l$ are called \textit{weight} and \textit{length} of the multi-index $\mathbf{s}$, respectively.

Following the early observations that products of two zeta values are $\mathbb{Q}$-linear combinations of zeta and double zeta values, and that products of more than two zeta values are analogously expressed in terms of multiple zeta values of higher length, linear relations among MZVs have been the object of a more and more extensive investigation by many mathematicians, including Brown, Cartier, Deligne, Drinfeld, \'{E}calle, Goncharov, Hain, Hoffman, Kontsevich, Terasoma, Zagier, Broadhurst and Kreimer. Indeed, the $\mathbb{Q}$-linear relations among multiple zeta values directly provide insights on the widely sought-after algebraic relations among Riemann zeta values.

The $\mathbb{Q}$-vector space spanned by multiple zeta values forms an algebra under the so-called \textit{stuffle product}.
Analytic methods, like partial fraction expansions, provide only a few of the known relations among MZVs. Many more are obtained, although conjecturally, by performing extensive numerical experiments, as described by Bl\"{u}mlein et al~\cite{BBV10}. However, enormous progress followed the analytic discovery of a crucial feature of multiple zeta values, that is, beside their representation as infinite series, they admit an alternative representation as iterated integrals over simplices of weight-dimension. Let~$\Delta^p= \{(t_1,\dots,t_p) \in \mathbb{R}^p \,|\, 1 \ge t_1 \ge t_2 \ge \dots \ge t_p \ge 0 \}$ and define the measures on the open interval $(0,1)$
\begin{gather}
\omega_0(t)=\frac{{\rm d}t}{t} , \qquad \omega_1(t) = \frac{{\rm d}t}{1-t}.
\end{gather}
If $\mathbf{s}$ is an admissible multi-index, write $r_i=s_1+\dots+s_i$ for each $i=1,\dots,l$ and set $r_0=0$. Define the measure $\omega_{\mathbf{s}}$ on the interior of the simplex $\Delta^{\mathop{\rm wt}(\mathbf{s})}$ by
\begin{gather}
\omega_{\mathbf{s}} = \prod_{i=1}^l \underbrace{\omega_0(t_{r_{i-1}+1}) \cdot \cdot \cdot \omega_0(t_{r_{i}-1})}_{s_i - 1 \text{ times }}\omega_1(t_{r_i}).
\end{gather}
The following insight is due to Kontsevich.
\begin{Proposition}
Let $\mathbf{s}=(s_1,\dots,s_l)$ be an admissible multi-index. The multiple zeta value~$\zeta(\mathbf{s})$ can be obtained by the convergent improper integral
\begin{gather} \label{eq: zint}
\zeta(\mathbf{s}) = \zeta(s_1,\dots,s_l) = \int_{\Delta^{\mathop{\rm wt}(\mathbf{s})}} \omega_{\mathbf{s}}.
\end{gather}
\end{Proposition}

\looseness=1
This different way of writing multiple zeta values yields a new algebra structure associated with the so-called \textit{shuffle product}. Many other linear relations among MZVs are obtained systematically in this alternative framework.
However, relations are also and most interestingly derived by the comparison of the two representations given by~\eqref{eq: zsum} and~\eqref{eq: zint}.
The coexistence of the stuffle and shuffle algebra structures on the $\mathbb{Q}$-vector space of MZVs proved to be the most productive source of information about these numbers.
For a more detailed discussion of the classical theory of multiple zeta values we refer to Fres\'an and Burgos Gil~\cite{FG}.
Making concrete use of the many $\mathbb{Q}$-linear relations that MZVs are known to satisfy is not an easy task, particularly at high weights. For example, there is no algorithm that leads to the reduction of~any given MZV into a chosen $\mathbb{Q}$-basis. A~boost in our understanding originated from the exact-numerical decomposition algorithm by Brown~\cite{Brown2012}. Developed from the non-classical perspective of the theory of motives, it conjecturally provides a general strategy to handle MZVs, and more generally polylogarithmic numbers, by converting them into the so-called \textit{f-alphabet}.\footnote{The elements of the shuffle algebra on the $\mathbb{Q}$-vector space of MZVs are interpreted as words in letters of certain weights. Precisely, there is one letter for each odd weight greater than 1. Words in these letters span finite-dimensional subspaces of definite weight. Notice, however, that the conversion into the $f$-alphabet depends on the choice of algebra basis.}

\begin{figure}[htb!]
\centering
\subfloat[][\emph{$l=5$}]
 {\includegraphics[scale=0.3]{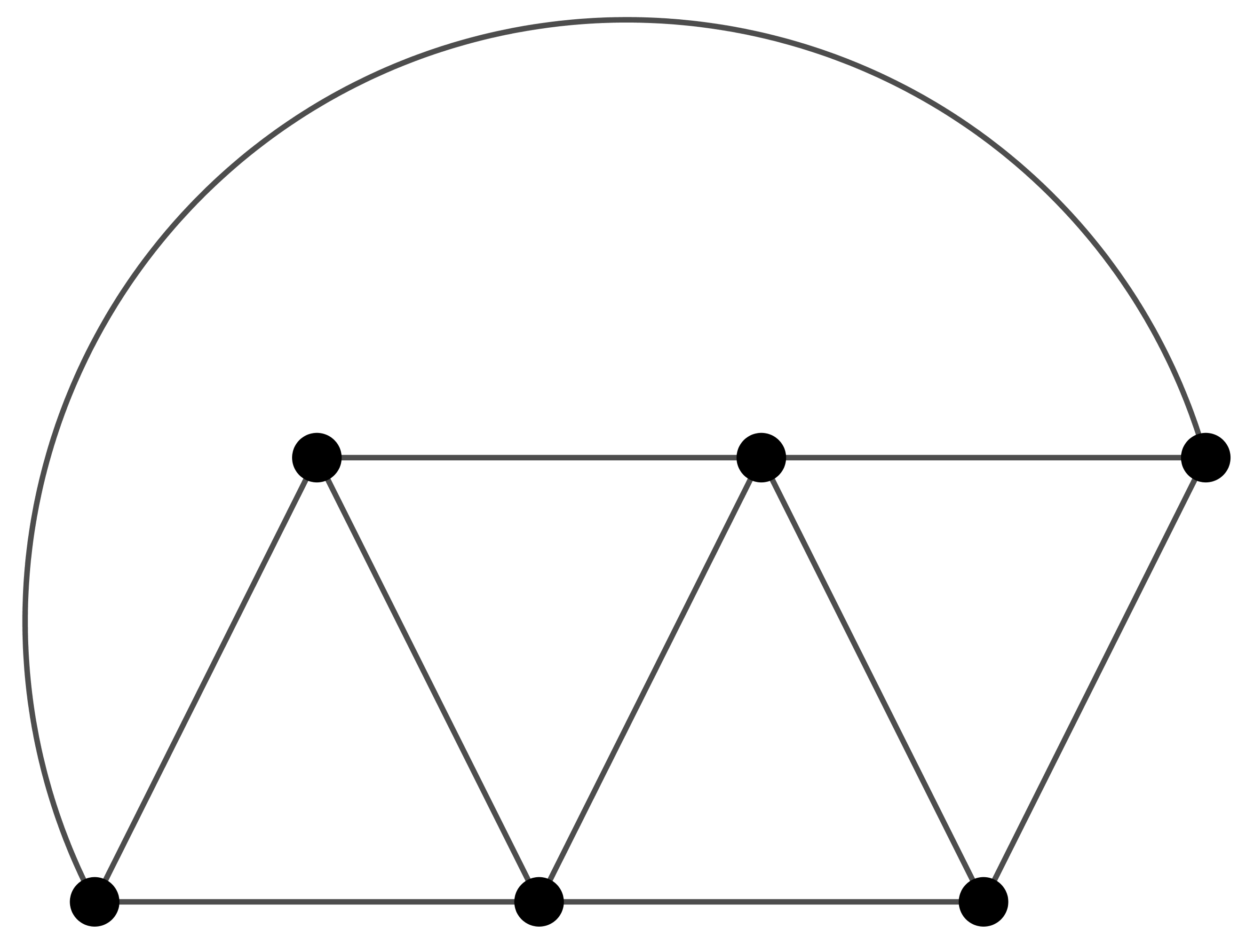}} \qquad\qquad
\subfloat[][\emph{$l=6$}]
 {\includegraphics[scale=0.3]{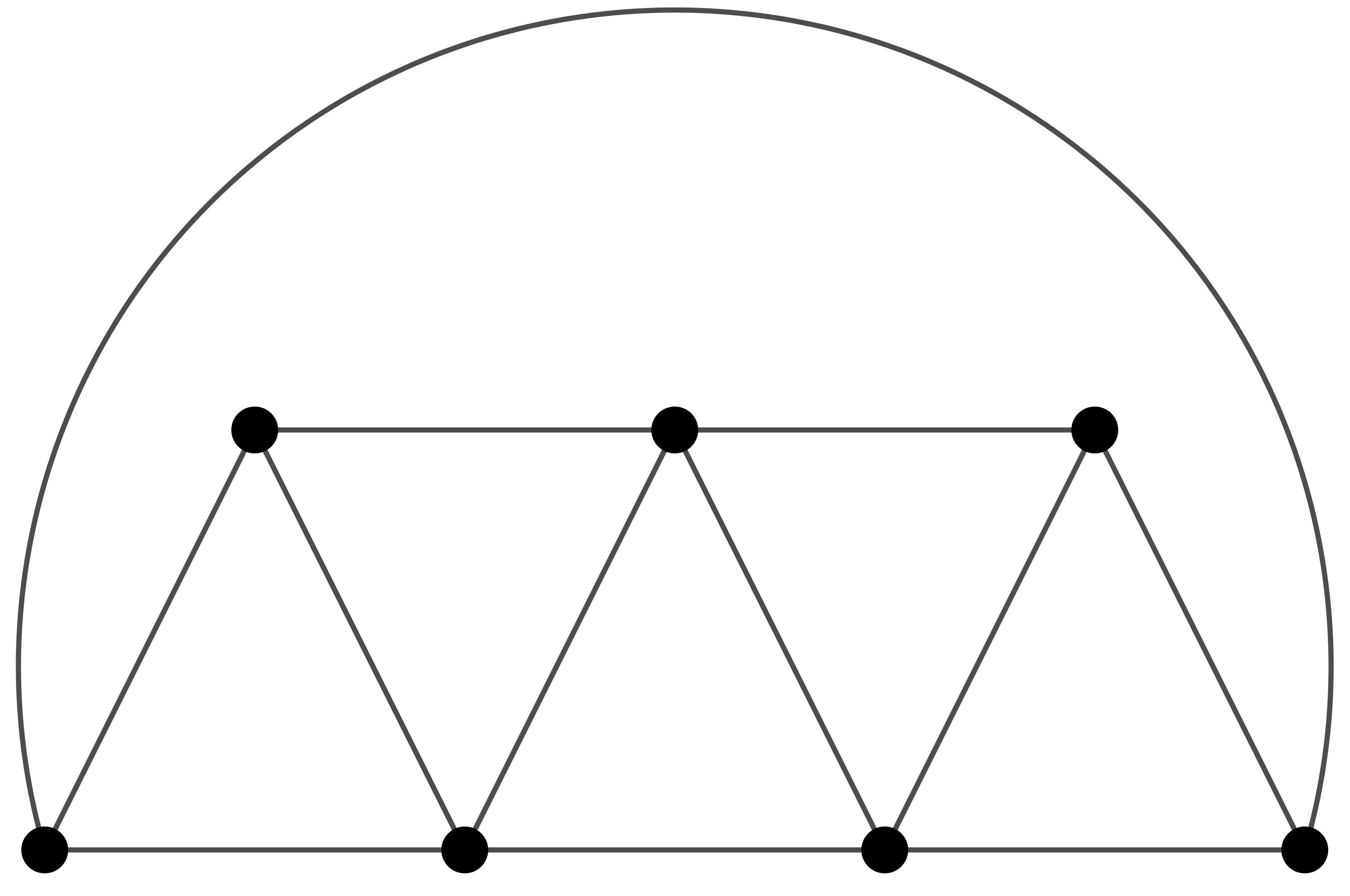}}
\caption{Examples of zig-zag graphs with 5 and 6 loops.}
\label{fig: zigzag}
\end{figure}

\begin{figure}[htb!]
\centering
\subfloat[Tree-level contribution]{\includegraphics[scale=.55]{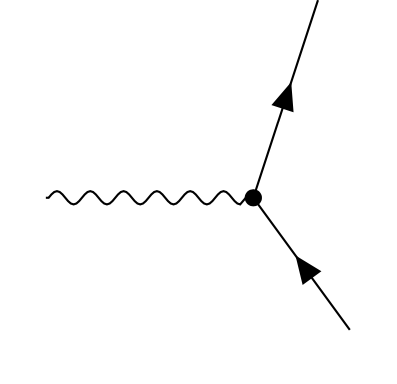}} \qquad\qquad
\subfloat[One-loop contribution]{\includegraphics[scale=.55]{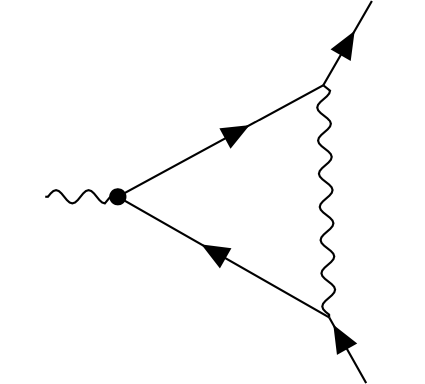}}%
\caption{Up to one-loop Feynman diagrams contributing to the anomalous magnetic moment of the electron.}
\label{fig: g}
\end{figure}

We observe the remarkable fact that $\mathbb{Q}$-linear combinations of multiple zeta values are ubiquitous in the evaluation of Feynman amplitudes in perturbative quantum field theories. It~was conjectured by Broadhurst and Kreimer~\cite{BK95} and then proved by Brown and Schnetz~\cite{BS12} that Feynman integrals of the infinite family of \textit{zig-zag graphs} in $\phi^4$ theory (see Fig.~\ref{fig: zigzag}) are certain known rational multiples of the odd values of the Riemann zeta function.
\begin{Theorem}
Let $Z_l$ be the zig-zag graph with $l$ loops. Its Feynman integral is
\begin{gather}
I_{Z_l} = 4 \frac{(2l-2)!}{l! (l-1)!} \left( 1 - \frac{1 - (-1)^l}{2^{2l-3}} \right) \zeta(2l-3).
\end{gather}
\end{Theorem}
Another example is given by the anomalous magnetic moment of the electron in quantum electrodynamics.
The tree level Feynman diagram representing a slow-moving electron emitting a photon is depicted in Fig.~\ref{fig: g} along with its one-loop correction.
The two-loop correction comes from the contributions of seven distinct two-loop diagrams. The total two-loop Feynman amplitude has been evaluated by Petermann~\cite{Pet57}, giving $\frac{197}{144} + \frac{1}{2} \zeta(2) - 3 \zeta(2) \log(2)+ \frac{3}{4} \zeta(3)$, which involves the logarithm of 2 and again values of the Riemann zeta function.

Many more examples are given by Broadhurst~\cite{Bro13}.
Due to a vast amount of evidence, it was believed for a long time that all primitive amplitudes of the form~\eqref{eq: I_G_last} in massless $\phi^4$ theory should be $\mathbb{Q}$-linear combinations of MZVs.
Only recently this conjectural statement was proved false in the motivic setup\footnote{Outside of the motivic framework, the statement relies on transcendentality conjectures.} by Brown and Schnetz~\cite{BS2012}.
Explicit examples of $\phi^4$-amplitudes at high loop orders not expressible in terms of multiple zeta values have been found by Panzer and Schnetz~\cite{PS17}. In the same work, explicit computation of all $\phi^4$-amplitudes with loop order up to $7$ suggests that not all MZVs appear among them. For example, no $\phi^4$-graph is known to evaluate to $\zeta(2)$ or $\zeta(2)^2$.
Remarkably, the integral representation of MZVs partially clarify the presence of these numbers in perturbative calculations in quantum field theory. Indeed, both expressions~\eqref{eq: I_G_last} and~\eqref{eq: zint} are suitably interpreted as \textit{periods} of algebraic varieties.

\section{Cohomology theory of algebraic varieties} \label{sec: geometry}

\subsection{Singular homology}
We follow the expositions by Weibel~\cite{Wei94} and Hartshorne~\cite{Har77}. Let~$M$ be a topological space.
For each integer $k \ge 0$, the standard $k$-simplex is
\begin{gather}
\Delta^k_{st} = \bigg\{ (t_0,\dots,t_k) \in \mathbb{R}^{k+1} \,\bigg|\, \sum_{i=0}^k t_i = 1, \; t_i \ge 0, \; i=0,\dots,k \bigg\}.
\end{gather}
For each $i=0,\dots,k$, the \textit{face map} $\delta_i^k \colon \Delta^{k-1}_{st} \rightarrow \Delta^k_{st}$ is defined by
\begin{gather}
\delta^k_i (t_0,\dots,t_{k-1}) = (t_0,\dots,t_{i-1},0,t_i,\dots,t_{k-1}).
\end{gather}
A \textit{singular k-chain} in $M$ is a continuous\footnote{If $M$ is a differentiable manifold, we can assume the singular chains to be piecewise smooth, or smooth, without altering the homology groups.} map $\sigma \colon \Delta^k_{st} \rightarrow M$.
For each $k \ge 0$, let
\begin{gather}
C_k(M) = \bigoplus_{\sigma} \mathbb{Z} \sigma
\end{gather}
be the free abelian group generated by singular $k$-chains. Elements of $C_k(M)$ are finite $\mathbb{Z}$-linear combinations of the continuous maps $\sigma\colon \Delta^k_{st} \rightarrow M$.
For each $k \ge 1$, the \textit{boundary map} $\partial_k \colon C_k(M) \rightarrow C_{k-1}(M)$ is defined by
\begin{gather}
\partial_k (\sigma) = \sum_{i=0}^k (-1)^i \big(\sigma \circ \delta_i^k\big),
\end{gather}
where the alternating signs in the sum guarantee that boundary maps satisfy the condition $\partial_{k-1} \circ \partial_k = 0$.
The pair $(C_{\bullet}(M), \partial_{\bullet})$ is called a \textit{homological chain complex} and is graphically represented as
\begin{gather}
\dots \xlongrightarrow{\partial_{k+1}} C_k(M) \xlongrightarrow{\partial_{k}} C_{k-1}(M) \xlongrightarrow{\partial_{k-1}} \dots \xlongrightarrow{\partial_{2}} C_1(M) \xlongrightarrow{\partial_{1}} C_0(M).
\end{gather}

\begin{Definition} \label{def: ho}
The \textit{singular homology} of the topological space $M$ is the homology of the complex $(C_{\bullet}(M), \partial_{\bullet})$, that is
\begin{gather}
H_k^{\text{s}}(M, \mathbb{Z}) = \begin{cases}
 C_0(M)/\Im(\partial_1), & k=0,\\
 \mathop{\rm Ker}(\partial_k)/\Im(\partial_{k+1}), & k \ge 1.
 \end{cases}
\end{gather}
In degree $k$, chains in the kernel of the boundary map $\partial_k$ are called \textit{$($closed$)$ cycles} and chains in the image of the boundary map $\partial_{k+1}$ are called \textit{$($exact$)$ boundaries}.
\end{Definition}

\begin{Example}\label{ex: ex1}
Let $M= \mathbb{C}^*$ be the punctured complex plane. The singular chains
\begin{gather}
\gamma_0\colon\quad \Delta^0_{st} \rightarrow \mathbb{C}^*, \qquad 1 \mapsto 1, \\
\gamma_2 \colon\quad \Delta^1_{st} \rightarrow \mathbb{C}^*, \qquad (t, 1-t) \mapsto {\rm e}^{2 \pi {\rm i} t}
\end{gather}
generate the singular homology groups $H_0^{\text{s}}(\mathbb{C}^*,\mathbb{Z})$ and $H_1^{\text{s}}(\mathbb{C}^*,\mathbb{Z})$, respectively. These are both free groups of rank one. All the other homology groups vanish.
\end{Example}

For each $k \ge 0$, the free abelian group of \textit{singular n-cochains} is defined by
\begin{gather}
C^k(M) = \mathop{\rm Hom}(C_k(M), \mathbb{Z}).
\end{gather}
Analogously, applying vector duality, we introduce the \textit{coboundary maps} $d^k\!\colon C^k(M) \!\rightarrow\! C^{k+1}(M)$, which satisfy the condition $d^{k+1} \circ d^k = 0$.
The corresponding \textit{cohomological chain complex} $(C^{\bullet}(M), d^{\bullet})$ is graphically represented as
\begin{gather}
\cdots \xlongleftarrow{d^{k+1}} C^{k+1}(M) \xlongleftarrow{d^{k}} C^{k}(M) \xlongleftarrow{d^{k-1}} \cdots \xlongleftarrow{d^{1}} C^1(M) \xlongleftarrow{d^{0}} C^0(M).
\end{gather}
\begin{Definition} \label{def: coho}
The \textit{singular cohomology} of the topological space $M$ is the cohomology of the complex $(C^{\bullet}(M), d^{\bullet})$, that is
\begin{gather}
H^k_{\text{s}}(M, \mathbb{Z}) = \begin{cases}
 \mathop{\rm Ker}\big(d^0\big), & k=0,\\
 \mathop{\rm Ker}\big(d^k\big)/\Im(d^{k-1}), & k \ge 1.
 \end{cases}
\end{gather}
\end{Definition}
Definitions~\ref{def: ho} and~\ref{def: coho} of singular homology and cohomology of topological spaces, given here with respect to $\mathbb{Z}$, extend naturally to other coefficient rings.
For our purposes, we assume the ring of coefficients to be $\mathbb{Q}$. This allows us to identify singular cohomology groups with the vector duals of the corresponding singular homology groups\footnote{This isomorphism is true for real or complex coefficients as well, but it does not hold for integer coefficients.}
\begin{gather}
H^k_{\text{s}}(M, \mathbb{Q}) \simeq \mathop{\rm Hom}(H_k^{\text{s}}(M,\mathbb{Q}),\mathbb{Q}),
\end{gather}
that is, classes of a cohomology group can be interpreted as classes of linear functionals on the corresponding homology group.
The singular cohomology of the topological space underlying a~complex algebraic variety is of particular interest.
\begin{Definition} \label{def: betti}
Let $X$ be an algebraic variety over a subfield $\mathbb{K}$ of $\mathbb{C}$. Its set of complex points $X(\mathbb{C})$ canonically carries the complex analytic topology, and the corresponding topological space\footnote{Equipped with the canonical structure sheaf, $X^{\rm an}$ is a complex analytic space, called the \textit{analytification} of $X$. The relationship between algebraic spaces over the complex numbers and complex analytic spaces is described by a series of results, known as GAGA-type theorems. These developments followed the work by Serre~\cite{GAGA} on the existence and faithfulness of the analytification of a complex algebraic variety.} is written as $X^{\rm an}$.
The \textit{Betti cohomology} of $X$ is the singular cohomology of the underlying topo\-lo\-gi\-cal space $X^{\rm an}$, that is
\begin{gather}
H^k_{{\rm B}}(X,\mathbb{Q})=H^k_{\text{s}}\big(X^{\rm an},\mathbb{Q}\big)
\end{gather}
for $k \ge 0$.
\end{Definition}

\begin{Example} \label{ex: ex2}
Let $\mathbb{G}_m = \mathop{\rm Spec} \mathbb{Q}[x,1/x]$ be the multiplicative group. $\mathbb{G}_m$ is an algebraic variety over $\mathbb{Q}$ and its underlying topological space of complex points is $\mathbb{G}_m^{\rm an} = \mathbb{C}^*$. For each $k \ge 0$, the $k$-th Betti cohomology group of $\mathbb{G}_m$ is $H_{\rm B}^k(\mathbb{G}_m,\mathbb{Q}) = H_{\text{s}}^k(\mathbb{C}^*, \mathbb{Q})$.
\end{Example}

\subsubsection{Some properties of homology}
We briefly recall some properties of singular homology and cohomology.
\begin{itemize}\itemsep=0pt
\item[($a$)] \textit{Homotopy invariance}. If~$M_1$ and $M_2$ are homotopically equivalent topological spaces, then $H_k^{\text{s}}(M_1,\mathbb{Q}) \simeq H_k^{\text{s}}(M_2, \mathbb{Q})$ for each $k \ge 0$. An analogous statement holds for singular cohomology.
\item[($b$)] \textit{Mayer--Vietoris sequences}. For any two open subspaces $U,V \subseteq M$ of a given topological space $M$, such that $M = U \cup V$, there is a long exact sequence of the following form:
\begin{equation}
\begin{tikzcd}
 \cdots \arrow[r]
 & H^s_k(U \cap V,\mathbb{Q}) \arrow[r]
 	\arrow[d, phantom, ""{coordinate, name=Z}]
 & H^s_k(U,\mathbb{Q}) \oplus H^s_k(V,\mathbb{Q}) \arrow[dll,
 "",
rounded corners,
to path={ -- ([xshift=2ex]\tikztostart.east)
|- (Z) [near end]\tikztonodes
-| ([xshift=-2ex]\tikztotarget.west) -- (\tikztotarget)}] \\
 H^s_{k}(M,\mathbb{Q}) \arrow[r]
 & H^s_{k-1}(U \cap V,\mathbb{Q}) \arrow[r]
 & \cdots.
\end{tikzcd}
\end{equation}
An analogous statement holds for singular cohomology.
\item[($c$)] \textit{K\"unneth formula}. For any two topological spaces $M_1$, $M_2$, for each $k \ge 0$, there is a natural isomorphism
\begin{gather}
H_k^{\text{s}}(M_1 \times M_2, \mathbb{Q}) \simeq \bigoplus_{i+j=k} H_i^{\text{s}}(M_1, \mathbb{Q}) \otimes H_j^{\text{s}}(M_2, \mathbb{Q}).
\end{gather}
An analogous statement holds for singular cohomology.
\item[($d$)] \textit{Push-forward}. Let~$f\colon M_1 \rightarrow M_2$ be a continuous map between two topological spa\-ces~$M_1$,~$M_2$. Then, $f$ induces a morphism of chain complexes
\begin{gather}
f_*\colon\ C_{\bullet}(M_1) \rightarrow C_{\bullet}(M_2)
\end{gather}
called \textit{push-forward}, sending $\sigma_1 \in C_k(M_1)$ to $\sigma_2 = f \circ \sigma_1 \in C_k(M_2)$. Equivalently, the following diagram:
\begin{equation}
\begin{tikzcd}
\Delta_{st}^k \arrow[r, "\sigma_1"] \arrow[rd, "\sigma_2"'] & M_1 \arrow[d, "f"] \\
 & M_2
\end{tikzcd}
\end{equation}
commutes. Hence, $f$ induces also a group homomorphism between the corresponding singular homology groups
\begin{gather}
f_* \colon\ H_k^{\text{s}}(M_1, \mathbb{Q}) \rightarrow H_k^{\text{s}}(M_2, \mathbb{Q})
\end{gather}
for each $k \ge 0$.
\item[($e$)] \textit{Pull-back.} Let $f\colon M_1 \rightarrow M_2$ be a continuous map between two topological spa\-ces~$M_1$,~$M_2$. Then, $f$ induces a morphism of cochain complexes
\begin{gather}
f^*\colon\ C^{\bullet}(M_2) \rightarrow C^{\bullet}(M_1)
\end{gather}
called \textit{pull-back}, sending $\omega_2 \in C^k(M_2)$ to $\omega_1 = \omega_2 \circ f_* \in C^k(M_1)$. Equivalently, the following diagram:
\begin{equation}
\begin{tikzcd}
C_k(M_1) \arrow[r, "\omega_1"] \arrow[d, "f_*"'] & \mathbb{Q} \\
C_k(M_2) \arrow[ur, "\omega_2"'] &
\end{tikzcd}
\end{equation}
commutes. Hence, $f$ induces also a group homomorphism between the corresponding singular cohomology groups
\begin{gather}
f^*\colon\ H^k_{\text{s}}(M_2, \mathbb{Q}) \rightarrow H^k_{\text{s}}(M_1, \mathbb{Q})
\end{gather}
for each $k \ge 0$.
\end{itemize}

\subsubsection{Relative singular homology}
Let $M$ be a topological space and $\iota\colon N \hookrightarrow M$ the canonical inclusion of a topological subspace $N \subseteq M$.
Denote by $(C_{\bullet}(N), \partial^N_{\bullet})$ and $(C_{\bullet}(M), \partial^M_{\bullet})$ their homological chain complexes, and by $\iota_*\colon C_{\bullet}(N) \rightarrow C_{\bullet}(M)$ the corresponding injective morphism obtained via push-forward.
For each $k \ge 1$, we define the total chain complex $C_{\bullet}(M,N)$ to be the \textit{mapping cone}\footnote{Note that, for any morphism of chain complexes $f_*\colon C_{\bullet}(M_1) \rightarrow C_{\bullet}(M_2)$, the mapping cone $C_k(M_2,M_1) = C_{k-1}(M_1) \oplus C_k(M_2)$ can be defined. However, injectivity of the morphism $\iota_*$ implies that the cone $C_{\bullet}(M,N)$ is quasi-isomorphic to the quotient $C_{\bullet}(M) / C_{\bullet}(N)$.} of the morphism~$\iota_*$, that is
\begin{gather}
C_k(M,N) = C_{k-1}(N) \oplus C_k(M),
\end{gather}
and the differential $\partial_k\colon C_k(M,N) \rightarrow C_{k-1}(M,N)$ to act as
\begin{gather}
\partial_k (\sigma_N,\sigma_M) = \big({-}\partial_{k-1}^N (\sigma_N), - \iota_*(\sigma_N) + \partial_k^M (\sigma_M)\big),
\end{gather}
where $(\sigma_N,\sigma_M) \in C_k(M,N)$.
\begin{Definition}
The \textit{relative homology} of the pair of topological spaces $(M,N)$ is the homology of the total chain complex $(C_{\bullet}(M,N), \partial_{\bullet})$. For $k \ge 1$, we denote the relative singular homology groups as $H_k^{\text{s}}(M,N,\mathbb{Q})$.
\end{Definition}
Relative homology fits into the following long exact sequence:
\begin{equation} \label{eq: long}
\begin{tikzcd}
 \cdots \arrow[r]
 & H_k^{\text{s}}(M,\mathbb{Q}) \arrow[r]
 	\arrow[d, phantom, ""{coordinate, name=Z}]
 & H_k^{\text{s}}(M,N,\mathbb{Q}) \arrow[dll,
 "",
rounded corners,
to path={ -- ([xshift=2ex]\tikztostart.east)
|- (Z) [near end]\tikztonodes
-| ([xshift=-2ex]\tikztotarget.west) -- (\tikztotarget)}] \\
 H_{k-1}^{\text{s}}(N,\mathbb{Q}) \arrow[r]
 & H_{k-1}^{\text{s}}(M,\mathbb{Q}) \arrow[r]
 & H_{k-1}^{\text{s}}(M,N,\mathbb{Q}) \arrow[r]
 & \cdots,
\end{tikzcd}
\end{equation}
where the connecting morphisms are the push-forward maps $\iota_*\colon H_k^{\text{s}}(N,\mathbb{Q}) \rightarrow H_k^{\text{s}}(M,\mathbb{Q})$ induced by the inclusion $\iota\colon N \hookrightarrow M$.
Consider an element of the relative homology group $H_k^{\text{s}}(M,N,\mathbb{Q})$. This is represented by a pair $(\sigma_N, \sigma_M)$ of singular chains $\sigma_N \in C_{k-1}(N)$ and $\sigma_M \in C_k(M)$ satisfying
\begin{gather}
\partial_{k-1}^N \sigma_N = 0 , \qquad \partial_k^M \sigma_M = \iota_* \sigma_N.
\end{gather}
Note that, since $\iota_*$ is injective, the latter condition implies the former.
Thus, relative homology classes are represented by chains in $M$ whose boundary is contained in $N$.
Relative cohomology groups $H^k_{\text{s}}(M,N, \mathbb{Q})$ are defined similarly.

\begin{Example} \label{ex: ex0}
Let $M = \mathbb{C}^*$ be the punctured complex plane and $N = \{ p,q \} \subset M$ be the subspace consisting of two points $p,q \in \mathbb{C}^*$ with $p \ne q$. Let~$\gamma_1\colon \Delta^1_{st} \rightarrow M$ be any continuous map\footnote{When it does not pass through the origin, the oriented segment starting at $p$ and ending at $q$ is an example of such a map.} such that $\gamma_1(0, 1) = p$, $\gamma_1(1, 0) = q$, and it does not encircle the origin. Then
\begin{gather}
\partial_1^M \gamma_1 = p-q \in C_0(N).
\end{gather}
Consequently, $\gamma_1$ defines a relative chain. It~follows from the long exact sequence~\eqref{eq: long} that the only non-trivial relative homology group is $H_1^{\text{s}}(M,N,\mathbb{Q})$. A~basis of this group is given by the chain $\gamma_1$ and the chain $\gamma_2$, introduced in Example~\ref{ex: ex1}, consisting of a counterclockwise circle containing the origin. Such a basis is graphically represented in Fig.~\ref{fig: base}.
\begin{figure}[htb!]
\centering
\includegraphics[scale=1.2]{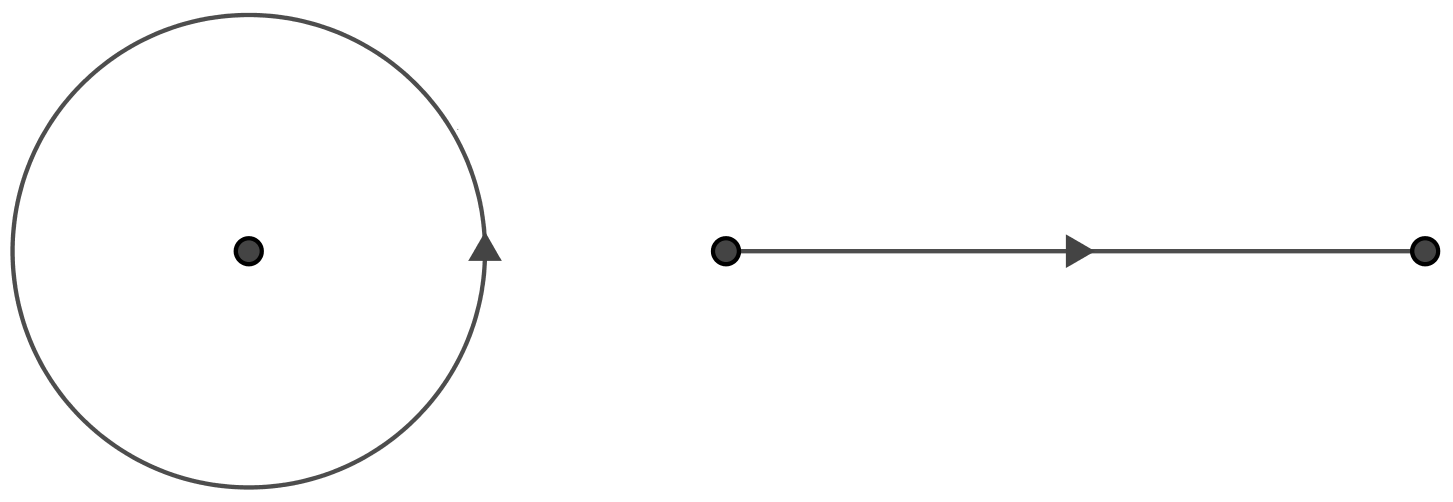}
\put(-145,55){\makebox(0,0)[lb]{\small $\gamma_2$}}
\put(-180,24){\makebox(0,0)[lb]{\small $O$}}
\put(-108,25){\makebox(0,0)[lb]{\small $p$}}
\put(-60,25){\makebox(0,0)[lb]{\small $\gamma_1$}}
\put(-9,25){\makebox(0,0)[lb]{\small $q$}}
\caption{Basis of $H_1^{\text{s}}(\mathbb{C}^*,\{ p,q \},\mathbb{Q})$.}
\label{fig: base}
\end{figure}
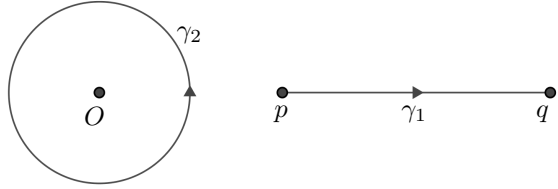
\end{Example}

\subsection{De Rham cohomology} \label{sec: smooth}
We start by reviewing a classical construction in differential geometry. Let~$M$ be a differentiable manifold of dimension $n$. A~\textit{differential $p$-form} on $M$ is written in local coordinates as
\begin{gather}
\sum_{1 \le i_1 < \dots < i_p \le n} f_{i_1,\dots,i_p} \; {\rm d}x_{i_1} \wedge \dots \wedge {\rm d}x_{i_p},
\end{gather}
where $f_{i_1,\dots,i_p}$ are $\mathcal{C}^{\infty}$-functions. Let~$\Omega^p(M)$ denote the $\mathbb{R}$-vector space of differential $p$-forms on~$M$ and define the space of differential forms on $M$ as
\begin{gather}
\Omega(M) = \bigoplus_{p=0}^n \Omega^p(M).
\end{gather}
The \textit{exterior derivative} ${\rm d}\colon \Omega(M) \rightarrow \Omega(M)$ is the unique $\mathbb{R}$-linear map which sends $p$-forms into $(p+1)$-forms and satisfies the following axioms:
\begin{itemize}\itemsep=0pt
\item[(1)] Let $f$ be a smooth function. Then, ${\rm d}f = \sum_{i=1}^n \frac{\partial f}{\partial x_i} {\rm d}x_i$ is the ordinary differential of $f$.
\item[(2)] ${\rm d} \circ {\rm d}=0$.
\item[(3)] Let $\alpha$ be a $p$-form on $M$ and $\beta$ any differential form in $\Omega(M)$. Denote by $\alpha \wedge \beta$ their exterior product. Then, ${\rm d}(\alpha \wedge \beta) = {\rm d}\alpha \wedge \beta + (-1)^p \alpha \wedge {\rm d}\beta$.
\end{itemize}
The associated cochain complex is
\begin{gather}
0 \rightarrow \Omega^0(M) \xrightarrow{\rm d} \Omega^1(M) \xrightarrow{\rm d} \cdots \xrightarrow{\rm d} \Omega^n(M) \rightarrow 0
\end{gather}
and its cohomology, denoted $H^{\bullet}_{\rm dR}(M, \mathbb{R})$, is called the \textit{smooth de Rham cohomology} of $M$.
A~differential $p$-form $\omega$ is \textit{closed} if ${\rm d}\omega=0$ and it is \textit{exact} if there exists a differential $(p-1)$-form $\eta$ such that $\omega = {\rm d}\eta$.
A~classical theorem\footnote{De Rham's theorem was first presented in his PhD thesis, published in 1931, when cohomology groups had not been introduced yet. He did not state the theorem in the way it is described today, but gave an equivalent version involving Betti numbers and integration of closed differential forms over cycles.} by de Rham~\cite{Der31} asserts that the singular cohomology $H^{\bullet}_{\text{s}}(M, \mathbb{R})$ can be computed using differential forms.\footnote{We refer to Bott and Tu~\cite{BT82} for a comprehensive investigation of differential forms in algebraic topology.}
\begin{Theorem} \label{th: derham}
Let $M$ be a differentiable manifold of dimension $n$. For $0 \le k \le n$, the map
\begin{align}
H^k_{\mathrm{dR}}(M,\mathbb{R}) &\longrightarrow H^k_{\mathrm{s}}(M,\mathbb{R}) \simeq \mathop{\rm Hom}\big(H_k^{\text{s}}(M,\mathbb{R}), \mathbb{R}\big),
\\
{}[\omega] &\longmapsto \int \omega,
\end{align}
which sends the class of a differential form $\omega$ to the integration functional
\begin{align}
\int \omega\colon\ H_k^{\text{s}}(M,\mathbb{R}) &\longrightarrow \mathbb{R},
\\
{}[\gamma] &\longmapsto \displaystyle\int_{\gamma} \omega,
\end{align}
is an isomorphism.
\end{Theorem}

\subsubsection{Algebraic de Rham cohomology} \label{sec: algdR}
A notion of de Rham cohomology for general algebraic varieties over fields of characteristic zero has been introduced by Grothendieck~\cite{Gro66}. Let~$\mathbb{K}$ be a subfield of $\mathbb{C}$ and let $X$ be an algebraic variety over $\mathbb{K}$.
\begin{Definition}
Consider $U \subseteq X$ an open affine subset in the Zariski topology. The ring of~regular functions on $U$, denoted by $\mathcal{O}(U)$, is a finitely-generated $\mathbb{K}$-algebra, and precisely a~quotient of a polynomial ring over $\mathbb{K}$. We~say that $X$ is \textit{smooth} or \textit{nonsingular} of dimension~$n$ if, for every closed point $x \in X$, the limit $\lim_{\substack{\\ \overrightarrow{U \ni x} }} \mathcal{O}(U)$, indexed over all Zariski open affine subsets $U \subseteq X$ containing $x$, and with ordering defined by reverse inclusion, is a regular local ring of dimension $n$.
\end{Definition}

Let $X$ be smooth of dimension $n$ and affine. We~can write $X = \mathop{\rm Spec} R$, where $R = \mathcal{O}(X)$ is the ring of regular functions on $X$. A~\textit{$\mathbb{K}$-linear algebraic $p$-form} on $X$ is a differential $p$-forms on $X$ with coefficients in $R$.
In a local coordinate chart, it is given by an expression of the form
\begin{gather} \label{eq: local}
\sum_{1 \le i_1 < \dots < i_p \le n} f_{i_1,\dots,i_p} \; {\rm d}x_{i_1} \wedge \dots \wedge {\rm d}x_{i_p},
\end{gather}
where $f_{i_1,\dots,i_p}$ are $\mathbb{K}$-polynomial functions on $X$. We~denote by $\Omega^p(X)$ the $\mathbb{K}$-vector space of~algebraic $p$-forms on $X$ and we define the space of algebraic forms on $X$ as
\begin{gather}
\Omega(X) = \bigoplus_{p=0}^n \Omega^p(X).
\end{gather}
A derivation ${\rm d}\colon \Omega(X) \rightarrow \Omega(X)$, satisfying properties that are analogous to the ones described in~Section~\ref{sec: smooth} for the exterior derivative, can be defined. It~canonically yields a cochain complex
\begin{gather}
0 \rightarrow R \simeq \Omega^0(X) \xrightarrow{\rm d} \Omega^1(X) \xrightarrow{\rm d} \cdots \xrightarrow{ \rm d} \Omega^n(X) \rightarrow 0
\end{gather}
called the \textit{algebraic de Rham complex} of $X$. The associated cohomology, denoted $H^{\bullet}_{\rm dR}(X, \mathbb{K})$, is called the \textit{algebraic de Rham cohomology} of $X$.

\begin{Remark}
If $X$ is smooth of dimension $n$, but not necessarily affine, at each closed point $x \in X$, we can choose some Zariski open affine neighbourhood $U$ of $x$ and some regular functions $x_1, \dots, x_n \in \mathcal{O}(U)$ in such a way to define a system of \textit{local parameters}\footnote{If we do not assume $X$ to be smooth, then we can find local coordinates in an affine open neighbourhood $U$ of a closed point $x \in X$ if and only if the rank of the Jacobian matrix at $x$ is equal to the dimension of $U$.} at $x$.
Viewed as a subvariety of the affine $\mathbb{K}$-space $\mathbb{A}^n$, $U$ inherits its local coordinate structure.
Intuitively, by choosing a covering of $X$ composed of Zariski open affine subsets, the algebraic variety is charted with affine spaces.
Observe that the morphism $U \rightarrow \mathbb{A}^n$ defined by the local coordinates $x_1, \dots, x_n$ is always an \'etale map,\footnote{\'Etale maps can be interpreted as the algebraic analogue of local isomorphisms in the complex analytic topology. However, open sets in the Zariski topology are not small enough for \'etale maps to be local isomorphisms.} but not generally an embedding.\footnote{For complex neighbourhoods, local coordinates define local isomorphisms. Indeed, smooth algebraic varieties over $\mathbb{C}$ can be locally embedded as submanifolds of the complex affine space.}
Conceptually, the $\mathbb{K}$-linear algebraic forms of degree $p$ on $X$ are obtained by suitably gluing\footnote{The assignment of algebraic forms to smooth affine varieties via local coordinates is well-behaved under gluing, and hence it globalises.} the algebraic $p$-forms defined locally, as in~\eqref{eq: local}, on each subset $U$ of an affine open covering of $X$. The notion of algebraic de Rham cohomology thus \v Ceck-style generalises to arbitrary smooth algebraic $\mathbb{K}$-varieties.
Such an intuition does not, however, capture the full picture. The algebraic substitute for the smooth differential form is rigorously defined through the notions of K\"ahler differential and exterior power, while the rigorous construction of the algebraic de Rham cohomology of any smooth algebraic $\mathbb{K}$-variety requires the use of sheaf cohomology and hypercohomology. We~do not present these concepts here, since an intuitive understanding is sufficient to our purpose, but we refer to Kashiwara and Schapira~\cite{KS06}, and Hartshorne~\cite{Har77}. Moreover, we mention that several constructions are available to adapt the definition of algebraic de Rham cohomology to the case of singular varieties giving well-behaved theories. Details are reported by Huber and M\"uller-Stach~\cite{HM17}.
\end{Remark}

\begin{Example} \label{ex: ex4}
Consider $X= \mathbb{G}_m= \mathop{\rm Spec} \mathbb{Q}[x, 1/x]$. The only non-vanishing spaces of $\mathbb{Q}$-linear algebraic forms are
\begin{gather}
\Omega^0(\mathbb{G}_m) = \mathbb{Q}[x, 1/x],
\\
\Omega^1(\mathbb{G}_m) = \mathbb{Q}[x, 1/x] \cdot {\rm d}x.
\end{gather}
Consequently, the two groups
\begin{gather}
H^0_{\rm dR}(\mathbb{G}_m, \mathbb{Q}) = \mathbb{Q} ,
\\
H^1_{\rm dR}(\mathbb{G}_m, \mathbb{Q}) = \frac{\mathbb{Q}[x, 1/x] \cdot {\rm d}x}{{\rm d} \mathbb{Q}[x, 1/x]} = \mathbb{Q}\bigg[ \frac{{\rm d}x}{x} \bigg]
\end{gather}
are the only non-trivial algebraic de Rham cohomology groups of $X$.
\end{Example}

\subsubsection{Relative de Rham cohomology}
The definition of algebraic de Rham cohomology extends to the relative setting. Let~$\mathbb{K}$ be a~subfield of $\mathbb{C}$ and let $X$ be a smooth algebraic variety over $\mathbb{K}$ of dimension $n$. Recall the following definition.
\begin{Definition}
A codimension-1 closed subvariety $D \subset X$ is called a \textit{divisor with normal crossings} if, for every point $x \in D$, there is an open affine neighbourhood $U \subseteq X$ of $x$ and some local coordinates $x_1, \dots, x_n$ on $U$ such that:
\begin{itemize}\itemsep=0pt
\item[(1)] The morphism $U \rightarrow \mathbb{A}^n$ defined by $x_1, \dots, x_n$ is \'etale.
\item[(2)] The restriction $D_{|U}$ is locally described by an equation of the form $x_1 \cdot x_2 \cdots x_r$ for some $1 \le r \le n$.
\end{itemize}
Moreover, $D$ is called a \textit{divisor with simple normal crossings}\footnote{$D$ looks locally like a collection of coordinate hyperplanes.} if, in addition, its irreducible components are smooth.
\end{Definition}

For simplicity,\footnote{We illustrate here the construction of relative algebraic de Rham cohomology in a particularly simple framework. The construction can, however, be adapted for the general case of a closed subvariety of a smooth algebraic $\mathbb{K}$-variety. For a general discussion, we refer to Huber and M\"uller-Stach~\cite{HM17}.} let $X$ be affine and $D \subset X$ a divisor with simple normal crossings. Denote by $D_i$, for $i=1,\dots,r$, the smooth irreducible components of $D$. For $I \subseteq \{0,\dots,r\}$, we set
\begin{gather}
D_I = \bigcap_{i \in I} D_i , \qquad
D^p = \begin{cases}
 X, & p=0,
 \\
 \coprod_{|I|=p} D_I, & p \ge 1.
 \end{cases}
\end{gather}
The associated double cochain complex of $\mathbb{K}$-vector spaces $K^{p,q}=\Omega^q(D^p)$ is graphically represented as
\begin{equation} \label{eq: double}
\begin{tikzcd}
\cdots & \cdots & \cdots & \\
\Omega^2(X) \arrow[u, "{\rm d}"] \arrow[r] & \bigoplus_i \Omega^2(D_i) \arrow[u, "-{\rm d}"] \arrow[r] & \bigoplus_{i<j} \Omega^2(D_i \cap D_j) \arrow[u, "{\rm d}"] \arrow[r] & \cdots \\
\Omega^1(X) \arrow[u, "{\rm d}"] \arrow[r] & \bigoplus_i \Omega^1(D_i) \arrow[u, "-{\rm d}"] \arrow[r] & \bigoplus_{i<j} \Omega^1(D_i \cap D_j) \arrow[u, "{\rm d}"] \arrow[r] & \cdots \\
\Omega^0(X) \arrow[u, "{\rm d}"] \arrow[r] & \underbrace{\bigoplus_i \Omega^0(D_i)}_{|I|=1} \arrow[u, "-{\rm d}"] \arrow[r] & \underbrace{\bigoplus_{i<j} \Omega^0(D_i \cap D_j)}_{|I|=2} \arrow[u, "{\rm d}"] \arrow[r] & \cdots,
\end{tikzcd}
\end{equation}
where the vertical differential ${\rm d}_{\rm ver}\colon K^{p,q} \rightarrow K^{p,q+1}$ is given by
\begin{gather}
{\rm d}_{\rm ver} = (-1)^p {\rm d}
\end{gather}
and the horizontal differential ${\rm d}_{\rm hor}\colon K^{p,q} \rightarrow K^{p+1,q}$ is given by
\begin{gather}
{\rm d}_{\rm hor} = \bigoplus_{\substack{|I|=p\\ |J|=p+1\\ I \subset J}}(-1)^l {\rm d}_{IJ},
\end{gather}
where $J = \{ j_0, \dots, j_p\}$ with $j_0 < \dots < j_p$, $I = \big\{j_0, \dots, \widehat{j_l}, \dots, j_p \big\}$, and ${\rm d}_{IJ}\colon \Omega^q(D_I) \rightarrow \Omega^q(D_J)$ is the restriction map.\footnote{We observe that the two-row sequence in~\eqref{eq: double} is exact. For the vertical lines $K^{p,q} \rightarrow K^{p,q+1}$, it follows from the property ${\rm d} \circ {\rm d} = 0$ of the differential ${\rm d}$. For the horizontal lines $K^{p,q} \rightarrow K^{p+1,q}$, it follows from the fact that the differential ${\rm d}_{\rm hor}\colon K^{p,q} \rightarrow K^{p+1,q}$ is surjective for even values of $p$ and trivial for odd values of $p$, as a~consequence of the surjectivity of the restriction maps ${\rm d}_{IJ}$.}
Note that the sign factor $(-1)^p$ in the definition of ${\rm d}_{\rm ver}$ implies that the vertical and horizontal differentials anticommute.
Moreover, since $D_I$ has dimension equal to $n-|I|$, the double complex is trivial for $p+q > n$. We~denote by $(\Omega^{\bullet}(X,D), \delta)$ the total cochain complex associated to $K^{p,q}$, that is
\begin{gather}
\Omega^{\bullet}(X,D) = \bigoplus_{p+q= \bullet} K^{p,q}, \qquad
\delta = {\rm d}_{\rm ver} + {\rm d}_{\rm hor}.
\end{gather}
For each $k \ge 0$, the space $\Omega^k(X,D)$ corresponds to the direct sum of the spaces on the $k$-th diagonal of the double cochain complex $K^{p,q}$ represented in~\eqref{eq: double}.
The total complex is indeed explicitly written as
\begin{gather}
\Omega^0(X,D) \simeq \Omega^0(X) \xlongrightarrow{\delta^0} \Omega^1(X,D) \simeq \Omega^1(X) \oplus \bigoplus_i \Omega^0(D_i) \xlongrightarrow{\delta^1} \cdots.
\end{gather}
The relative algebraic de Rham cohomology $H_{\rm dR}^{\bullet}(X,D,\mathbb{K})$ is the cohomology of the total cochain complex $\Omega^{\bullet}(X,D)$, that is
\begin{gather}
H^k_{\rm dR}(X,D,\mathbb{K}) = \begin{cases}
 \mathop{\rm Ker}\big(\delta^0\big), & k=0,
 \\
 \mathop{\rm Ker}\big(\delta^k\big)/\Im\big(\delta^{k-1}\big), & k \ge 1.
 \end{cases}
\end{gather}
The following proposition is a consequence of the surjectivity of the restriction maps ${\rm d}_{IJ}$.
\begin{Proposition} \label{prop: topdeg}
Let $X$ be a smooth affine variety over $\mathbb{K}$ of dimension $n$ and $D \subset X$ a divisor with simple normal crossings.
Each class in the top-degree cohomology group $H_{\rm dR}^n(X,D,\mathbb{K})$ has a representative in $\Omega^n(X)$.
\end{Proposition}

\begin{Example} \label{ex: exx}
Let $X=\mathbb{G}_m=\mathop{\rm Spec} \mathbb{Q}[x,1/x]$ and $D=\{ 1, z \}$ with $z \in \mathbb{Q}$, $z \ne 1$. The cor\-res\-ponding double algebraic de Rham complex is
\begin{equation}
\begin{tikzcd}
0 & & \\
\displaystyle\mathbb{Q}\bigg[ x, \frac{1}{x} \bigg] {\rm d}x \arrow[u, "{\rm d}"] \arrow[r] & 0 & \\
\displaystyle\mathbb{Q}\bigg[ x, \frac{1}{x} \bigg] \arrow[u, "{\rm d}"] \arrow[r] & \mathbb{Q} \oplus \mathbb{Q} \arrow[u, "-{\rm d}"] \arrow[r] & 0,
\end{tikzcd}
\end{equation}
where the only non-trivial horizontal differential is the evaluation map
\begin{align}
\mathbb{Q}\bigg[ x, \frac{1}{x} \bigg] & \longrightarrow \mathbb{Q} \oplus \mathbb{Q},
\\
f & \longmapsto (f(1), f(z)).
\end{align}
The corresponding total complex is
\begin{align}
\mathbb{Q}\bigg[x,\frac{1}{x}\bigg] & \xlongrightarrow{\delta^0} \mathbb{Q}\bigg[x, \frac{1}{x}\bigg] {\rm d}x \oplus \mathbb{Q} \oplus \mathbb{Q},
\\
f(x) & \longmapsto (f'(x){\rm d}x, f(1), f(z)),
\end{align}
where the only non-trivial differential is explicitly written.
The non-trivial relative algebraic de Rham cohomology groups are
\begin{gather}
H^0_{\rm dR}(X,D,\mathbb{Q}) = \mathop{\rm Ker}\big(\delta^0\big) = 0 ,
\\
H^1_{\rm dR}(X,D,\mathbb{Q}) = \mathop{\rm coKer}\big(\delta^0\big) = \frac{\mathbb{Q}\left[ x, \frac{1}{x} \right] {\rm d}x \oplus \mathbb{Q} \oplus \mathbb{Q}}{\Im(\delta^0)}
\end{gather}
and a basis of $H^1_{\rm dR}(X,D)$ is given by the classes $\big[ \big( \frac{{\rm d}x}{x},0,0 \big) \big] = \big[\frac{{\rm d}x}{x} \big]$ and $ \big[ \big( \frac{{\rm d}x}{z-1},0,0 \big) \big] = \big[\frac{{\rm d}x}{z-1}\big]$.
\end{Example}

\subsection{Comparison isomorphism}
The following fundamental theorem is due to Grothendieck~\cite{Gro66}.
\begin{Theorem} \label{th: comp}
Let $\mathbb{K}$ be a subfield of $\mathbb{C}$ and let $X$ be a smooth algebraic variety over $\mathbb{K}$. There is a canonical isomorphism
\begin{gather}\label{eq: comp_is}
\mathop{\rm comp}\colon\ H^{\bullet}_{\rm dR} (X, \mathbb{K}) \otimes_{\mathbb{K}} \mathbb{C} \xlongrightarrow{\sim} H^{\bullet}_{{\rm B}}(X,\mathbb{Q}) \otimes_{\mathbb{Q}} \mathbb{C}
\end{gather}
known as \textit{comparison isomorphism}. Moreover, if $Y \subset X$ is closed subvariety, we have
\begin{gather}
\mathop{\rm comp}\colon\ H^{\bullet}_{\rm dR} (X, Y, \mathbb{K}) \otimes_{\mathbb{K}} \mathbb{C} \xlongrightarrow{\sim} H^{\bullet}_{{\rm B}}(X, Y,\mathbb{Q}) \otimes_{\mathbb{Q}} \mathbb{C}.
\end{gather}
\end{Theorem}

As mentioned in Definition~\ref{def: betti}, any algebraic $\mathbb{K}$-variety $X$ canonically yields an analytic complex space $X^{\rm an}$ associated with the space of complex points $X(\mathbb{C})$. If~$X$ is smooth, then~$X^{\rm an}$ is an analytic complex manifold. The classical theory of de Rham cohomology, discussed in Section~\ref{sec: smooth} for differentiable manifolds, extends to complex geometry as well.
\begin{Definition}
Let $M$ be a complex manifold of dimension $n$.
For $p,q \ge 0$, a differential form of \textit{holomorphic degree $p$} and \textit{antiholomorphic degree $q$} on $M$, also called a \textit{differential $(p,q)$-form}, is written in local analytic coordinates as
\begin{gather} \label{eq: diffC}
\sum_{I, J} f_{IJ}\, {\rm d}z_{i_1} \wedge \dots \wedge {\rm d}z_{i_p} \wedge {\rm d}\bar{z}_{j_1} \wedge \dots \wedge {\rm d}\bar{z}_{j_q},
\end{gather}
where the sum runs over the index subsets $I=\{ i_1,\dots,i_p \}, J = \{ j_1,\dots,j_q \} \subseteq \{1,\dots,n\}$ and $f_{IJ}$ are $\mathcal{C}^{\infty}$-functions. Let~$\Omega^{p,q}(M)$ denote the $\mathbb{C}$-vector space of differential $(p,q)$-forms on~$M$.
The standard de Rham differential splits as ${\rm d} = \partial + \bar{\partial}$, where $\partial\colon \Omega^{p,q} \rightarrow \Omega^{p+1,q}$ and $\bar{\partial}\colon \Omega^{p,q} \rightarrow \Omega^{p,q+1}$.
The \textit{Dolbeault cohomology} of $M$, denoted by $H^{\bullet, \bullet}_{{\rm D}}(M, \mathbb{C})$, is the cohomology of the double cochain complex $(\Omega^{\bullet, \bullet}(M), \bar{\partial})$, called the \textit{Dolbeault complex} of $M$.
\end{Definition}
\begin{Definition}
A \textit{holomorphic $t$-form} on $M$ is a finite sum of differential $(p,q)$-forms on $M$ with $p+q=t$ that are locally expressed as in~\eqref{eq: diffC} with the coefficient functions $f_{IJ}$ being holomorphic.
The \textit{holomorphic de Rham cohomology}\footnote{A singular version of the holomorphic de Rham complex, called logarithmic de Rham complex, is obtained by considering meromorphic forms which are holomorphic in the bulk, but admit logarithmic poles towards the compactification boundaries of $X$.} of $M$, denoted by $H^{\bullet}_{\rm dR}(M, \mathbb{C})$, is the cohomology of the cochain complex associated with the graded $\mathbb{C}$-vector space of holomorphic forms on $M$ and the holomorphic component $\partial$ of the de Rham differential.\footnote{For a formally rigorous definition of holomorphic de Rham cohomology, using the tools of sheaf cohomology and hypercohomology, see Voisin~\cite{Voi02}.}
\end{Definition}

The following proposition is the complex analogue of Theorem~\ref{th: derham} by de Rham.
\begin{Proposition} \label{prop: derhamC}
Let $M$ be a complex manifold of dimension $n$. For $0 \le k \le n$, there is an isomorphism
\begin{gather} \label{eq: derhamC1}
H^k_{\rm dR}(M,\mathbb{C}) \xlongrightarrow{\sim} H^k_{\text{s}}(M, \mathbb{Q}) \otimes_{\mathbb{Q}} \mathbb{C}.
\end{gather}
In particular, if $M=X^{\rm an}$, where $X$ is a smooth algebraic $\mathbb{K}$-variety of dimension $n$, we have
\begin{gather} \label{eq: derhamC2}
H^k_{\rm dR}(X^{\rm an},\mathbb{C}) \xlongrightarrow{\sim} H^k_{{\rm B}}(X,\mathbb{Q}) \otimes_{\mathbb{Q}} \mathbb{C}
\end{gather}
for $0 \le k \le n$.
\end{Proposition}

\begin{Remark}
The holomorphic de Rham complex of $X^{\rm an}$ is equivalently obtained by analytification of the algebraic de Rham complex of $X$, and an equivalent notion of holomorphic de Rham cohomology of $X^{\rm an}$ follows canonically. Thus, the procedure of analytification of an algebraic variety and the properties of holomorphic de Rham cohomology provide the conceptual link between algebraic de Rham cohomology and Betti cohomology, which underlies Grothendieck's comparison isomorphism.
\end{Remark}

{\sloppy\begin{Remark}
An important observation follows from combining Grothendieck's and de Rham's theorems. Given a smooth algebraic variety $X$ over a subfield $\mathbb{K}$ of $\mathbb{C}$, the holomorphic de Rham cohomology of the underlying topological space $X^{\rm an}$, equivalent to its singular cohomology, is isomorphic to the algebraic de Rham cohomology of $X$ after complexification, that~is
\begin{gather} \label{eq: combo}
H^k_{\rm dR}(X^{\rm an}, \mathbb{C}) \simeq H^k_{\rm dR}(X, \mathbb{K}) \otimes_{\mathbb{K}} \mathbb{C}
\end{gather}
for $k \ge 0$. The holomorphic de Rham cohomology can therefore be computed considering algebraic forms only. In this way, a purely algebraic definition of cohomology is obtained.
\end{Remark}

}

\subsection{Pure Hodge structures} \label{sec: pure}
As a consequence of Theorem~\ref{th: comp}, the Betti cohomology of an algebraic variety is endowed with a richer structure than the singular cohomology of a generic topological space.
Recall the following definition.
\begin{Definition}
Let $H$ be a finite-dimensional $\mathbb{Q}$-vector space and let $H_{\mathbb{C}}= H \otimes_{\mathbb{Q}} \mathbb{C}$. Assume that $H_{\mathbb{C}}$ possesses a bigrading
\begin{gather}
H_{\mathbb{C}} = \bigoplus_{p+q=k} H^{p,q}
\end{gather}
for some integer $k$, satisfying the property $\overline{H^{p,q}} = H^{q,p}$, called \textit{Hodge symmetry}. $H$ is called a~\textit{pure Hodge structure of weight $k$} and the given direct sum decomposition of its complexification~$H_{\mathbb{C}}$ is called a \textit{Hodge decomposition}.
\end{Definition}

\begin{Remark}
An equivalent definition of pure Hodge structure of weight $k$ is obtained by~observing that the data encoded in the Hodge decomposition is equivalent to a finite decreasing filtration $F^{\bullet}$ of $H_{\mathbb{C}}$, called \textit{Hodge filtration}, such that
\begin{gather}
F^pH_{\mathbb{C}} \oplus \overline{F^{k-p+1}H_{\mathbb{C}}} = H_{\mathbb{C}}
\end{gather}
for all integers $p$. The two equivalent descriptions are related by
\begin{gather}
H^{p,q} = F^pH_{\mathbb{C}} \cap \overline{F^qH_{\mathbb{C}}} , \qquad
F^pH_{\mathbb{C}} = \bigoplus_{i \ge p} H^{i, k-i}
\end{gather}
for $p$, $q$ integers such that $p+q = k$.
\end{Remark}

Let $M$ be a compact K\"ahler\footnote{A K\"ahler manifold is a manifold with a complex structure, a Riemannian structure, and a symplectic structure which are mutually compatible.} manifold. For $p,q \ge 0$, its Dolbeault cohomology classes in~bidegree $(p,q)$ uniquely correspond\footnote{This result, true for any compact hermitian complex manifold, is known as Hodge isomorphism.} to the \textit{harmonic $(p,q)$-forms} on $M$, and there are canonical maps
\begin{gather}
H^{p, q}_{{\rm D}}(M, \mathbb{C}) \rightarrow H^{p+q}_{\rm dR}(M,\mathbb{C}) \simeq H^{p+q}_{\text{s}}(M,\mathbb{C}).
\end{gather}
The following theorem by Hodge~\cite{Hod41} marks the beginning of what is currently known as \textit{Hodge theory}.
\begin{Theorem} \label{th: dec}
Let $M$ be a compact K\"ahler manifold.
The following direct sum decomposition
\begin{gather} \label{eq: hodge}
H^k_{\rm dR}(M,\mathbb{C}) = \bigoplus_{p+q=k} H^{p,q}_{{\rm D}}(M, \mathbb{C})
\end{gather}
holds for $k \ge 0$.
\end{Theorem}
\begin{Remark}
Note that the complex conjugate of $H^{p,q}(M)$ is $H^{q,p}(M)$. Following equation~\eqref{eq: derhamC1}, the ordinary cohomology group $H^k_{\text{s}}(M,\mathbb{Q})$ is a pure Hodge structure of weight $k$, and Hodge's theorem gives a decomposition\footnote{The Hodge decomposition of a compact K\"ahler manifold is independent of the choice of K\"ahler metric, although there is no analogous decomposition for arbitrary compact complex manifolds.} of its complexification as a direct sum of $\mathbb{C}$-vector spaces.
\end{Remark}

Let $X$ be a smooth projective variety defined over a subfield $\mathbb{K}$ of $\mathbb{C}$.
$X^{\rm an}$ is a compact K\"ahler manifold, and thus the Hodge decomposition and filtration are defined on $H^k_{\rm dR}(X^{\rm an}, \mathbb{C})$ for $k \ge 0$. Following equation~\eqref{eq: derhamC2}, the Betti cohomology groups of $X$ are then pure Hodge structures of~weights equal to their degrees.
Moreover, it can be proven\footnote{This result follows non-trivially from the interpretation of the Hodge filtration in terms of hypercohomology with coefficients in the complex of holomorphic forms, and the GAGA theorem. We~refer to Voisin~\cite{Voi02}.} that the Hodge filtration $F^{\bullet}$, which makes the Betti cohomology group $H^k_{{\rm B}}(X,\mathbb{Q})$ into a pure Hodge structure of~weight $k$, directly acts on the algebraic de Rham cohomology groups of $X$ over $\mathbb{K}$. Precisely, there is an~integer $n$ such that $F^{\bullet}$ is a finite decreasing filtration on $H_{\mathbb{K}} = H^k_{\rm dR} (X, \mathbb{K})$ satisfying
\begin{gather}
F^pH_{\mathbb{K}} \oplus \overline{F^{n-p+1}H_{\mathbb{K}}} = H_{\mathbb{K}}
\end{gather}
for all integers $p$.
To keep track of all these structures, we define the formal assignment $X \mapsto H^{\bullet}(X)$, where $H^k(X)$ is the triple of data given by
\begin{gather}
H^k(X) = \big(\big(H^k_{\rm dR}(X, \mathbb{K}), F^{\bullet}\big), H^k_{{\rm B}}(X, \mathbb{Q}), {\rm comp}\big)
\end{gather}
for $k \ge 0$. We~call $H^k(X)$ a \textit{(pure) de Rham and Betti system of realisations}, or shortly a \textit{(pure) $H$-system}, of weight $n$ over $\mathbb{K}$. Observe that the weight of $H^k(X)$ is defined by the action of the Hodge filtration on the algebraic de Rham cohomology, and it does not generally equal the degree $k$.

\begin{Definition}
Let $X, X'$ be smooth projective $\mathbb{K}$-varieties.
Write $H = H^k(X)$ and $H'= H^{k'}(X')$, where $k,k' \ge 0$. A~\textit{morphism of pure $H$-systems} $f \colon H \rightarrow H'$ is a pair $f = (f_{\rm dR}, f_{{\rm B}})$ consisting of a $\mathbb{K}$-linear map $f_{\rm dR}\colon H_{\rm dR} \rightarrow H_{\rm dR}'$ and a $\mathbb{Q}$-linear map $f_{{\rm B}}\colon H_{{\rm B}} \rightarrow H_{{\rm B}}'$ such that:
\begin{itemize}\itemsep=0pt
\item[(1)] $f_{\rm dR}$ is filtered\footnote{Let $\mathbb{K}$ be a subfield of $\mathbb{C}$ and $(V,F)$, $(V',F)$ be filtered $\mathbb{K}$-vector spaces. A~morphism $f\colon V \rightarrow V'$ is called \textit{filtered} if $f(F^pV) \subseteq F^pV'$ for each $p \ge 0$.} with respect to the Hodge filtration, that is
\begin{gather}
f_{\rm dR}(F^{\bullet}H_{\rm dR}) \subseteq F^{\bullet}H_{\rm dR}'.
\end{gather}
\item[(2)] The following diagram commutes:
\begin{equation}
\begin{tikzcd}[column sep=large]
H_{\rm dR} \otimes_{\mathbb{K}} \mathbb{C} \arrow[r, "\mathop{\rm comp}"] \arrow[d, "f_{\rm dR} \otimes_{\mathbb{K}} \text{Id}_{\mathbb{C}}"'] & H_{{\rm B}} \otimes_{\mathbb{Q}} \mathbb{C} \arrow[d, "f_{{\rm B}} \otimes_{\mathbb{Q}} \text{Id}_{\mathbb{C}}"]\\
H_{\rm dR}' \otimes_{\mathbb{K}} \mathbb{C} \arrow[r, "\mathop{\rm comp}'"'] & H_{{\rm B}}' \otimes_{\mathbb{Q}} \mathbb{C}.
\end{tikzcd}
\end{equation}
\end{itemize}
\end{Definition}
Observe that, if $H$ and $H'$ have different weights, then every morphism between them is zero.
The following variant of Theorem~\ref{th: dec} implies that pure $H$-systems are functorial for morphisms of smooth projective varieties.
\begin{Theorem} \label{th: smoothHf}
Let $X$, $X'$ be smooth projective varieties over $\mathbb{K}$. For any morphism $f\colon X \rightarrow X'$, the induced map on cohomology $f^*\colon H^{\bullet}(X') \rightarrow H^{\bullet}(X)$ is a morphism of pure $H$-systems.
\end{Theorem}

\begin{Example} \label{ex: hodgetate}
For each $n \in \mathbb{Z}$, we define
\begin{gather}
\mathbb{Q}(n) = ((\mathbb{K}, F^{\bullet}), \mathbb{Q}, \mathop{\rm comp}),
\end{gather}
where the filtration yields $\mathbb{K} = F^{-n} \mathbb{K} \supseteq F^{-n+1} \mathbb{K} = 0$ and the isomorphism $\mathop{\rm comp}\colon \mathbb{C} \rightarrow \mathbb{C}$ is given by multiplication by $(2 \pi {\rm i} )^{-n}$. $\mathbb{Q}(n)$ is a one-dimensional pure $H$-system of weight $-2n$ over $\mathbb{K}$ and is called a \textit{Tate--Hodge system}. As an example, $\mathbb{Q}(-1)$ is isomorphic to $H^1(\mathbb{G}_m)= \big(\big(H^1_{\rm dR}(\mathbb{G}_m), F^{\bullet}\big), H^1_{{\rm B}}(\mathbb{G}_m), {\rm comp}\big)$, where $F^{\bullet}$ is the trivial filtration concentrated in degree $1$. Observe that $\mathbb{Q}(-1)$ is a pure $H$-system of weight~$2$, although $H^1(\mathbb{G}_m)$ has degree~$1$.
\end{Example}

\subsection{Mixed Hodge structures} \label{sec: mixed}
The Betti cohomology in degree $k$ of a smooth projective $\mathbb{K}$-variety $X$ carries canonically a~pure Hodge structure of weight $k$. However, this is no longer true when $X$ fails to be smooth or~projective.
The generalisation of the notion of pure Hodge structure to the case of quasi-projective varieties is due to Deligne~\cite{Del71_1,Del71_2,Del74}, who proved that the Betti cohomology of a~quasi-projective variety over a~subfield $\mathbb{K}$ of $\mathbb{C}$ is an \textit{iterated extension} of pure Hodge structures.
Recall the following definition.
\begin{Definition} \label{def: mixH}
Let $H$ be a finite-dimensional $\mathbb{Q}$-vector space and let $H_{\mathbb{C}}= H \otimes_{\mathbb{Q}} \mathbb{C}$. Assume that $H$ possesses a finite increasing filtration $W_{\bullet}$, called \textit{weight filtration}, and that $H_{\mathbb{C}}$ possesses a finite decreasing filtration $F^{\bullet}$, called \textit{Hodge filtration}, such that, for all integers $m$, the $m$-th graded quotient of $H$ with respect to $W_{\bullet}$
\begin{gather}
{\rm Gr}^W_m H = W_m / W_{m-1}
\end{gather}
together with the filtration induced by $F^{\bullet}$ on its complexification
\begin{gather}
F^{\bullet} {\rm Gr}^W_m H = ( F^{\bullet} \cap W_m \otimes \mathbb{C} + W_{m-1} \otimes \mathbb{C}) / W_{m-1} \otimes \mathbb{C}
\end{gather}
is a pure Hodge structure of weight $m$. $H$ is called a \textit{mixed Hodge structure}.
\end{Definition}
\begin{Remark}
Let $H$ be a mixed Hodge structure. For all integers $m$, there is a short exact sequence
\begin{gather}
0 \rightarrow W_{m-1} \rightarrow W_m \rightarrow {\rm Gr}^W_m H \rightarrow 0.
\end{gather}
Take $m = h$ to be the highest weight of $H$, defined by $W_h = H$. The short exact sequence above gives $H$ as an extension of the pure Hodge structure ${\rm Gr}^W_h H$ by $W_{h-1}$. Analogously, taking $m = h-1$, $W_{h-1}$ is an extension of ${\rm Gr}^W_{h-1} H$ by $W_{h-2}$, which in turn is an extension of ${\rm Gr}^W_{h-2} H$ by $W_{h-3}$, and so on. In this way, mixed Hodge structures are explicitly realised as iterated extensions of pure ones.
\end{Remark}

\begin{Theorem}
Let $X$ be a quasi-projective variety over a subfield $\mathbb{K}$ of $\mathbb{C}$.
\begin{itemize}\itemsep=0pt
\item[$(1)$] For $k \ge 0$, its Betti cohomology group $H^k_{{\rm B}}(X,\mathbb{Q})$ is a mixed Hodge structure with respect to a weight filtration $W_{\bullet}$ and a Hodge filtration $F^{\bullet}$ which satisfy
\begin{gather}
W_{-1} = 0 \subseteq W_0 \subseteq W_1 \subseteq \dots \subseteq W_{2k} = H^k_{{\rm B}}(X,\mathbb{Q}) ,
\\
F^0 = H^k_{{\rm B}}(X,\mathbb{Q}) \otimes_{\mathbb{Q}} \mathbb{C} \supseteq F^1 \supseteq\dots \supseteq F^k \supseteq F^{k+1} = 0.
\end{gather}
{\sloppy If $X$ is smooth, then ${\rm Gr}^W_m H^k_{{\rm B}}(X,\mathbb{Q}) = 0$ for all $m < k$. If~$X$ is projective, then ${\rm Gr}^W_m H^k_{{\rm B}}(X,\mathbb{Q})= 0$ for all $m > k$.
\item[$(2)$] The Hodge filtration $F^{\bullet}$ acts on the algebraic de Rham cohomology groups of $X$ over $\mathbb{K}$, and the weight filtration $W^{{\rm B}}_{\bullet} = W_{\bullet}$ induces a corresponding weight filtration $W_{\bullet}^{\rm dR}$ on the algebraic de Rham cohomology groups of $X$ over $\mathbb{K}$.

 }
\item[$(3)$] The comparison isomorphism is filtered with respect to the weight filtration
\begin{gather}
\mathop{\rm comp}( W^{\rm dR}_{\bullet} \otimes_{\mathbb{K}} \mathbb{C}) = W^{{\rm B}}_{\bullet} \otimes_{\mathbb{Q}} \mathbb{C}.
\end{gather}
\end{itemize}
\end{Theorem}
Again, to keep track of the several structures that we have introduced, we define a formal assignment $X \mapsto H^{\bullet}(X)$, where $H^k(X)$ is the triple of data given by
\begin{gather}
H^k(X) = \big(\big(H^k_{\rm dR}(X, \mathbb{K}), F^{\bullet}, W_{\bullet}^{\rm dR}\big), \big(H^k_{{\rm B}}(X, \mathbb{Q}), W^{{\rm B}}_{\bullet}\big), \mathop{\rm comp}\!\big)
\end{gather}
for $k \ge 0$. We~call $H^k(X)$ a \textit{(mixed) de Rham and Betti system of realisations}, or shortly a~\textit{(mixed) $H$-system}, over $\mathbb{K}$. Observe that, for each integer $m$, the triple of data
\begin{gather}
 {\rm Gr}^W_m H =\big(\big({\rm Gr}^W_m H_{\rm dR}, F^{\bullet}\big), {\rm Gr}^W_m H_{{\rm B}}, \mathop{\rm comp}\!\big),
\end{gather}
where $H = H^k(X)$, is a pure $H$-system of weight $m$.

\begin{Definition}
Let $X$, $X'$ be quasi-projective $\mathbb{K}$-varieties.
Write $H = H^k(X)$ and $H'= H^{k'}(X')$, where $k,k' \ge 0$. A~\textit{morphism of mixed $H$-systems} $f\colon H \rightarrow H'$ is a pair $f = (f_{\rm dR}, f_{{\rm B}})$ consisting of a $\mathbb{K}$-linear map $f_{\rm dR}\colon H_{\rm dR} \rightarrow H_{\rm dR}'$ and a $\mathbb{Q}$-linear map $f_{{\rm B}}\colon H_{{\rm B}} \rightarrow H_{{\rm B}}'$ such that:
\begin{itemize}\itemsep=0pt
\item[$(1)$] $f_{{\rm B}}$ is filtered with respect to the weight filtration, that is
\begin{gather}
f_{{\rm B}}(W_{\bullet}^{{\rm B}} H_{{\rm B}}) \subseteq W_{\bullet}^{{\rm B}} H_{{\rm B}}'.
\end{gather}
\item[$(2)$] $f_{\rm dR}$ is filtered with respect to the weight and Hodge filtrations, that is
\begin{gather}
f_{\rm dR}\big(W_{\bullet}^{\rm dR} H_{\rm dR}\big) \subseteq W_{\bullet}^{\rm dR} H_{\rm dR}' .
\\
f_{\rm dR}\big(F^{\bullet}H_{\rm dR}\big) \subseteq F^{\bullet}H_{\rm dR}',
\end{gather}
\item[(3)] $f_{{\rm B}}$ and $f_{\rm dR}$ are compatible with the comparison isomorphism, that is
\begin{gather}
(f_{{\rm B}} \otimes_{\mathbb{Q}} \text{Id}_{\mathbb{C}}) \circ \mathop{\rm comp} = \mathop{\rm comp}\nolimits' \circ (f_{\rm dR} \otimes_{\mathbb{K}} \text{Id}_{\mathbb{C}}).
\end{gather}
\end{itemize}
\end{Definition}
The following analogue of Theorem~\ref{th: smoothHf} holds.
\begin{Theorem}
Let $X$, $X'$ be quasi-projective varieties over $\mathbb{K}$. For any morphism $f\colon X \rightarrow X'$, the induced map on cohomology $f^*\colon H^{\bullet}(X') \rightarrow H^{\bullet}(X)$ is a morphism of mixed $H$-systems.
\end{Theorem}

We denote by $\mathbf{MHSy}(\mathbb{Q})$ the category\footnote{Further aspects of de Rham and Betti systems of realisations are discussed by Brown~\cite{Bro17_1}.} of mixed $H$-systems over $\mathbb{Q}$.
Deligne~\cite{Del71_2} proved that $\mathbf{MHSy}(\mathbb{Q})$ is an abelian category.
Moreover, it is naturally endowed with two forgetful functors
\begin{gather}
\omega_{{\rm B}}\colon\ \mathbf{MHSy}(\mathbb{Q}) \rightarrow \text{Vec}_{\mathbb{Q}}, \qquad
\omega_{\rm dR}\colon\ \mathbf{MHSy}(\mathbb{Q}) \rightarrow \text{Vec}_{\mathbb{Q}}
\end{gather}
called \textit{Betti} and \textit{de Rham functors}, sending the mixed system of realisations $H \in \mathbf{MHSy}(\mathbb{Q})$ into the $\mathbb{Q}$-vector spaces $H_{{\rm B}}$ and $H_{\rm dR}$, respectively.

\section{Periods of motives} \label{sec: periods}

\subsection{Periods} \label{sec: num}
The following elementary definition was introduced by Kontsevich and Zagier~\cite{KZ01}.
\theoremstyle{definition}
\begin{Definition} \label{def: period}
A \textit{period} is a complex number whose real and imaginary parts are values of absolutely convergent integrals of the form
\begin{gather} \label{eq: per_int}
\int_{\sigma} f(x_1,\dots,x_n)\, {\rm d}x_1 \cdots {\rm d}x_n,
\end{gather}
where the integrand $f$ is a rational function with rational coefficients and the domain of integration $\sigma \subseteq \mathbb{R}^n$ is defined by finite unions and intersections of domains of the form $\{ g(x_1,\dots,x_n)$ $\ge 0 \}$
with $g$ a rational function with rational coefficients.
\end{Definition}
If rational functions and coefficients are replaced in Definition~\ref{def: period} by algebraic functions and coefficients, the same set of numbers is obtained. Indeed, algebraic functions in the integrand can be substituted by rational functions by enlarging the set of variables. Note that, because the integral of any real-valued function is equivalent to the volume subtended by its graph, any period admits a representation as the volume of a domain defined by polynomial inequalities with rational coefficients. Thus, the integrand can always be assumed to be the constant function~1. However, this extremely simplified framework does not prove to be particularly useful. Quite the opposite, in what follows, we mostly work with an even more general description of periods than the one given in Definition~\ref{def: period}. We~denote by $\mathcal{P}$ the set of periods.
Being $\bar{\mathbb{Q}} \subset \mathcal{P} \subset \mathbb{C}$, periods are generically transcendental numbers and nonetheless they contain only a finite amount of information, which is captured by the integrand and domain of integration of its integral representation as in~\eqref{eq: per_int}.
Indeed, just like $\bar{\mathbb{Q}}$, $\mathcal{P}$ is countable.
Many famous numbers belong to the class of periods. Here are some examples:
\begin{itemize}\itemsep=0pt
\item[$(a)$] Algebraic numbers are periods, e.g.,
\begin{gather}
\sqrt{2} = \int_{2x^2 \le 1} {\rm d}x.
\end{gather}
\item[$(b)$] Logarithms of algebraic numbers are periods, e.g.,
\begin{gather} \label{eq: log}
\log 2= \int_{1}^{2} \frac{{\rm d}x}{x}.
\end{gather}
\item[$(c)$] The transcendental number $\pi$ is a period, as given by
\begin{gather}
\pi = \int_{-1}^1 \frac{{\rm d}x}{\sqrt{1-x^2}} = \int\displaylimits_{-\infty}^{+\infty} \frac{{\rm d}x}{1+x^2} = \iint\displaylimits_{x^2 + y^2 \le 1} {\rm d}x {\rm d}y
\end{gather}
and alternatively by
\begin{gather} \label{eq: pi}
2 \pi {\rm i} = \oint_{\gamma_2} \frac{{\rm d}z}{z},
\end{gather}
where $\gamma_2$ is a closed path encircling the origin in the complex plane.
\item[$(d)$] Values of the beta function at positive rational arguments are periods, as given by
\begin{gather}
B(u,v) = \int_{0}^1 t^{u-1} (1-t)^{v-1} {\rm d}t, \qquad \mathop{\rm Re}(u),\, \mathop{\rm Re}(v) > 0,
\end{gather}
and values of the gamma function at positive rational arguments satisfy\footnote{The statement follows from the relation between the gamma and the beta functions $\Gamma(a_1) \cdots \Gamma(a_n) = \Gamma(a_1+\dots+ a_n) \prod_{i=1}^{n-1} B(a_1+\dots+ a_{i-1}, a_i)$ with $\mathop{\rm Re}(a_k)>0$, $k=1, \dots, n$.}
\begin{gather}
\Gamma \bigg( \frac{p}{q} \bigg)^q \in \mathcal{P}, \qquad p, q \in \mathbb{N}.
\end{gather}
\item[$(e)$] The elliptic integral
\begin{gather}
2 \int_{-b}^{b} \sqrt{1 + \frac{a^2 x^2}{b^4 - b^2 x^2}}\, {\rm d}x
\end{gather}
representing the perimeter of an ellipse with radii $a$ and $b$, is a period. Note that it is not an algebraic function of $\pi$ for $a \neq b$, $a, b \in \mathbb{Q}_{> 0}$,
\item[$(f)$] Values of the Riemann zeta function at integer arguments $s \ge 2$ are periods, e.g.,
\begin{gather}
\zeta (3) = 1 + \frac{1}{2^3} + \frac{1}{3^3} + \dots = \sum_{n=1}^{\infty} \frac{1}{n^3} = \iiint\displaylimits_{0<x<y<z<1} \frac{{\rm d}x {\rm d}y {\rm d}z}{(1-x)yz},
\end{gather}
and more generally multiple zeta values are periods by means of their integral representation~\eqref{eq: zint}.
\item[$(g)$] Convergent Feynman integrals, as in~\eqref{eq: I_G_last}, are periods. Moreover, removing the convergence requirement, the statement suitably extends to a wider class of Feynman integrals.\footnote{Under some assumptions, Bogner and Weinzierl~\cite{BWdiv} showed that the coefficients appearing in the Laurent series of any scalar multi-loop integral are periods.}
\item[$(h)$] Special values at algebraic arguments of hypergeometric functions, values of modular forms at suitable arguments, and values of various kinds of L-functions are periods.
\end{itemize}

Because the integral representation of a period is not unique, it is possible that a certain integral of a transcendental function admits a representation as a period as well.
For example, $\log(2)$ is a period, and yet it can be written as the following integral of a transcendental function
\begin{gather}
\int\displaylimits_{0}^{1} \frac{x}{\log \frac{1}{1-x}}\, {\rm d}x.
\end{gather}
Indeed, there seems to be no general principle able to predict if a certain infinite sum or integral of a transcendental function is a period according to Definition~\ref{def: period}, or able to determine whether two periods, given by explicit integrals, are equal or different. A~number in $\bar{\mathbb{Q}}$ also admits apparently different expressions, but those same techniques that work for checking the equality of algebraic numbers do not in general work for periods. In fact, two different periods may be~numerically very close and yet be distinct.\footnote{For example, the approximation $\pi = \frac{6}{\sqrt{3502}} \log(2 u) + 7.37 \times 10^{-82}$, where $u$ is the product of four quartic units, has been found by Shanks~\cite{Sha82}.}
However, the following conjecture is presented by~Kontsevich and Zagier~\cite{KZ01}.

\begin{Conjecture} \label{conj: per1}
If a period has two different integral representations, then one expression can be transformed into the other by application of the three integral transformation rules of~addi\-ti\-vity, change of variables, and Stokes' formula, in which all integrands and domains of~integration are algebraic with algebraic coefficients.
\end{Conjecture}

We note that even a proof of Conjecture~\ref{conj: per1} would not solve the additional problem of finding an algorithm to determine whether or not two given numbers in $\mathcal{P}$ are equal, or whether or not a~given real number belongs to $\mathcal{P}$. Another fundamental open problem in the theory of periods is to explicitly exhibit one number which does not belong to $\mathcal{P}$. Such numbers must exist, because~$\mathcal{P}$ is a countable subset of $\mathbb{C}$, but the concrete identification of one of such numbers has only been proposed conjecturally. Indeed, the basis of natural logarithms $e$ and the Euler--Mascheroni constant $\gamma$ are conjecturally not periods.
Several further questions on the arithmetic nature and transcendence of periods are open or only conjecturally answered.\footnote{See Waldschmidt~\cite{Wal06} for an overview of the topic.}

Before moving to a more sophisticated definition of periods written in the language of algebraic geometry, which is essential to subsequent developments, we mention the fruitful interplay between the theory of periods and the theory of linear differential equations. When the inte\-g\-rands or the domains of integration depend on some set of parameters, the integrals, as functions of these parameters, usually satisfy linear differential equations with algebraic coefficients. The solutions of these differential equations generate periods when evaluated at algebraic arguments. The differential equations occurring in this way are called \textit{Picard--Fuchs differential equations}.
The relation between periods and Picard--Fuchs equations has proved to be particularly productive in the case of elliptic curves, hypergeometric functions, modular forms and L-functions.

\subsection{Algebra of motivic periods} \label{sec: mot_alg}
The theory of periods is alternatively developed within the formalism of algebraic geometry. We~refer to Huber and M\"{u}ller-Stach~\cite{HM17}.
\begin{Definition} \label{def: period2}
Let $X$ be a smooth quasi-projective variety defined over $\bar{\mathbb{Q}}$, and $Y \subset X$ a~clo\-sed subvariety. A~\textit{period} is a complex number which can be expressed as an integral of the form $\int_{\gamma} \omega \in \mathbb{C}$, where $\omega$ is a closed algebraic differential $k$-form on $X$ vanishing on $Y$, and $\gamma$ is a~singular $k$-chain on the complex manifold $X^{\rm an}$ with boundary contained in $Y^{\rm an}$ for some integer $k \ge 0$.
\end{Definition}
The equivalence of Definitions~\ref{def: period2} and~\ref{def: period} follows from the observation that the algebraic chain $\gamma$ can be deformed to a semi-algebraic chain and then broken up into small pieces, which can be bijectively projected onto open domains in $\mathbb{R}^n$ with algebraic boundary.
Without loss of generality, we work with coefficients in $\mathbb{Q}$ instead of $\bar{\mathbb{Q}}$. We~note that, like Definition~\ref{def: period}, Definition~\ref{def: period2} also contains redundancy. The integral $\int_{\gamma} \omega$ can be formally decomposed into the quadruple
\begin{gather}
(X, Y, \omega, \gamma)
\end{gather}
and different quadruples can give the same resulting number.
To get rid of this redundancy, the various forms of topological invariance of the integral must be suitably accounted for.
Following Stokes' theorem, the integral is insensitive to the individual cycle and form, being instead determined by the homology and cohomology classes of these. Let~us associate to $\omega$ its cohomology class in the $k$-th algebraic de Rham cohomology group of $X$ relative to $Y$, and to $\gamma$ its homology class in the $k$-th Betti homology group of $X$ relative to $Y$.
Then, the first step towards a unique algebraic description of periods consists of the following substitutions
\begin{gather}
\omega \longrightarrow [\omega] \in H^k_{\rm dR} (X, Y,\mathbb{Q}) ,
\\
\gamma \longrightarrow [\gamma] \in H_k^{{\rm B}}(X,Y,\mathbb{Q})
\end{gather}
into the quadruple $(X,Y,\omega,\gamma)$.
The problem of the coexistence of distinct, but similarly behaved, cohomologies associated to the same variety, which seems to imply an arbitrary choice here and in many other situations, has been tackled by Grothendieck\footnote{
Grothendieck proposed the notion of a motive in a letter to Serre in 1964. He himself did not author any publication on motives, although he mentioned them frequently in his correspondence. The first formal expositions of the theory of motives are due to Demazure~\cite{Dem71} and Kleiman~\cite{Kle72}, who based their work on Grothendieck's lectures on the topic.}~\cite{CS01} with the introduction of the \textit{theory of motives}.
He suggested that there should exist a universal cohomology theory taking values in a $\mathbb{Q}$-category of motives.
The notion of a motive is thus proposed to capture the intrinsic cohomological essence of a variety.
Without delving into the category-theoretic details of the theory of motives\footnote{For a thorough introduction to the theory of motives, we refer to Voevodsky~\cite{Voe00}, Andr\'e~\cite{And04}, Deligne and Goncharov~\cite{DG05}, and Murre et al~\cite{MNP13}.}, we give here a conceptual introduction to its specific application to the theory of periods. Further discussion on the fundamental features of motives, as they appear in the study of periods, is presented in Section~\ref{sec: motives} in a more rigorous formalism.
Recall from Theorem~\ref{th: comp} that there is a comparison isomorphism
\begin{gather}
\mathop{\rm comp}\colon\ H^k_{\rm dR} (X, Y,\mathbb{Q}) \otimes_{\mathbb{Q}} \mathbb{C} \xlongrightarrow{\sim} H^k_{{\rm B}}(X, Y,\mathbb{Q}) \otimes_{\mathbb{Q}} \mathbb{C}
\end{gather}
induced by the pairing
\begin{align}
H^k_{\rm dR} (X, Y,\mathbb{Q}) \times H_k^{\text{s}} \big(X^{\rm an}, Y^{\rm an}, \mathbb{Q}\big) & \longrightarrow \mathbb{C},
\\[1ex]
([\omega] , [\gamma]) & \longmapsto \int_{\gamma} \omega.
\end{align}
Momentarily neglecting the presence of filtrations for simplicity, the de Rham and Betti system of realisations of $X$ relative to $Y$ in degree $k$ is
\begin{gather}
H^k(X,Y) = \big(H^k_{\rm dR} (X, Y,\mathbb{Q}), H^k_{{\rm B}}(X, Y,\mathbb{Q}), {\rm comp}\big).
\end{gather}
In the same way that the cohomology class of a differential form singles out its cohomological behaviour, the $H$-system of an algebraic variety intuitively selects the content shared by its coexisting algebraic de Rham and Betti cohomologies, and it filters out everything else. It~is, therefore, a first approximation towards the realisation of Grothendieck's idea of a motive. We~define the motivic version of the period $\int_{\gamma} \omega$ as the triple
\begin{gather}
\big[H^k(X,Y), [\omega], [\gamma]\big]^{\rm m},
\end{gather}
where $\text{m}$ in the superscript stands for motivic. We~call a period in this guise a \textit{motivic period}.
This has proved to be the most profitable reformulation of the original notion of a period.
However, a second source of redundancy in the description of periods via the integral formulation in Definition~\ref{def: period2}, corresponding to the same transformation rules in Conjecture~\ref{conj: per1}, has yet to be factored out.
\begin{Definition} \label{def: algebra_p}
The space $\mathcal{P}^{\rm m}$ of motivic periods is defined as the $\mathbb{Q}$-vector space\footnote{In what follows, we no longer display the field $\mathbb{Q}$ among the arguments of the cohomology groups for simplicity of notation.} generated by the symbols $[H^{\bullet}(X,Y), [\omega], [\gamma]]^{\rm m}$ after factorisation modulo the following three equivalence relations:
\begin{itemize}\itemsep=0pt
\item[(1)] \textit{Bilinearity}. $[H^{\bullet}(X,Y), [\omega], [\gamma]]^{\rm m}$ is bilinear in $[\omega]$ and $[\gamma]$.
\item[(2)] \textit{Change of variables}. If~$f\colon (X_1,Y_1) \rightarrow (X_2,Y_2)$ is a $\mathbb{Q}$-morphism of pairs of algebraic varieties, $[\gamma_1] \in H^{{\rm B}}_{\bullet}(X_1,Y_1)$ and $[\omega_2] \in H^{\bullet}_{\rm dR} (X_2,Y_2)$, then
\begin{gather}
[H^{\bullet}(X_1,Y_1), f^* [\omega_2], [\gamma_1]]^{\rm m} = [H^{\bullet}(X_2,Y_2), [\omega_2], f_* [\gamma_1]]^{\rm m},
\end{gather}
where $f^*$ and $f_*$ are the pull-back and the push-forward of $f$, respectively.
\item[(3)] \textit{Stokes' formula}. Assume for simplicity that $X$ is a smooth affine algebraic variety over~$\mathbb{Q}$ of dimension $d$ and $D \subset X$ is a simple normal crossing divisor.
The normalisation\footnote{$\tilde{D}$ is the disjoint union of the irreducible components of $D$.}~$\tilde{D}$ of~$D$ contains a simple normal crossing divisor $\tilde{D}_1$ coming from double points in $D$. If~$[\omega] \in H^{d-1}_{\rm dR} \big(\tilde{D}, \tilde{D}_1\big)$ and $[\gamma] \in H^{{\rm B}}_d(X,D)$, then
\begin{gather}
[H^d(X,D), \delta [\omega], [\gamma]]^{\rm m} = \big[H^{d-1}(\tilde{D}, \tilde{D}_1), [\omega], \partial [\gamma]\big]^{\rm m},
\end{gather}
where $\delta\colon H^{d-1}_{\rm dR} \big(\tilde{D}, \tilde{D}_1\big) \rightarrow H^d_{\rm dR} (X,D)$ is the coboundary operator acting on the algebraic de Rham cohomology and $\partial\colon H_d^{{\rm B}} (X, D) \rightarrow H_{d-1}^{{\rm B}} \big(\tilde{D}, \tilde{D}_1\big)$ is the boundary operator acting on the Betti homology.
\end{itemize}
\end{Definition}
We observe that the space of motivic periods $\mathcal{P}^{\rm m}$ is naturally endowed with an algebra structure. Indeed, new periods are obtained by taking sums and products of known ones.

\subsection{Period map} \label{sec: periodmap}
We call \textit{period map} the evaluation homomorphism
\begin{align} \label{eq: per_map}
\mathop{\rm per}\nolimits\colon\ \mathcal{P}^{\rm m} &\longrightarrow \mathcal{P} ,
\\
{}\big[H^k(X,Y), [\omega], [\gamma]\big]^{\rm m} &\longmapsto [\gamma] \circ \mathop{\rm comp} \circ [ \omega] = \int_{\gamma} \omega.
\end{align}
Following the construction in Section~\ref{sec: mot_alg}, the period map is explicitly surjective. Its injectivity is, on the other hand, not proven.
Indeed, a period has a unique motivic realisation only conjecturally.
Conjecture~\ref{conj: per1} is equivalent to the \textit{period conjecture} below.
\begin{Conjecture} \label{conj: per2}
The period map $\mathop{\rm per}\nolimits\colon \mathcal{P}^{\rm m} \rightarrow \mathcal{P}$ is an isomorphism.
\end{Conjecture}
Let us briefly discuss the key idea underlying the period conjecture. A~$\mathbb{Q}$-morphism $f\colon$ $(X_1,Y_1) \rightarrow (X_2,Y_2)$ between two pairs of algebraic varieties induces a \textit{change of coordinates} between the corresponding algebraic de Rham cohomologies by pull-back, that is
\begin{equation}
\begin{tikzcd}
(X_1,Y_1)\arrow[r, squiggly] \arrow[d, "f"]
 & H^{\bullet}_{\rm dR}(X_1,Y_1) \\
(X_2,Y_2) \arrow[r, squiggly]
	& H^{\bullet}_{\rm dR}(X_2,Y_2). \arrow[u, "f^*"]
\end{tikzcd}
\end{equation}
The same morphism $f$ acts on the topological spaces of complex points underlying the given algebraic varieties, and it induces a change of coordinates between the corresponding singular homologies by push-forward, that is
\begin{equation}
\begin{tikzcd}
\big(X_1^{\rm an},Y_1^{\rm an}\big)\arrow[r, squiggly] \arrow[d, "f"]
 & H_{\bullet}^{\text{s}}\big(X_1^{\rm an},Y_1^{\rm an}\big) \arrow[d, "f_*"] \\
\big(X_2^{\rm an},Y_2^{\rm an}\big) \arrow[r, squiggly]
	& H_{\bullet}^{\text{s}}\big(X_2^{\rm an},Y_2^{\rm an}\big) .
\end{tikzcd}
\end{equation}
By means of such changes of coordinates, one can easily derive two distinct integral representations of the same period.
For example, taking $[\gamma_1] \in H_{\bullet}^{\text{s}}\big(X_1^{\rm an},Y_1^{\rm an}\big)$ and $[\omega_2] \in H^{\bullet}_{\rm dR}(X_2,Y_2)$, we~have
\begin{gather}
\int_{[\gamma_1]} f^*[\omega_2] = \int_{f_* [\gamma_1]} [\omega_2].
\end{gather}
The corresponding two motivic representations of the same period
\begin{gather}
[H^{\bullet}(X_1,Y_1), f^*[\omega_2], [\gamma_1]]^{\rm m}, \qquad
[H^{\bullet}(X_2,Y_2), [\omega_2], f_* [\gamma_1]]^{\rm m}
\end{gather}
could a priori be different motivic periods. However, they are identified with each other by change of variables.
Indeed, the period conjecture corresponds to the statement that, whenever different motivic representations of the same period arise, they can always be interrelated by~the three equivalence relations in Definition~\ref{def: algebra_p}.

\begin{Definition} \label{def: matrix}
Let $X$ be a smooth quasi-projective $\mathbb{Q}$-variety, $Y \subset X$ a closed subvariety, and $H = H^{\bullet}(X,Y)$ the $H$-system of $X$ relative to $Y$.
Assume that $\{[\omega_j]\}_{j=1}^{n}$ is a basis of the algebraic de Rham cohomology $H^{\bullet}_{\rm dR}(X, Y)$, and that $\{[\gamma_i]\}_{i=1}^{n}$ is a basis of the Betti homology $H_{\bullet}^{{\rm B}}(X, Y)$. We~denote by $\mathop{\rm per}\nolimits|_H$ the period map restricted to the motivic periods in $\mathcal{P}^{\rm m}$ that are built on the given Hodge structure $H$.
Observe that $\mathop{\rm per}\nolimits|_H$ is fully determined by the values that it takes when evaluated at $[H, [\omega_j], [\gamma_i]]^{\rm m}$, which are
\begin{gather}
\mathop{\rm per}\nolimits|_H([H, [\omega_j], [\gamma_i]]^{\rm m}) = \int_{\gamma_i} \omega_j
\end{gather}
for each pair of indices $(i,j)$ with $i,j = 1,\dots,n$. We~define the \textit{period matrix} of $H$ as the $n \times n$-matrix with complex entries $(p_{ij})_{i,j = 1,\dots,n}$ given by
\begin{gather}
p_{ij} = \int_{\gamma_i} \omega_j.
\end{gather}
\end{Definition}
The period matrix expresses in a different guise the same information contained in the period map, once it has been restricted to a specific $H$-system.

\begin{Remark}
For a given mixed $H$-system $H = (H_{\rm dR}, \, H_{{\rm B}}, \, \mathop{\rm comp})$, there is a canonical choice of bases on $H_{\rm dR}$ and $H^{{\rm B}}$ which is compatible\footnote{See Brown~\cite{Bro17_1} for details.} with the comparison isomorphism and with the Hodge and weight filtrations in $H$. We~often implicitly assume to work in the canonical bases when writing the period matrix. Let~us denote by $\{ e_i \}_{i = 1}^n$ and $\{ f_i \}_{i = 1}^n$ the canonical bases on the de Rham and Betti realisations of $H$, respectively.
For $i=1, \dots, n$, the action of the comparison isomorphism on the $i$-th element of the canonical basis of $H_{\rm dR}$ is given by
\begin{align}
\mathop{\rm comp}\colon\ H_{\rm dR} \otimes_{\mathbb{Q}} \mathbb{C} & \xlongrightarrow{\sim} H_{{\rm B}} \otimes_{\mathbb{Q}} \mathbb{C},
\\
e_i \otimes 2 \pi {\rm i} & \longmapsto f_i^{\vee} \otimes 2 \pi {\rm i},
\end{align}
where $f_i^{\vee}$ denotes the standard vector dual basis element of $f_i$ on $H_{{\rm B}}$. For $i, j=1, \dots, n$, the pairing map gives
\begin{align}
H_{\rm dR} \times H^{{\rm B}} & \longrightarrow \mathbb{C},
\\
(e_j , f_i) & \longmapsto \int_{f_i} e_j = p_{ij}.
\end{align}
For $i = 1, \dots, n$, we define the vector dual $e_i^{\vee}$ of the basis element $e_i$ to be $e_i^{\vee} = f_i$. Observe that, since we cannot easily make sense of a notion of de Rham homology, the dual of a basis of~$H_{\rm dR}$ is defined to be a basis of $H^{{\rm B}}$.
\end{Remark}

\begin{Example} \label{ex: ex10}
Let $H= H^1(\mathbb{G}_m, \{ 1, z \})$ with $z \in \mathbb{Q} \backslash \{0, 1\}$. As shown in Examples~\ref{ex: ex0} and~\ref{ex: exx}, a basis of the Betti homology group $H_1^{{\rm B}}(\mathbb{G}_m, \{ 1, z \}) \simeq H^{\text{s}}_1(\mathbb{C}^*, \{ 1, z \})$ is given by $[\gamma_1]$, where $\gamma_1$ is a continuous oriented map from $1$ to $z$ which does not encircle the origin, and $[\gamma_2]$, where $\gamma_2$ is a counterclockwise cycle encircling the origin. A~basis of the algebraic de Rham cohomology group $H^1_{\rm dR}(\mathbb{G}_m, \{ 1, z \})$ is given by $[\omega_1] = \big[ \frac{{\rm d}x}{z-1} \big]$ and $[\omega_2] = \big[ \frac{{\rm d}x}{x} \big]$. Such a choice of bases is indeed canonical, and the period matrix of $H$ is
\begin{gather}
\begin{pmatrix}
1 & \log(z) \\
0 & 2 \pi {\rm i}
\end{pmatrix}\!.
\end{gather}
\end{Example}

\subsection{Examples} \label{sec: examples}
\subsubsection[Motivic $2 \pi {\rm i}$]{Motivic $\boldsymbol{2 \pi {\rm i}}$}
The period $2 \pi {\rm i}$ is given by the contour integral
\begin{gather} \label{eq: int_2pii}
2 \pi {\rm i} = \oint\displaylimits_{\gamma_2} \frac{{\rm d}x}{x},
\end{gather}
where $\gamma_2$ is a counterclockwise cycle encircling the origin in the punctured complex plane $\mathbb{C}^*$.
As observed in Example~\ref{ex: ex2}, the complex manifold $\mathbb{C}^*$ is isomorphic to the topological space of complex points $\mathbb{G}_m^{\rm an}$ underlying the algebraic variety $\mathbb{G}_m$ over $\mathbb{Q}$.
As shown in Examples~\ref{ex: ex1} and~\ref{ex: ex4}, we have that
\begin{gather}
H_1^{{\rm B}}(\mathbb{G}_m) = \mathbb{Q}[\gamma_2], \qquad
H^1_{\rm dR}(\mathbb{G}_m) = \mathbb{Q} \bigg[ \frac{{\rm d}x}{x} \bigg].
\end{gather}
Recalling that $H^1(\mathbb{G}_m) = \big(H^1_{\rm dR}(\mathbb{G}_m), H^1_{{\rm B}}(\mathbb{G}_m), {\rm comp}\big)$, a motivic version of $2 \pi {\rm i} $ is
\begin{gather} \label{eq: 2pii_1}
(2 \pi {\rm i} )^{\rm m} = \bigg[ H^1(\mathbb{G}_m), \bigg[ \frac{{\rm d}x}{x} \bigg], [\gamma_2] \bigg]^{\rm m},
\end{gather}
which is alternatively represented by the pairing
\begin{align}
H^1_{\rm dR}(\mathbb{G}_m) \times H_1^{{\rm B}}(\mathbb{G}_m) & \longrightarrow \mathbb{C} ,
\\
\bigg( \bigg[ \frac{{\rm d}x}{x} \bigg], [\gamma_2] \bigg) & \longmapsto
\oint_{\gamma_2} \frac{{\rm d}x}{x} = 2 \pi {\rm i}.
\end{align}
A second integral representation of $2 \pi {\rm i}$ is given by
\begin{gather} \label{eq: int2}
2 \pi {\rm i} = \int_{\mathbb{P}^1(\mathbb{C})} \frac{{\rm d}z \wedge {\rm d}\bar{z}}{(1 + z \bar{z})^2},
\end{gather}
where $\frac{{\rm d}z \wedge {\rm d}\bar{z}}{(1 + z \bar{z})^2}$ is a closed algebraic $2$-form over the projective manifold $\mathbb{P}^{1, \text{an}}$.
Because $\mathbb{P}^{1, \text{an}}$ is compact and K\"ahler, Theorem~\ref{th: dec} applies, giving the Hodge decomposition
\begin{gather} \label{eq: H2pi}
H^2_{\rm dR}(\mathbb{P}^1) \otimes_{\mathbb{Q}} \mathbb{C} = \bigoplus_{p+q=2} H^{p,q}_{{\rm D}}\big(\mathbb{P}^1, \mathbb{C}\big)
\end{gather}
which implies that the pure $H$-system $H^2(\mathbb{P}^1)$ has weight 2.
Recalling that the differential forms in $H^{p,q}_{{\rm D}}$ contain $p$ copies of the differential ${\rm d}z$ and $q$ copies of the anti-holomorphic differential ${\rm d}\bar{z}$, we have that $\big[ \frac{{\rm d}z \wedge {\rm d}\bar{z}}{(1 + z \bar{z})^2} \big] \in H^{1,1}_{{\rm D}}\big(\mathbb{P}^1, \mathbb{C}\big)$. Therefore, the integral~\eqref{eq: int2} corresponds to the motivic period\footnote{Note that we are here using the intuitive definition of algebraic de Rham cohomology of non-affine varieties given in Section~\ref{sec: algdR}. Although $\frac{{\rm d}z \wedge {\rm d}\bar{z}}{(1 + z \bar{z})^2}$ is not a global algebraic $2$-form on $\mathbb{P}^1(\mathbb{C})$, and indeed there are no non-zero global algebraic $2$-forms on $\mathbb{P}^1(\mathbb{C})$ for dimension reasons, one can still rigorously make sense of $2 \pi {\rm i}$ as a period of $H^2(\mathbb{P}^1)$ via the \v Ceck construction mentioned in Section~\ref{sec: algdR}, that is, choosing a Zariski open affine covering of $\mathbb{P}^1$ and computing the algebraic de Rham cohomology as a hypercohomology of sheaves.}
\begin{gather} \label{eq: 2pii_2}
(2 \pi {\rm i} )^{\rm m} = \bigg[H^2(\mathbb{P}^1), \bigg[\frac{{\rm d}z \wedge {\rm d}\bar{z}}{(1 + z \bar{z})^2} \bigg], \left[\mathbb{P}^{1, \text{an}}\right] \bigg]^{\rm m}.
\end{gather}

\begin{Remark}
The two apparently different motivic periods in~\eqref{eq: 2pii_1} and~\eqref{eq: 2pii_2} are the same, thus preserving the period conjecture.
To show this, define
\begin{gather}
A = \mathbb{P}^{1, \text{an}} \backslash \{ \infty \} \cong \mathbb{C} \subset \mathbb{P}^{1, \text{an}} , \qquad
B = \mathbb{P}^{1, \text{an}} \backslash \{ 0 \} \cong \mathbb{C} \subset \mathbb{P}^{1, \text{an}}
\end{gather}
which satisfy the relations
\begin{gather}
A \cap B \simeq \mathbb{C}^* \simeq \mathbb{G}_m^{\rm an} , \qquad
A \cup B = \mathbb{P}^{1, \text{an}}.
\end{gather}
By the Mayer--Vietoris theorem applied to the singular homology groups, the following long exact sequence holds
\hfsetfillcolor{red!0}
\hfsetbordercolor{red!70!black}
\begin{equation}
\begin{tikzcd}
 0 \arrow[r]
 & H^{\text{s}}_0(A \cup B) \arrow[r]
 & H^{\text{s}}_0(A) \oplus H^{\text{s}}_0(B) \arrow[d] \\
 \underbrace{H^{\text{s}}_1(A) \oplus H^{\text{s}}_1(B)}_{ \simeq 0} \arrow[d]
 & H^{\text{s}}_1(A \cup B) \arrow[l]
 & H^{\text{s}}_0(A \cap B) \arrow[l] \\
		 H^{\text{s}}_1(A \cap B) \arrow[r] & H^{\text{s}}_2(A \cup B) \arrow[r]
	 & \underbrace{H^{\text{s}}_2(A) \oplus H^{\text{s}}_2(B)}_{ \simeq 0}.
\end{tikzcd}
\end{equation}
Hence, the step $ H^{\text{s}}_1(A \cap B) \rightarrow H^{\text{s}}_2(A \cup B)$ is an isomorphism, giving
\begin{gather}
H^{\text{s}}_1(\mathbb{G}_m^{\rm an}) \simeq H^{\text{s}}_2\big(\mathbb{P}^{1, \text{an}}\big).
\end{gather}
Similarly, one can prove that the whole $H$-systems $H^1(\mathbb{G}_m)$ and $H^2(\mathbb{P}^1)$ are isomorphic and that the change of coordinates occurring between them relates the cohomology classes $\big[\frac{{\rm d}z \wedge {\rm d}\bar{z}}{(1 + z \bar{z})^2} \big]$ and $\big[ \frac{{\rm d}x}{x} \big]$ and the homology classes $[\gamma_0]$ and $\left[\mathbb{P}^{1, \text{an}} \right]$ via pull-back and push-forward maps, respectively.
\end{Remark}

\subsubsection[Motivic $\log(z)$]{Motivic $\boldsymbol{\log(z)}$} \label{sec: log}
Recall the integral representation of $\log(z)$, $z \in \mathbb{Q} \backslash \{ 0, 1 \}$, given by
\begin{gather} \label{eq: log_num}
\log(z) = \int_1^z \frac{{\rm d}x}{x}.
\end{gather}
As in the case of $2 \pi {\rm i}$, this is an integral over the punctured complex plane $\mathbb{C}^* = \mathbb{G}_m^{\rm an}$.
However, contrary to the case of $2 \pi {\rm i}$, where the integration path $\gamma_2$ is closed, integral~\eqref{eq: log_num} is performed on an open path, precisely any continuous oriented path $\gamma_1 \subset \mathbb{C}^*$ starting at $1$ and ending at $z$, which does not encircle the origin. The integration path being open requires us to work in the framework of relative homology. Let~$\mathbb{G}_m$ be the ambient variety. Then, $\mathbb{C}^*$ is the underlying topological space, and $\{ 1, z \} \subset \mathbb{C}^*$ with $z \in \mathbb{Q} \backslash \{ 0,1 \}$ is a simple normal crossing divisor.
As~shown in Examples~\ref{ex: ex0} and~\ref{ex: exx}, we have
\begin{gather}
H_1^{{\rm B}}(\mathbb{G}_m, \{1, z\}) = \mathbb{Q}[\gamma_1, \gamma_2], \qquad
H^1_{\rm dR}(\mathbb{G}_m, \{1, z\}) = \mathbb{Q} \bigg[\frac{{\rm d}x}{z-1}, \frac{{\rm d}x}{x}\bigg].
\end{gather}
Observe that we can write $\big[ \big(\frac{{\rm d}x}{z-1},0,0 \big), \big(\frac{{\rm d}x}{x},0,0 \big) \big] = \big[ \frac{{\rm d}x}{z-1}, \frac{{\rm d}x}{x} \big]$ as a consequence of Proposition~\ref{prop: topdeg}.
Setting as usual $H^1(\mathbb{G}_m, \{1, z\}) = \big(H^1_{\rm dR}(\mathbb{G}_m, \{ 1, z \}), H^1_{{\rm B}}(\mathbb{G}_m, \{ 1, z \}), {\rm comp}\big)$, a motivic version of $\log(z)$ is
\begin{gather}
\log(z)^{\rm m} = \bigg[ H^1(\mathbb{G}_m, \{ 1, z \}), \bigg[\frac{{\rm d}x}{x} \bigg], [ \gamma_1 ] \bigg]^{\rm m}
\end{gather}
which is alternatively represented by the pairing
\begin{align}
H^1_{\rm dR}(\mathbb{G}_m, \{1, z\}) \times H_1^{{\rm B}}(\mathbb{G}_m, \{1, z\}) & \longrightarrow \mathbb{C} ,
\\[1ex]
\bigg(\bigg[ \frac{{\rm d}x}{x} \bigg], [\gamma_1] \bigg) & \longmapsto \int_{\gamma_1} \frac{{\rm d}x}{x} = \log(z).
\end{align}

\subsubsection{Elementary relations}
Elementary relations among periods are often simply recast in the formalism of motivic periods. In fact, de Rham and Betti systems of realisations conjecturally capture all algebraic relations among periods.

\begin{Example}
For $a,b \in \mathbb{Q} \backslash \{ 0, 1 \}$, such that $ab \ne 1$, consider the following injective morphisms of pairs of $\mathbb{Q}$-spaces:
\begin{gather}
(\mathbb{G}_m, \{ 1, a \}) \hookrightarrow (\mathbb{G}_m, \{ b, ab \}) \hookrightarrow (\mathbb{G}_m, \{ 1, b, ab \}),
\\
(\mathbb{G}_m, \{ 1, b \}) \hookrightarrow (\mathbb{G}_m, \{ 1, b, ab \}),
\\
 (\mathbb{G}_m, \{ 1, ab \}) \hookrightarrow (\mathbb{G}_m, \{ 1, b, ab \}).
\end{gather}
Since the differential form $\frac{{\rm d}x}{x}$ is invariant under rescaling of $x$, we have the motivic representations
\begin{gather}
\log(a)^{\rm m} = \bigg[H^1(\mathbb{G}_m, \{ 1, b, ab \}), \bigg[\frac{{\rm d}x}{x} \bigg], [b, ab] \bigg]^{\rm m} ,
\\
\log(b)^{\rm m}= \bigg[H^1(\mathbb{G}_m, \{ 1, b, ab \}), \bigg[\frac{{\rm d}x}{x} \bigg], [1, b] \bigg]^{\rm m},
\\
\log(ab)^{\rm m} =\bigg [H^1(\mathbb{G}_m, \{ 1, b, ab \}), \bigg[\frac{{\rm d}x}{x} \bigg], [1, ab] \bigg]^{\rm m},
\end{gather}
where $[z, w]$ denotes the class of a continuous oriented map in $\mathbb{C}^*$ which goes from $z$ to $w$ and it does not encircle the origin in the Betti homology group $H_1^{{\rm B}}(\mathbb{G}_m, \{ 1, b, ab \})$ for $z,w \in \{ 1, b, ab \}$.
Additivity of the Betti homology classes $[b, ab] \cup [1, b] = [1, ab]$ implies that motivic logarithms satisfy the expected relation
\begin{gather}
\log(ab)^{\rm m} = \log(a)^{\rm m} + \log(b)^{\rm m}.
\end{gather}
\end{Example}

\begin{Example}
Consider $H = H^1(\mathbb{G}_m, \{ 1, z \})$ for $z \in \mathbb{Q} \backslash \{ 0, 1 \}$. Let~$\gamma$ be the union of the paths $\gamma_1$ and $\gamma_2$ in the punctured complex plane, as shown in Fig.~\ref{fig: new_base}.
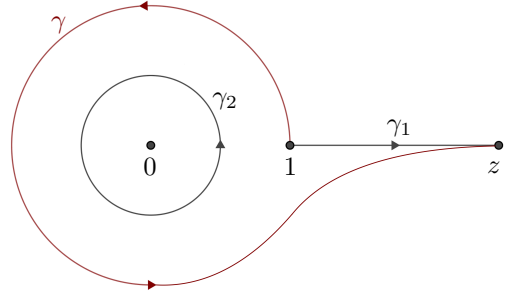
\begin{figure}[htb!]
\centering
\includegraphics[scale=.37]{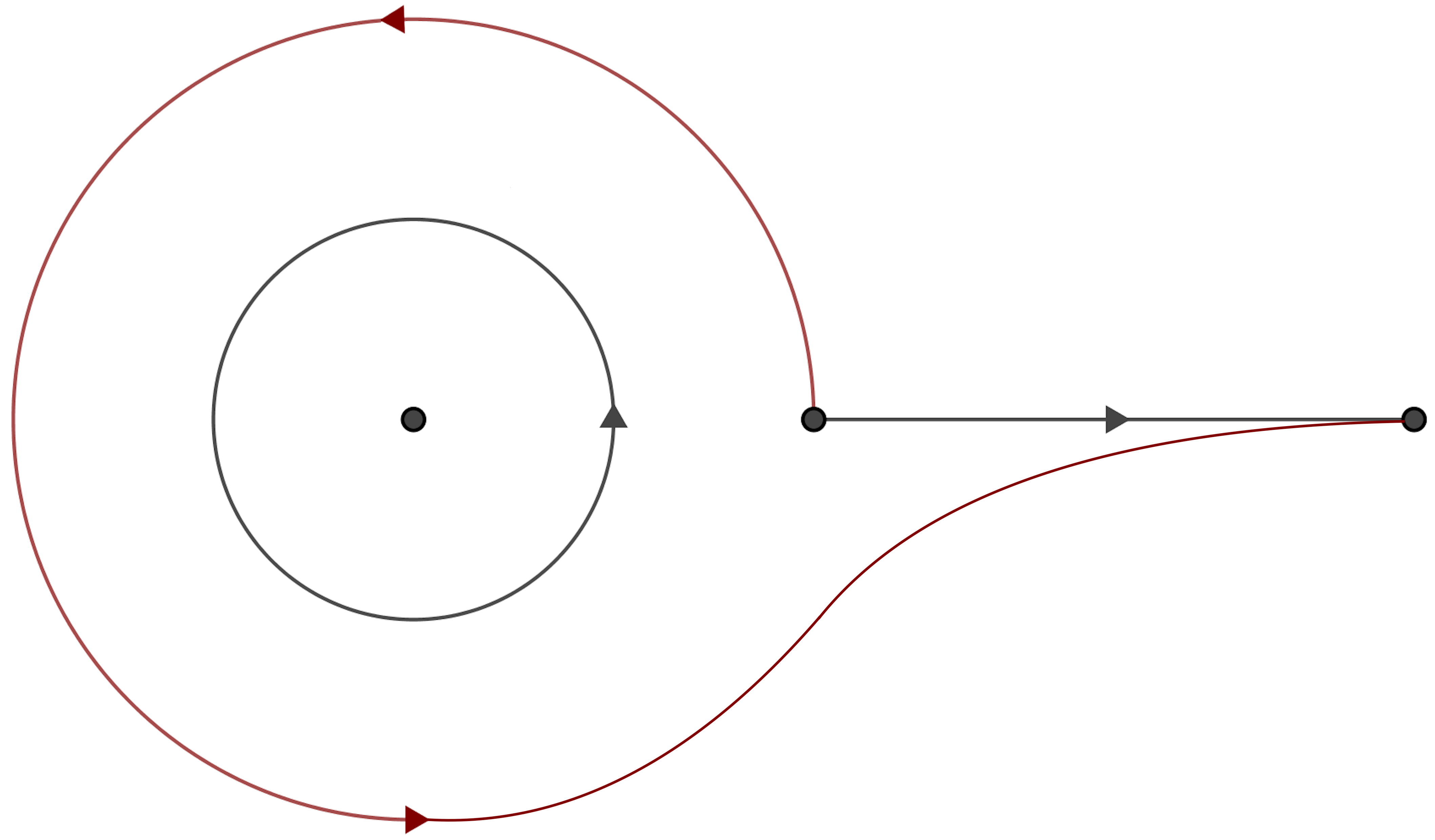}
\put(-172,98){\makebox(0,0)[lb]{\small \textcolor{ared}{$\gamma$}}}
\put(-111,69){\makebox(0,0)[lb]{\small $\gamma_2$}}
\put(-137,44){\makebox(0,0)[lb]{\small $0$}}
\put(-84,44){\makebox(0,0)[lb]{\small $1$}}
\put(-45,59){\makebox(0,0)[lb]{\small $\gamma_1$}}
\put(-7,45){\makebox(0,0)[lb]{\small $z$}}
\caption{The paths $\gamma_1$, $\gamma_2$, and $\gamma$ in $\mathbb{C}^*$.}
\label{fig: new_base}
\end{figure}

The period obtained by integrating $\omega_2$ along $\gamma$ is
\begin{gather}
\int_{\gamma}\omega_2 = \int_{\gamma_1 \cup \gamma_2} \omega_2 = \int_{\gamma_1} \omega_2 + \int_{\gamma_2} \omega_2 = \log(z) + 2 \pi {\rm i},
\end{gather}
which translates into the following relation among motivic periods
\begin{gather}
\begin{split}
(\log(z) + 2 \pi {\rm i})^{\rm m}
&= \left[H, [\omega_2], [\gamma] \right]^{\rm m}
=\left[ H, [\omega_2], [\gamma_1 \cup \gamma_2] \right]^{\rm m}\\
&= \left[H, [\omega_2], [\gamma_1] \right]^{\rm m} + \left[ H,[\omega_2], [\gamma_2] \right]^{\rm m}
\\
&= \log(z)^{\rm m} + (2 \pi {\rm i} )^{\rm m},
\end{split}
\end{gather}
where we have used the additivity of the Betti homology classes and the injective morphism $\mathbb{G}_m \hookrightarrow (\mathbb{G}_m, \{ 1, z \})$.
Because $ (\log(z) + 2 \pi {\rm i})^{\rm m} \in \mathop{\rm per}\nolimits^{-1} (\log(z) + 2 \pi {\rm i})$ and $\log(z)^{\rm m} + (2 \pi {\rm i} )^{\rm m} \in \mathop{\rm per}\nolimits^{-1} (\log(z)) + \mathop{\rm per}\nolimits^{-1} (2 \pi {\rm i})$, it follows that
\begin{gather}
\mathop{\rm per}\nolimits^{-1} (\log(z) + 2 \pi {\rm i}) \cap \big(\mathop{\rm per}\nolimits^{-1} (\log(z)) + \mathop{\rm per}\nolimits^{-1} (2 \pi {\rm i} )\big) \ne \varnothing.
\end{gather}
Note that $\mathop{\rm per}\nolimits^{-1} (\log(z) + 2 \pi {\rm i}) = \mathop{\rm per}\nolimits^{-1} (\log(z)) + \mathop{\rm per}\nolimits^{-1} (2 \pi {\rm i} )$ only holds conjecturally.
\end{Example}

Moreover, many new functional equations among motivic periods are found by means of the more abstract, and yet more powerful formalism that we discuss in Section~\ref{sec: motives}.
By the period conjecture, new relations among motivic periods automatically translates into new algebraic relations among the corresponding numbers.

\section{Feynman motives} \label{sec: motives}

\subsection{Singularities and the blow up}
Multiple zeta values and convergent Feynman integrals are periods by means of the integral representations~\eqref{eq: zint} and~\eqref{eq: I_G_last}, respectively.
In both cases, singularities of the integrand can be contained in the domain of integration, a feature that does not occur in the examples of $2 \pi {\rm i}$ and $\log(z)$.
Whenever singularities are present, they have to be taken care of with particular attention.

\begin{Example}
The period $\zeta(2)$ is given by the integral
\begin{gather} \label{eq: zeta2}
\zeta(2) = \int\displaylimits_{1 \ge x_1 \ge x_2 \ge 0} \frac{{\rm d}x_1}{x_1} \wedge \frac{{\rm d}x_2}{1-x_2}
\end{gather}
over the complex manifold $\mathbb{C}^2$. The domain of integration is the simplex
\begin{gather}
\sigma = \{(x_1,x_2) \in \mathbb{C}^2 \,|\, 1 \ge x_1 \ge x_2 \ge 0 \}
\end{gather}
and the integrand is the differential $2$-form
\begin{gather}
\omega = \frac{{\rm d}x_1}{x_1} \wedge \frac{{\rm d}x_2}{1-x_2}.
\end{gather}
Observing that $\mathbb{C}^2$ is isomorphic to the topological space of complex points $\mathbb{A}^2(\mathbb{C})$, underlying the affine\footnote{For any positive integer $n$, the $n$-dimensional affine variety over $\mathbb{Q}$ is defined as $\mathbb{A}^n = \mathop{\rm Spec} \mathbb{Q}[x_1, \dots, x_n]$. For any field extension $\mathbb{K} \supseteq \mathbb{Q}$, the space of $\mathbb{K}$-points of $\mathbb{A}^n$ is $\mathbb{A}^n(\mathbb{K}) = \mathbb{K}^n$. The multiplicative group $\mathbb{G}_m = \mathop{\rm Spec} \mathbb{Q}[x, \frac{1}{x}]$ satisfies $\mathbb{G}_m =\mathop{\rm Spec} \mathbb{Q}[x_1, x_2] / (1 - x_1 x_2) =\mathbb{A}^1 \backslash \{0\} \subset \mathbb{A}^2$, that is, $\mathbb{G}_m$ is a hyperbola in $\mathbb{A}^2$.} $\mathbb{Q}$-algebraic variety $\mathbb{A}^2 = \mathop{\rm Spec} \mathbb{Q}[x_1,x_2]$, we may try to build $\zeta(2)^{\rm m}$ as we did for the examples in Section~\ref{sec: examples}.
Consider the lines $l_0 = \{x_1 = 0 \}$ and $l_1 = \{x_2 = 1 \}$ in the affine plane $\mathbb{A}^2$. Since $L = l_0 \cup l_1$ is the locus of singular points of $\omega$, the latter is an algebraic $2$-form on $X = \mathbb{A}^2 \backslash L$.
Thus, $[\omega ]$ is a class in the second algebraic de Rham cohomology group of $X$ and, consequently, we may want to consider the integral~\eqref{eq: zeta2} as a period of $X$ relative to some divisor containing the boundary of $\sigma$.
In an attempt to do so, define the simple normal crossing divisor
\begin{gather}
D = \{ x_1 = x_2 \} \cup \{ x_1 = 1 \} \cup \{ x_2 = 0 \} \subset \mathbb{A}^2
\end{gather}
containing $\partial \sigma$.
Note that $D$ is not in $X$ because $D \cap L \ne \varnothing $. However, the divisor $D \backslash (D \cap L) \subset X$ does no longer contain $\partial \sigma$.
The problem arises from the fact that $\sigma$ itself is not contained in $X$, intersecting the singular locus $L$ in two points
\begin{gather}
p = (0,0) = \sigma \cap l_0 = D \cap l_0,
\qquad
q = (1,1) = \sigma \cap l_1 = D \cap l_1.
\end{gather}
Removing the singular points $p$, $q$ from $D$ and considering the second relative $H$-system $H^2(X, D$ $\backslash (D \cap L))$ does not solve the mentioned technical issue, because $[\sigma]$ is not a class in $H_2^{{\rm B}}(X, D\backslash (D \cap L))$.
See Fig.~\ref{fig: z2}.
\begin{figure}[htb!]
\centering
\includegraphics[scale=0.3]{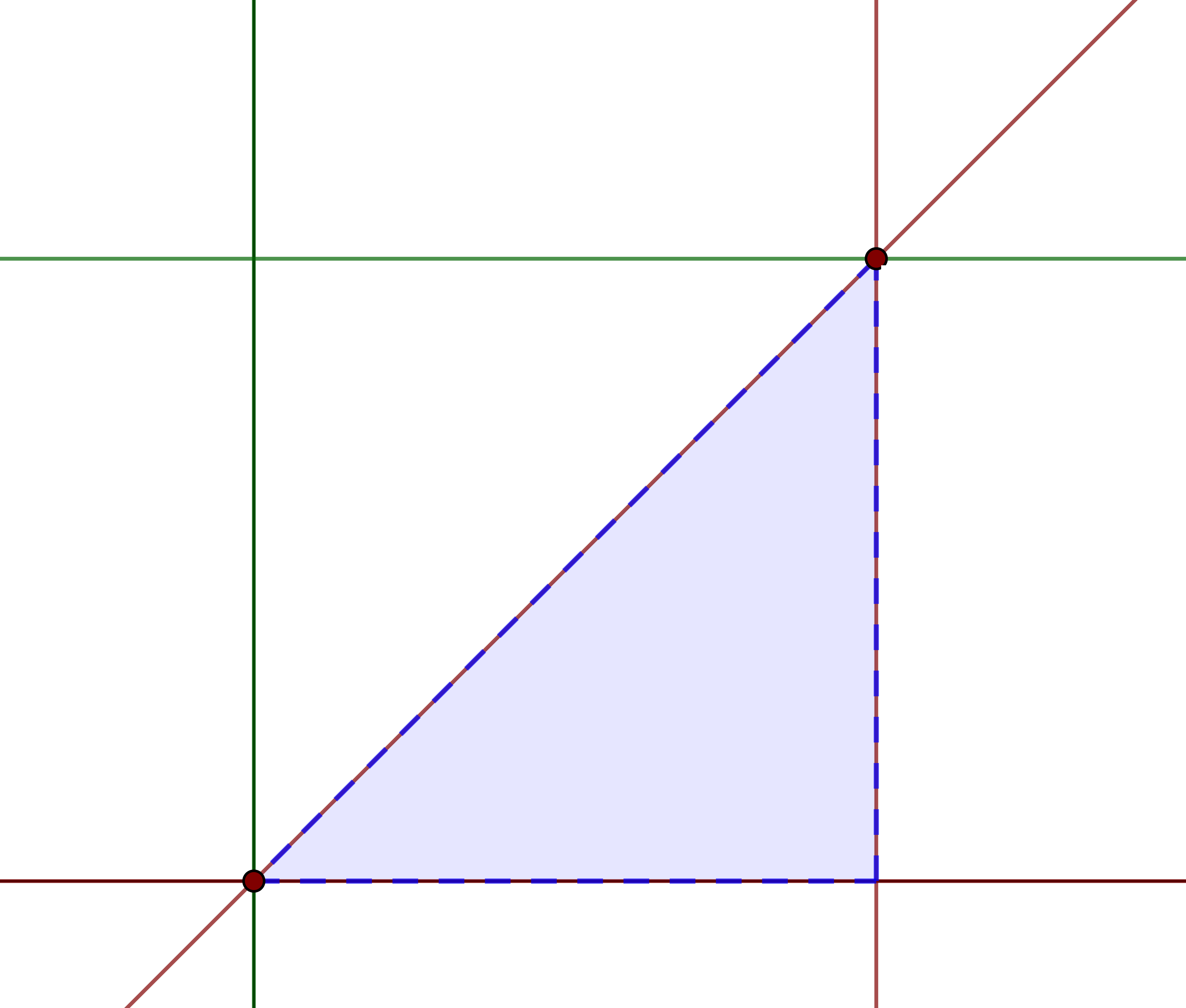}
\put(-40,120){\makebox(0,0)[lb]{\small \textcolor{ared}{$D$}}}
\put(-40,93){\makebox(0,0)[lb]{\small \textcolor{ared}{$q$}}}
\put(-125,8){\makebox(0,0)[lb]{\small \textcolor{ared}{$p$}}}
\put(-125,55){\makebox(0,0)[lb]{\small \textcolor{agreen}{$l_0$}}}
\put(-85,92){\makebox(0,0)[lb]{\small \textcolor{agreen}{$l_1$}}}
\put(-75,42){\makebox(0,0)[lb]{\small \textcolor{ablue}{$\sigma$}}}
\caption{Construction of $\zeta(2)^{\rm m}$ in the affine plane $\mathbb{A}^2$.}
\label{fig: z2}
\end{figure}
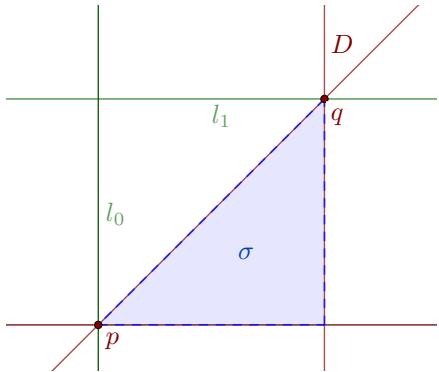
\end{Example}

\begin{figure}[htb!]
\centering
\subfloat[][{Before the blow up}]
 {\includegraphics[scale=.3]{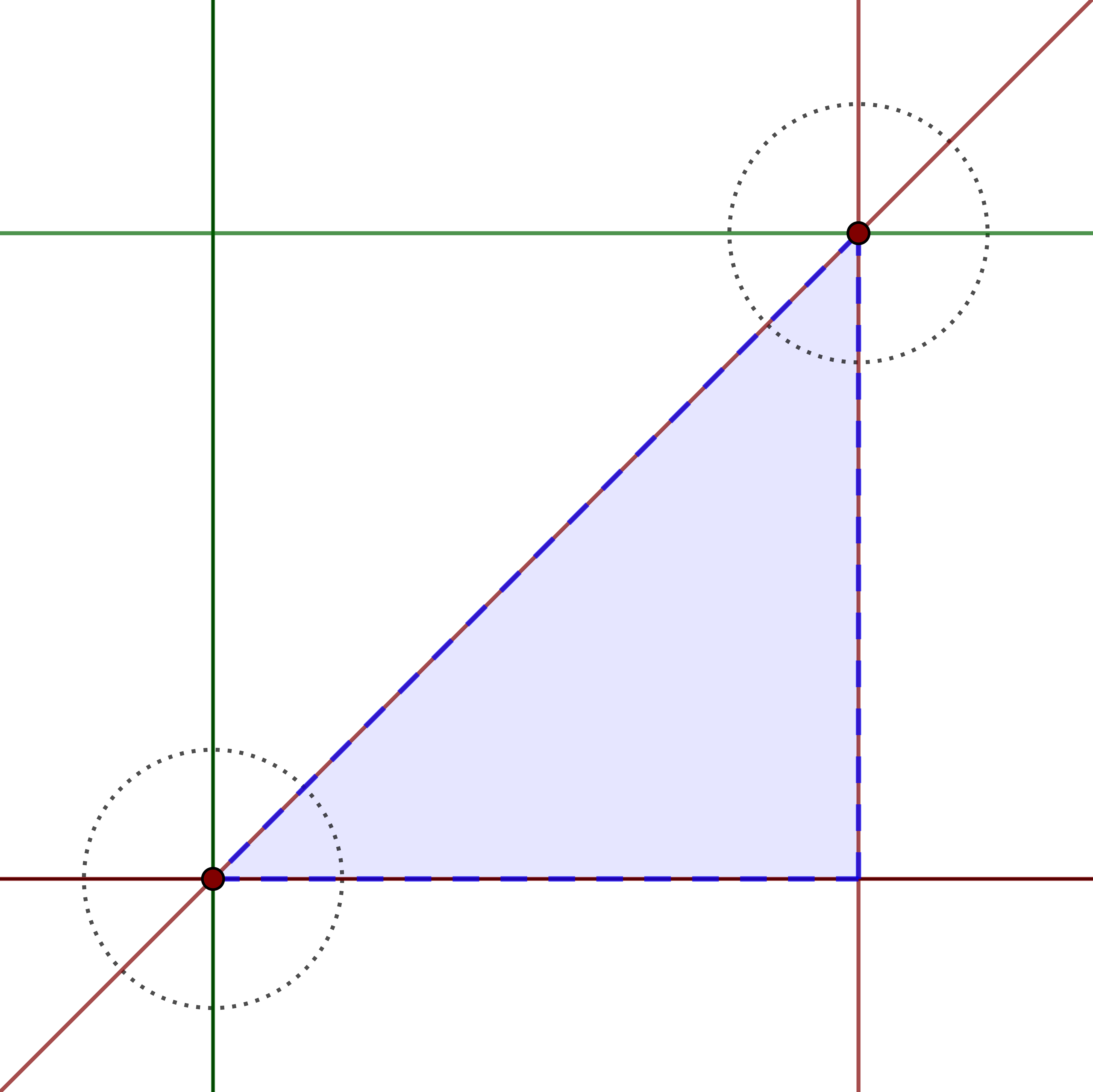}} \qquad
\subfloat[][{After the blow up}]
 {\includegraphics[scale=.3]{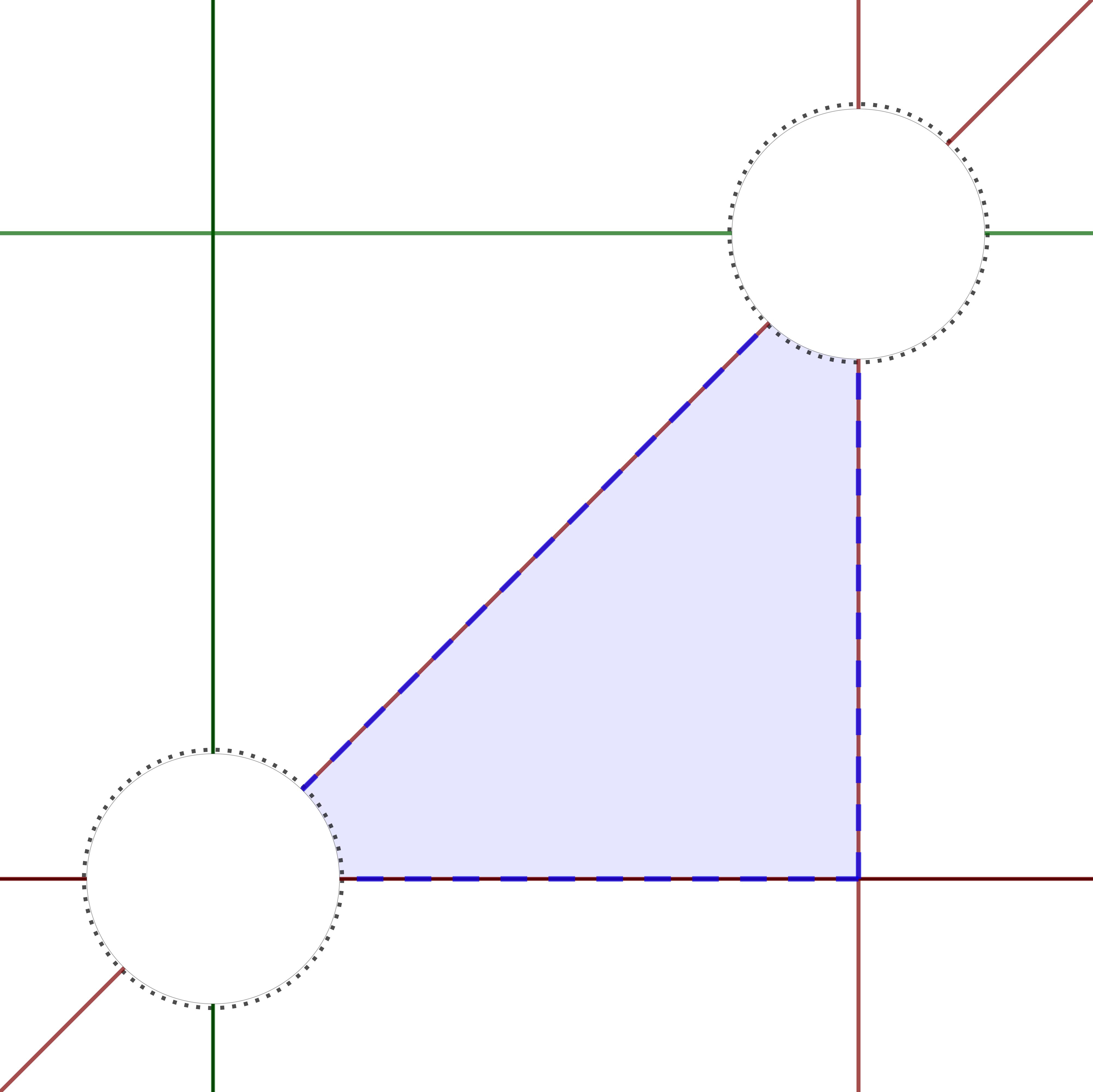}}
\caption{Qualitative illustration of the blow up of the singular points of $\zeta(2)$.}
\label{fig: blow}
\end{figure}

The example of $\zeta(2)$ shows how direct removal of singular points explicitly fails and motivates a more articulated geometric construction, called \textit{blow up}, which proves to be successful in the case of $\zeta(2)$ and many more examples.
Graphically, we may illustrate the procedure as the removal of a whole region of space centred at the singularity and the corresponding reshaping of the integration domain. See Fig.~\ref{fig: blow} for a qualitative representation of how the blow up of~the two singular points $p,q \in \mathbb{A}^2$ acts on $\sigma$ in the case of $\zeta(2)$.

\subsection{Motivic multiple zeta values}
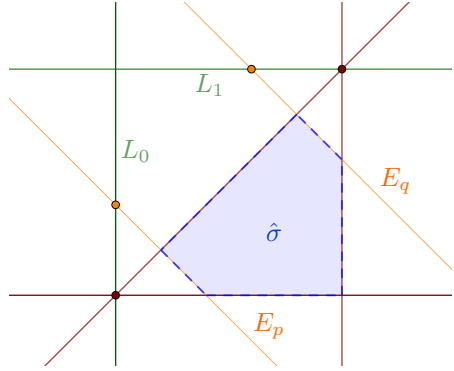
\begin{figure}[htb!]
\centering
\includegraphics[scale=.3]{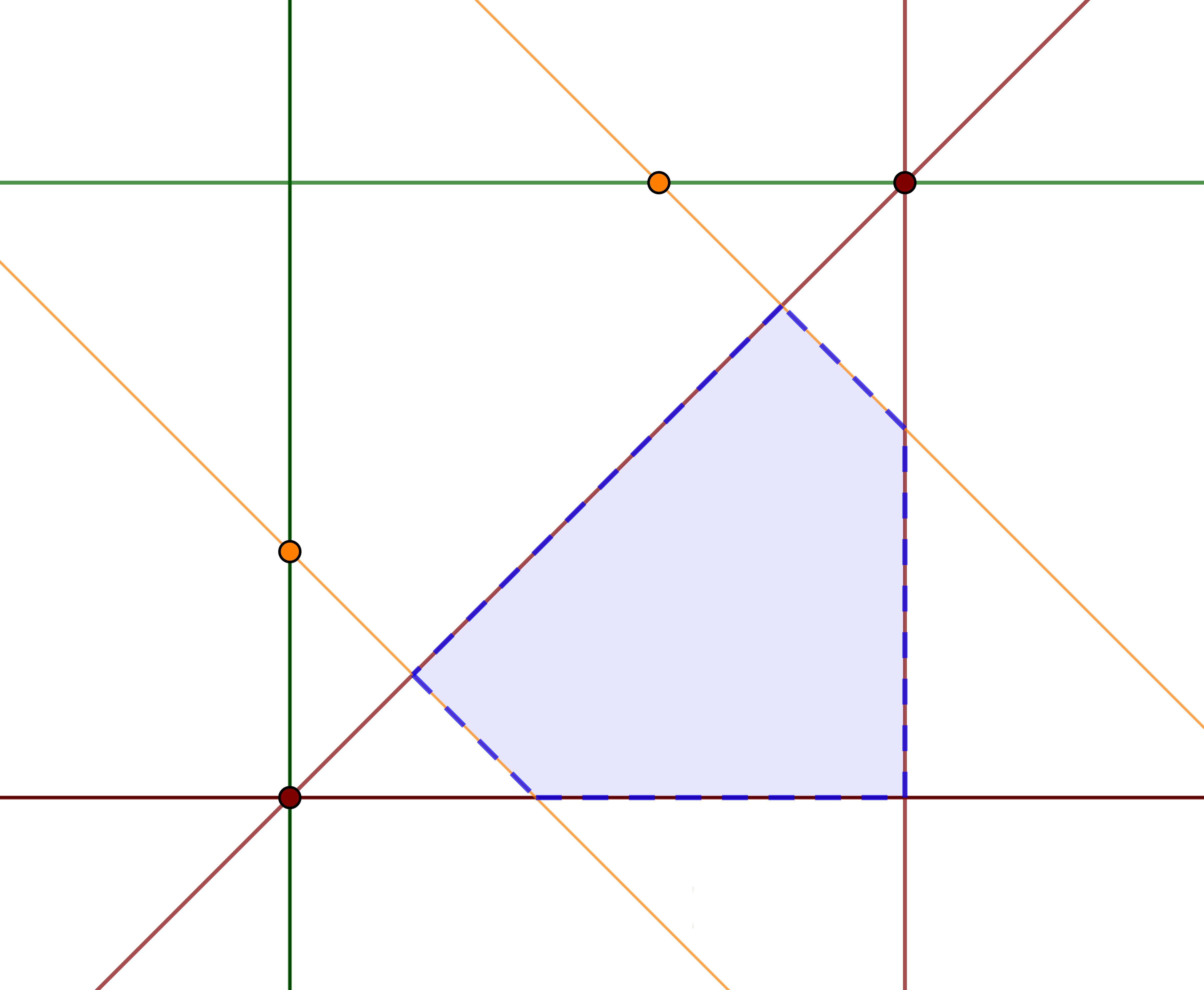}
\put(-125,77){\makebox(0,0)[lb]{\small \textcolor{agreen}{$L_0$}}}
\put(-97,102){\makebox(0,0)[lb]{\small \textcolor{agreen}{$L_1$}}}
\put(-70,47){\makebox(0,0)[lb]{\small \textcolor{ablue}{$\hat{\sigma}$}}}
\put(-27,65){\makebox(0,0)[lb]{\small \textcolor{aorange}{$E_q$}}}
\put(-75,10){\makebox(0,0)[lb]{\small \textcolor{aorange}{$E_p$}}}
\caption{The strict transform of $\sigma$ in the blow up $Y$.}
\label{fig: blow_last}
\end{figure}

Consider $\zeta(2)$ again. The blow up of the affine plane $\mathbb{A}^2$ along the singular points $p$, $q$ is defined as the closed subvariety
\begin{gather}
Y =\mathop{\rm Blow}_{p,q}\big(\mathbb{A}^2\big) \subset \mathbb{A}^2 \times \mathbb{P}^1 \times \mathbb{P}^1
\end{gather}
given by the equations
\begin{gather}
x_1 \alpha_1 = x_2 \beta_1,\qquad
(x_1 -1) \alpha_2 = (x_2 -1) \beta_2,
\end{gather}
where $[\alpha_i : \beta_i]$, $i=1,2$, are homogeneous coordinates on the two copies of $\mathbb{P}^1$.
The projection of $Y$ onto the first factor in $\mathbb{A}^2 \times \mathbb{P}^1 \times \mathbb{P}^1$ is the proper surjective map
\begin{align}
\pi\colon\ Y & \longrightarrow \mathbb{A}^2 ,
\\
(x_1, x_2) \times [\alpha_1 : \beta_1] \times [\alpha_2 : \beta_2] & \longmapsto (x_1,x_2).
\end{align}
Let us write $\pi^{-1}$ to denote the inverse image operator\footnote{Observe that $\pi^{-1}$ is not a map defined on the affine plane $\mathbb{A}^2$ because $\pi$ is not invertible.} under the projection $\pi$.
The inverse images of the singular points $p,q \in \mathbb{A}^2$ are the projective lines $E_p,E_q \subset Y$, called \textit{exceptional divisors}. Precisely, we have
\begin{gather}
\pi^{-1}(p) = \pi^{-1}(0,0) = (0,0) \times \mathbb{P}^1 \times [1:1] = E_p,
\\
\pi^{-1}(q) = \pi^{-1}(1,1) = (1,1) \times [1:1] \times \mathbb{P}^1 = E_q.
\end{gather}
Moreover, the restriction of $\pi$ to the complement in $Y$ of the exceptional divisors $E_p$, $E_q$
\begin{align}
\pi |_{Y \backslash (E_p \cup E_q)}\colon\ Y \backslash (E_p \cup E_q) & \longrightarrow \mathbb{A}^2 \backslash \{p,q\},
\\
(x_1,x_2) \times [1:1] \times [1:1] & \longmapsto (x_1,x_2)
\end{align}
is an isomorphism.
For any closed subset $C \subset \mathbb{A}^2$, the inverse image $\pi^{-1}(C)$ is called \textit{total transform} of $C$. The \textit{strict transform} of $C$, denoted $\hat{C}$, is instead the closed subset of $Y$ obtained by first removing the points $p,q$ if they belong to $C$, then taking the inverse image under $\pi$, and finally taking the Zariski closure, that is
\begin{gather} \label{eq: strictdef}
\hat{C} = \overline{\pi^{-1}(C \backslash \{p,q\})} \subseteq \pi^{-1}(C).
\end{gather}
It follows that the strict transforms of $l_0, l_1$ are the affine lines
\begin{gather}
L_0 = \hat{l}_0 = \big\{ (0,x_2) \times [1:0] \times [1-x_2:1] \,|\, x_2 \in \mathbb{A}^1 \big\},
\\
L_1 = \hat{l}_1 = \big\{ (x_1,1) \times [1:x_1] \times [0:1] \,|\, x_1 \in \mathbb{A}^1 \big\}
\end{gather}
and their total transforms are
\begin{gather}
\pi^{-1}(l_0) = L_0 \cup E_p,
\qquad
\pi^{-1}(l_1) = L_1 \cup E_q.
\end{gather}
We observe that $L_0$, $E_p$ and $L_1, E_q$ intersect in only one point each. Precisely
\begin{gather}
L_0 \cap E_p = \{ (0,0) \times [1:0] \times [1:1] \}, \\
L_1 \cap E_q = \{ (1,1) \times [1:1] \times [0:1] \}.\label{eq: LcapsE}
\end{gather}
Moreover, we have
\begin{gather}
L_1 \cap E_p = \varnothing = L_0 \cap E_q, \\
L_1 \cap L_0 = \{ (0,1) \times [1:0] \times [0:1] \}.
\end{gather}
In a similar way to~\eqref{eq: strictdef}, but taking the closure in the analytic topology, we define the strict transform $\hat{\sigma}$ of the domain of integration.
Observing that the closed points of $E_p$ can be interpreted as lines passing through $p$, and analogously that the closed points of $E_q$ can be interpreted as lines passing through $q$, we obtain
\begin{gather}
\hat{\sigma} \cap E_p = \{ (0,0) \times [t:1] \times [1:1] \,|\, 0 \le t \le 1 \}, \\
\hat{\sigma} \cap E_q = \{ (1,1) \times [1:1] \times [1:t] \,|\, 0 \le t \le 1 \}
\end{gather}
which, combined with~\eqref{eq: LcapsE}, imply that
\begin{gather} \label{eq: intersect}
\hat{\sigma} \cap L_0 = \varnothing , \qquad
\hat{\sigma} \cap L_1 = \varnothing.
\end{gather}
See Fig.~\ref{fig: blow_last} for a graphical representation of the blow up.

\begin{figure}[htb!]
\centering
\includegraphics[scale=.5]{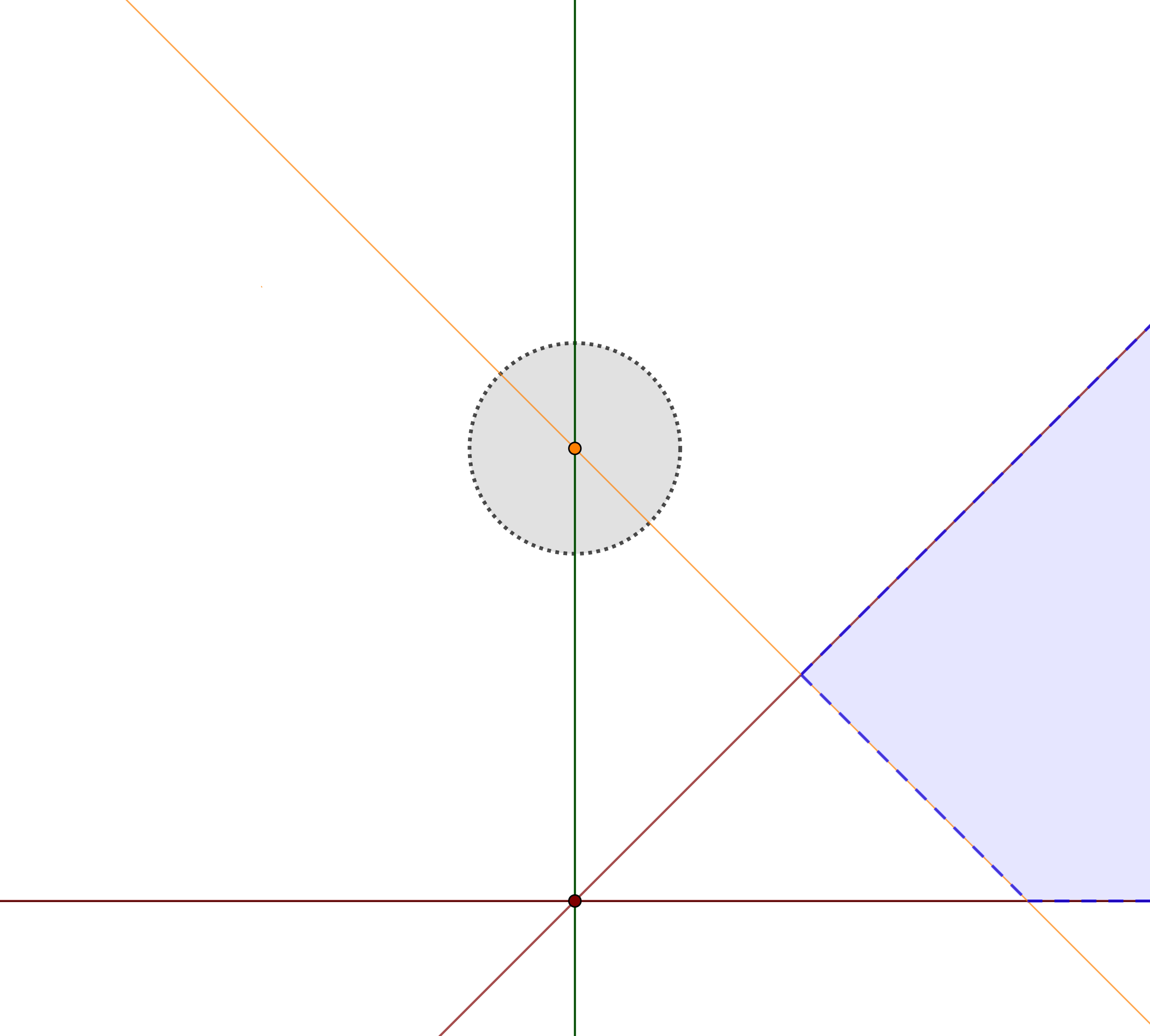}
\put(-70,105){\makebox(0,0)[lb]{\small \textcolor{agreen}{$L_0$}}}
\put(-115,95){\makebox(0,0)[lb]{\small \textcolor{aorange}{$E_p$}}}
\caption{Local patch of $Y$ around the intersection of $L_0$ and $E_p$.}
\label{fig: patch}
\end{figure}

While the inverse image $\pi^{-1}$ is applied to the ambient variety, giving the reshaped domain $\hat{\sigma}$, the differential form $\omega$ is replaced by its pull-back $\pi^*(\omega)$, denoted by $\hat{\omega}$. Let~us now show that the pull-back $\hat{\omega}$ is only singular\footnote{In principle, $\hat{\omega}$ might have singularities along the total transform of $l_0 \cup l_1$, i.e., $L_0 \cup L_1 \cup E_p \cup E_q$. However, in the case of $\zeta(2)$, it turns out that $\hat{\omega}$ has no singularities along the exceptional divisors. More generally, this condition determines whether the blow up prescription turns out to be successful or not for a given period.} on the strict transform $L = L_0 \cup L_1$. We~use local coordinates on the blow up $Y$. In particular, consider a patch of $Y$ around the point $L_0 \cap E_p$ as shown in~Fig.~\ref{fig: patch}.

Here, a local system of coordinates is explicitly given by
\begin{gather}
t = \frac{x_1}{x_2}=\frac{\beta_1}{\alpha_1}, \qquad s = x_2,
\end{gather}
where $L_0$ and $E_p$ have equations $t=0$ and $s=0$, respectively.
Applying this change of variables to $\hat{\omega}$, we have
\begin{gather}
\hat{\omega} = \frac{{\rm d} (st)}{st} \wedge \frac{{\rm d}s}{1-s} = \frac{{\rm d}s}{s} \wedge \frac{{\rm d}s}{1-s} + \frac{{\rm d}t}{t} \wedge \frac{{\rm d}s}{1-s} = \frac{{\rm d}t}{t} \wedge \frac{{\rm d}s}{1-s}.
\end{gather}
It follows that $\hat{\omega}$ is singular along the strict transform $L_0$, while it is smooth along the exceptional divisor $E_p$, because it has no pole at $s=0$.
Analogously, we find that $\hat{\omega}$ is singular along $L_1$, but not along $E_q$. Then, the singular locus of $\hat{\omega}$ is $L$.
Observe that the complement $Y \backslash L$ is the closed affine subvariety of $\mathbb{A}^2 \times \mathbb{A}^1 \times \mathbb{A}^1$ given by the equations
\begin{gather}
x_1 t = x_2, \qquad
x_1 -1 = (x_2 -1) s,
\end{gather}
where $t$, $s$ are affine coordinates on the two copies of $\mathbb{A}^1$. Therefore, the differential form $\hat{\omega}$ determines a class in $H^2_{\rm dR}(Y \backslash L)$.
Moreover, it follows from~\eqref{eq: intersect} that, moving from the original affine plane $\mathbb{A}^2$ to the blow up $Y$, the singular locus of the differential form $\hat{\omega}$ and the domain of integration $\hat{\sigma}$ are disjoint.
As usual, we may want to consider the integral~\eqref{eq: zeta2} as a period of~$Y \backslash L$ relative to some divisor containing the boundary of $\hat{\sigma}$. The blow up construction is thus successful for the period $\zeta(2)$ if $[\hat{\sigma}]$ turns out to be a class in the given relative Betti homology group.
To see this, recall that $\partial \sigma$ is contained in the union $D$ of the affine lines
\begin{gather}
m_1 = \{ x_1 = x_2 \}, \qquad m_2 = \{ x_1 = 1 \}, \qquad m_3 = \{ x_2 = 0 \}.
\end{gather}
Thus, we naturally consider the normal crossing divisor $M \subset Y$ defined by
\begin{gather}
M = \pi^{-1}(D) = \pi^{-1}(m_1 \cup m_2 \cup m_3) = E_p \cup E_q \cup M_1 \cup M_2 \cup M_3,
\end{gather}
where $M_i= \hat{m_i}$ denotes the strict transform of $m_i$ for $i=1,2,3$.
Note that $L \cap M$ is the union of the points $L_0 \cap E_p$ and $L_1 \cap E_q$ expressed in~\eqref{eq: LcapsE}.
Therefore, $\hat{\sigma}$ is contained in $Y \backslash L$ and $\partial \hat{\sigma}$ is contained in $M \backslash (M \cap L) \subset Y \backslash L$, implying
\begin{gather}
[\hat{\sigma}] \in H_2^{{\rm B}}(Y \backslash L, M \backslash (M \cap L)).
\end{gather}
Besides, the restriction of $\hat{\omega}$ to every irreducible component $M_i$, $i=1,2,3$, of $M$ gives zero, implying
\begin{gather}
[\hat{\omega}] \in H^2_{\rm dR}(Y \backslash L, M \backslash (M \cap L)).
\end{gather}
Setting $H = H^2(Y \backslash L, M \backslash (M \cap L))$, the resulting motivic version of $\zeta(2)$ is
\begin{gather}
\zeta(2)^{\rm m} = \left[ H, [\hat{\omega}], [\hat{\sigma}] \right]^{\rm m}.
\end{gather}
Indeed, the pairing of $[\hat{\sigma}]$ and $[\hat{\omega}]$ yields
\begin{gather}
\int_{\hat{\sigma}} \hat{\omega} = \int_{ \hat{\sigma}} \pi^*(\omega) = \int_{\pi_*(\hat{\sigma})} \omega = \int_{\sigma} \omega = \zeta(2)
\end{gather}
by the equivalence relation under change of variables in $\mathcal{P}^{\rm m}$.
Moreover, the whole canonical period matrix of $H$ is
\begin{gather} \label{eq: matrixZ2}
\begin{pmatrix}
1 & \zeta(2) \\[1ex]
0 & (2 \pi {\rm i})^2
\end{pmatrix}\!.
\end{gather}

\subsection{Motivic Feynman integrals} \label{sec: intGmot}
In an attempt to overcome singularity issues, the blow up procedure can be similarly applied to generic MZVs and other families of periods, such as convergent Feynman integrals. For an~expo\-sition of the general computation of the $H$-system of a blow up we refer to Voisin~\cite{Voi02}.

Let $G$ be a primitive log-divergent Feynman graph, $E_G$ the collection of its edges, and \mbox{$n_G = |E_G|$}, as in Section~\ref{sec: primitive}. Recall that $x_e$ denotes the Schwinger parameter associated to $e \in E_G$, and $\Psi_G$, $I_G$, and $X_G$ denote the first graph polynomial, the Feynman integral, and the graph hypersurface, as given in~\eqref{eq: Psi_G},~\eqref{eq: I_G_last}, and~\eqref{eq: X_G}, respectively.
Denote by $\omega_G$ and $\sigma$ the integrand and the domain of integration of $I_G$.
Since $\omega_G$ is a top-degree algebraic differential form on~$\mathbb{P}^{n_G-1} \backslash X_G$, and $\partial \sigma$ is contained in the union $D$ of the coordinate hyperplanes $\{x_e = 0$, $e \in E_G \}$, we may intuitively try to build the motive $I_G^{\rm m}$ on the relative $H$-system
\begin{gather}
H^{n_G-1}\big(\mathbb{P}^{n_G-1} \backslash X_G, D \backslash (D \cap X_G)\big).
\end{gather}
However, this na\"ive attempt fails whenever the hypersurface $X_G$ intersects the integration cycle~$\sigma$ non-trivially, implying the presence of non-negligible singularities. Whenever singularities are present, $\sigma$ does not define an element in the corresponding na\"ive relative Betti homology group.
To successfully build the motive $I_G^{\rm m}$ in the presence of singularities, the blow up technique is applied.

A linear subvariety $L \subset \mathbb{P}^{n_G-1}$ defined by the vanishing of a subset of the set of Schwinger parameters is called a \textit{coordinate linear space}, while its subspace of real points with non-negative coordinates is denoted by
\begin{gather}
L(\mathbb{R}_{\ge 0}) = \{ [x_e]_{e \in E_G} \in L \,|\, x_e \in \mathbb{R}_{\ge 0} \}.
\end{gather}
Since the coefficients of $\Psi_G$ are positive, the locus of problematic singularities is
\begin{gather}
\sigma \cap X_G(\mathbb{C}) = \bigcup_{L \subset X_G} L(\mathbb{R}_{\ge 0}),
\end{gather}
where the union is taken over all coordinate linear spaces $L \subset X_G$.

\begin{Remark}
The coordinate linear spaces $L \subset X_G$ are in one-to-one correspondence with the subgraphs $\gamma \subset G$ such that $l_{\gamma} > 0$. It~follows that
\begin{gather}
\sigma \cap X_G(\mathbb{C}) = \bigcup_{\gamma \subset G} L_{\gamma}(\mathbb{R}_{\ge 0}),
\end{gather}
where the union is taken over all subgraphs $\gamma \subset G$ with $l_{\gamma} > 0$, and $L_{\gamma}$ is the linear subvariety of $\mathbb{P}^{n_G-1}$ defined by the equations $\{ x_e = 0 , e \in E_{\gamma} \}$.
\end{Remark}

The following theorem is proven, and an explicit algorithmic construction of the blow ups is given, by Bloch, Esnault and Kreimer~\cite{BEK06}.
\begin{Theorem} \label{th: BEK_th}
Let $G$ be a primitive log-divergent Feynman graph. There exists a tower
\begin{gather}
\pi\colon\ P = P_r \rightarrow P_{r-1} \rightarrow \dots \rightarrow P_1 \rightarrow P_0 = \mathbb{P}^{n_G-1}
\end{gather}
such that, for each $i=1,\dots,r$, $P_i$ is obtained by blowing up $P_{i-1}$ along the strict transform of a coordinate linear space $L_i \subset X_G$, and the following conditions hold:
\begin{itemize}\itemsep=0pt
\item[$(1)$] The pulled-back differential $\hat{\omega}_G = \pi^* \omega_G$ has no poles along the exceptional divisors associated to the blow ups.
\item[$(2)$] Let $B$ be the total transform of $D$ in $P$, that is
\begin{gather}
B = \pi^{-1}(D) = \pi^{-1} \bigg( \bigcup_{e \in E_G} \{ x_e = 0 \} \bigg).
\end{gather}
Then, $B \subset P$ is a normal crossing divisor such that none of the non-empty intersections of its irreducible components is contained in the strict transform $Y_G$ of $X_G$ in $P$.
\item[$(3)$] The strict transform of $\sigma$ in $P$ does not meet $Y_G$, that is
\begin{gather}
\hat{\sigma} \cap Y_G(\mathbb{C}) = \varnothing.
\end{gather}
\end{itemize}
\end{Theorem}
As a consequence of Theorem~\ref{th: BEK_th}, the motive $I_G^{\rm m}$ associated to any primitive log-divergent Feynman graph $G$ can be written explicitly.
Being $\partial \hat{\sigma} \subset B \backslash (B \cap Y_G)$, the domain of integration defines the class
\begin{gather}
[\hat{\sigma}] \in H_{n_G-1}^{{\rm B}}(P \backslash Y_G, B \backslash (B \cap Y_G))
\end{gather}
called \textit{Betti framing}, while the integrand defines the class
\begin{gather}
[\hat{\omega}_G] \in H^{n_G-1}_{\rm dR}(P \backslash Y_G, B \backslash (B \cap Y_G))
\end{gather}
called \textit{de Rham framing}. Brown and Doryn~\cite{BD13} present a method for explicit computation of the framings on the cohomology of Feynman graph hypersurfaces.
Then, the de Rham and Betti system of realisations $H_G = H^{n_G-1}(P \backslash Y_G, B \backslash (B \cap Y_G))$ is called the \textit{graph $H$-system},\footnote{The graph $H$-system is also explicitly known in the general case of renormalised amplitudes of single-scale graphs due to the work of Brown and Kreimer~\cite{BK13}, who paved the way for the rigorous investigation of divergent Feynman graphs and their renormalised amplitudes from an algebro-geometric perspective.} and the motivic Feynman integral $I_G^{\rm m}$ is given by
\begin{gather}
I_G^{\rm m} =\big [H_G, [\hat{\omega}_G], [\hat{\sigma}]\big]^{\rm m}.
\end{gather}
Indeed, the pairing of the classes $[\hat{\omega}_G]$ and $[\hat{\sigma}]$ yields the period
\begin{gather}
\int_{\hat{\sigma}} \hat{\omega}_G = \int_{ \hat{\sigma}} \pi^*(\omega_G) = \int_{\pi_*(\hat{\sigma})} \omega_G = \int_{\sigma} \omega_G = I_G
\end{gather}
by the equivalence relation under change of variables in $\mathcal{P}^{\rm m}$.

\begin{Example}
Adopting the notation
\begin{gather}
\mathcal{P}_{\rm log} = \mathbb{Q} \langle I_G \,|\, G \text{ is a primitive log-divergent Feynman graph} \rangle, \\
\mathcal{P}_{\phi^4} = \mathbb{Q} \big\langle I_G \,|\, G \text{ is a primitive log-divergent Feynman graph in $\phi^4$ theory} \big\rangle,
\end{gather}
we observe that the sequence of inclusions $\mathcal{P}_{\phi^4} \subset \mathcal{P}_{\rm log} \subset \mathcal{P}$ is preserved after promoting periods to periods of motives, that is $\mathcal{P}_{\phi^4}^{\rm m} \subset \mathcal{P}_{\rm log}^{\rm m} \subset \mathcal{P}^{\rm m}$.
\end{Example}

Many concrete results on the structure of $\mathcal{P}_{\rm log}$ follow from the study of the corresponding motivic space $\mathcal{P}_{\rm log}^{\rm m}$. For example, the following proposition on graph $H$-systems is proven by~Brown~\cite{Bro17_2} within the formalism of motivic Feynman integrals.
\begin{Proposition} \label{prop: trivial}
Let $G$ be a primitive log-divergent Feynman graph. If~$G$ has a single vertex, that is $v_G = 1$, or if $G$ has a single loop, that is $l_G =1$, then its $H$-system $H_G$ is isomorphic to the pure Hodge--Tate system $\mathbb{Q}(0)$.
\end{Proposition}
We observe that Proposition~\ref{prop: trivial} makes no restriction on the physicality of the graph $G$, which can have arbitrary vertex-degrees, so that $I_G$ belongs to $\mathcal{P}_{\rm log}$, but not necessarily to $\mathcal{P}_{\phi^4}$.

\subsection{Tannakian formalism}
We briefly introduce the fundamentals of the theory of \textit{Tannakian categories}, following the more detailed and comprehensive exposition by Deligne et al~\cite{Detal82}.
The concept of a Tannakian category was first introduced by Saavedra Rivano~\cite{Saa72} to encode the properties of the category $\mathop{\rm Rep}\nolimits_{\mathbb{K}}(G)$ of the finite-dimensional $\mathbb{K}$-linear representations of an affine group scheme $G$ over a~field~$\mathbb{K}$. Let~us recall some preliminary notions in category theory. Let~$\mathbb{K}$ be a subfield of $\mathbb{C}$.
\begin{Definition}
A \textit{$\mathbb{K}$-linear category} $\mathcal{C}$ is an additive category such that, for each pair of objects~$X,Y \in \text{Ob}(\mathcal{C})$, the group $\mathop{\rm Hom}_{\mathcal{C}}(X,Y)$ is a $\mathbb{K}$-vector space and the composition maps are $\mathbb{K}$-bilinear.
\end{Definition}
\begin{Definition}
Let $\mathcal{C}$ be a $\mathbb{K}$-linear category endowed with a $\mathbb{K}$-bilinear functor $\otimes\colon \mathcal{C} \times \mathcal{C} \rightarrow \mathcal{C}$.
\begin{itemize}\itemsep=0pt
\item[$(a)$] An \textit{associativity constraint} for $(\mathcal{C}, \otimes)$ is a natural transformation
\begin{gather}
\phi = \phi_{\cdot , \cdot , \cdot} \colon\ \cdot \otimes (\cdot \otimes \cdot) \longrightarrow (\cdot \otimes \cdot ) \otimes \cdot
\end{gather}
such that the following two conditions hold:
\begin{itemize}\itemsep=0pt
\item[$(a.1)$] For all $X,Y,Z \in \text{Ob}(\mathcal{C})$, the map $\phi_{X,Y,Z}$ is an isomorphism.
\item[$(a.2)$] For all $X,Y,Z,T \in \text{Ob}(\mathcal{C})$, the following diagram commutes:
\begin{equation}
\begin{tikzcd}[column sep=-2.3em]
& & X \otimes (Y \otimes (Z \otimes T)) \arrow[dll, "\text{Id} \otimes \phi_{Y,Z,T}"'] \arrow[drr, "\phi_{X,Y,Z \otimes T}"] & &\\
X \otimes ((Y \otimes Z) \otimes T) \arrow[dr, "\phi_{X,Y \otimes Z,T}"'] & & & & (X \otimes Y) \otimes (Z \otimes T) \arrow[dl, "\phi_{X \otimes Y,Z,T}"]\\
& (X \otimes (Y \otimes Z)) \otimes T \arrow[rr, "\phi_{X,Y,Z} \otimes \text{Id}"'] & & ((X \otimes Y) \otimes Z) \otimes T. &
\end{tikzcd}
\end{equation}
\end{itemize}
\item[$(b)$] A \textit{commutativity constraint} for $(\mathcal{C}, \otimes)$ is a natural transformation
\begin{gather}
\psi = \psi_{\cdot , *}\colon \cdot \otimes * \longrightarrow * \otimes \cdot
\end{gather}
such that the following two conditions hold:
\begin{itemize}\itemsep=0pt
\item[$(b.1)$] For all $X,Y \in \text{Ob}(\mathcal{C})$, the map $\psi_{X,Y}$ is an isomorphism.
\item[$(b.2)$] For all $X,Y \in \text{Ob}(\mathcal{C})$, the following composition is the identity:
\begin{gather}
\psi_{Y,X} \circ \psi_{X,Y} \colon\ X \otimes Y \longrightarrow X \otimes Y.
\end{gather}
\end{itemize}
\item[$(c)$] An associativity constraint and a commutativity constraint are \textit{compatible} if, for all $X,Y,\allowbreak Z \in \text{Ob}(\mathcal{C})$, the following diagram commutes:
\begin{equation}
\begin{tikzcd}[column sep=+3.8em]
 & X \otimes (Y \otimes Z) \arrow[r, "\phi_{X,Y,Z}"] \arrow[dl, "\text{Id} \otimes \psi_{Y,Z}"'] & (X \otimes Y) \otimes Z \arrow[dr, "\psi_{X \otimes Y, Z}"] & \\
 X \otimes (Z \otimes Y) \arrow[dr, "\phi_{X,Z,Y}"'] & & & Z \otimes (X \otimes Y) \arrow[dl, "\phi_{X,Z,Y}"] \\
 & (X \otimes Z) \otimes Y \arrow[r, "\psi_{X,Z} \otimes \text{Id}"'] & (Z \otimes X) \otimes Y. &
\end{tikzcd}
\end{equation}
\item[($d)$] A pair $(U,u)$ consisting of an object $U \in \text{Ob}(\mathcal{C})$ and an isomorphism $u\colon U \rightarrow U \otimes U$ is an \textit{identity object} if the functor $X \mapsto U \otimes X$ is an equivalence of categories.
\end{itemize}
\end{Definition}

\begin{Definition}
A \textit{$\mathbb{K}$-linear tensor category} is a tuple $(\mathcal{C}, \otimes, \phi, \psi)$ consisting of a $\mathbb{K}$-linear cate\-gory $\mathcal{C}$, a $\mathbb{K}$-bilinear functor $\otimes \colon \mathcal{C} \times \mathcal{C} \rightarrow \mathcal{C}$, and compatible associativity and commutativity constraints $\phi$, $\psi$ such that $\mathcal{C}$ contains an identity object.
\end{Definition}

\begin{Definition}
An object $L \in \text{Ob}(\mathcal{C})$ is \textit{invertible} if the functor $X \mapsto L \otimes X$ is an equivalence of categories.
Equivalently, $L$ is invertible if and only if there exists an object $L' \in \text{Ob}(\mathcal{C})$ such that $L \otimes L' \simeq \mathbf{1}$. Then, $L'$ is also invertible.
\end{Definition}

\begin{Definition}
Let $(\mathcal{C}, \otimes)$ be a $\mathbb{K}$-linear tensor category, where we omit the constraints $\phi$, $\psi$ for simplicity, and let $X,Y \in \text{Ob}(\mathcal{C})$. Assume that there exists an object $Z \in \text{Ob}(\mathcal{C})$ such that, for all $T \in \text{Ob}(\mathcal{C})$, the functors $T \mapsto \mathop{\rm Hom}(T,Z)$ and $T \mapsto \mathop{\rm Hom}(T \otimes X, Y)$ admit a functorial isomorphism
\begin{gather}
\mathop{\rm Hom}(T,Z) \xlongrightarrow{\sim} \mathop{\rm Hom}(T \otimes X, Y).
\end{gather}
In this case, the functor $T \mapsto \mathop{\rm Hom}(T \otimes X, Y)$ is said to be \textit{representable} and the object $Z$ is called the \textit{internal Hom} between the objects $X$ and $Y$. It~is alternatively written as $\underline{\mathop{\rm Hom}}(X,Y)$ and it is unique up to isomorphism.
\end{Definition}

\begin{Definition}
The \textit{dual} of an object $X \in \text{Ob}(\mathcal{C})$ is defined as $X^{\vee} = \underline{\mathop{\rm Hom}}(X,\mathbf{1})$. If~$X^{\vee}$ and~${(X^{\vee})}^{\vee}$ exist, then there is a natural morphism $X \mapsto {\big(X^{\vee}\big)}^{\vee}$ and the object $X$ is \textit{reflexive} if such a morphism is an isomorphism.
\end{Definition}

\begin{Definition}
A $\mathbb{K}$-linear tensor category $(\mathcal{C}, \otimes)$ is \textit{rigid} if the following conditions hold:
\begin{itemize}\itemsep=0pt
\item[(1)] For all $X,Y \in \text{Ob}(\mathcal{C})$, $\underline{\mathop{\rm Hom}}(X,Y)$ exists.
\item[(2)] For all $X_1,X_2,Y_1,Y_2 \in \text{Ob}(\mathcal{C})$, the natural morphism
\begin{gather}
\underline{\mathop{\rm Hom}}(X_1,Y_1) \otimes \underline{\mathop{\rm Hom}}(X_2,Y_2) \longrightarrow \underline{\mathop{\rm Hom}}(X_1 \otimes X_2,Y_1 \otimes Y_2)
\end{gather}
is an isomorphism.
\item[(3)] All objects are reflexive.
\end{itemize}
\end{Definition}

\begin{Definition} \label{def: tannaka}
A \textit{Tannakian category} over the field $\mathbb{K}$ is a rigid abelian $\mathbb{K}$-linear tensor category $\mathcal{T}$ such that $\text{End}(\mathbf{1}) = \mathbb{K}$, and there exists an exact faithful $\mathbb{K}$-linear tensor functor $\omega\colon \mathcal{T} \rightarrow \text{Vec}_{\mathbb{K}}$, where $\text{Vec}_{\mathbb{K}}$ is the category of finite-dimensional vector spaces over $\mathbb{K}$. Any such functor is called a \textit{fibre functor}.
\end{Definition}

\begin{Example}
The category $\text{Vec}_{\mathbb{K}}$ of finite-dimensional $\mathbb{K}$-vector spaces, together with the identity functor, is a Tannakian category over $\mathbb{K}$.
\end{Example}

\begin{Example}
The category $\text{GrVec}_{\mathbb{K}}$ of finite-dimensional graded $\mathbb{K}$-vector spaces, together with the forgetful functor $\omega\colon \text{GrVec}_{\mathbb{K}} \rightarrow \text{Vec}_{\mathbb{K}}$, sending $(V, (V_n)_{n \in \mathbb{Z}})$ to $V$, is~a~Tannakian category over $\mathbb{K}$.
\end{Example}

\begin{Example}
The category $\mathop{\rm Rep}\nolimits_{\mathbb{K}}(G)$ of finite-dimensional $\mathbb{K}$-linear representations of an~abst\-ract group $G$, together with the functor $\omega\colon \mathop{\rm Rep}\nolimits_{\mathbb{K}}(G) \rightarrow \text{Vec}_{\mathbb{K}}$ that forgets the action of $G$, is~a~Tannakian category over $\mathbb{K}$.
\end{Example}

Let us fix a Tannakian category $\mathcal{T}$ over $\mathbb{K}$ and a fibre functor $\omega$ of $\mathcal{T}$. Let~$R$ be a $\mathbb{K}$-algebra. We~denote by $\underline{\mathop{\rm Aut}}^{\otimes}(\omega)(R)$ the collection of families $(\lambda_X)_{X \in \text{Ob}(\mathcal{T})}$ of $R$-linear automorphisms
\begin{gather}
\lambda_X\colon\ \omega(X) \otimes_{\mathbb{K}} R \longrightarrow \omega(X) \otimes_{\mathbb{K}} R
\end{gather}
which are compatible with the tensor structure and functorial.
Here, compatibility with the tensor structure and functoriality mean\footnote{In the given diagrams, all unlabelled tensor products are over $\mathbb{K}$ and all unlabelled arrows are the natural isomorphisms.} that:
\begin{itemize}\itemsep=0pt
\item[(1)] For all $X_1, X_2 \in \text{Ob}(\mathcal{T})$, the following diagram commutes:
\begin{equation}
\begin{tikzcd}[column sep=+4.0em]
\omega(X_1 \otimes X_2) \otimes R \arrow[r, "\lambda_{X_1 \otimes X_2}"] \arrow[d] & \omega(X_1 \otimes X_2) \otimes R \arrow[d] \\
\omega(X_1) \otimes \omega(X_2) \otimes R \arrow[d] & \omega(X_1) \otimes \omega(X_2) \otimes R \arrow[d] \\
(\omega(X_1) \otimes R) \otimes_R (\omega(X_2) \otimes R) \arrow[r,"\lambda_{X_1} \otimes_R \lambda_{X_2}"'] & (\omega(X_1) \otimes R) \otimes_R (\omega(X_2) \otimes R).
\end{tikzcd}
\end{equation}
\item[(2)] The following diagram commutes:
\begin{equation}
\begin{tikzcd}
\omega(\mathbf{1}) \otimes R \arrow[r, "\lambda_{\mathbf{1}}"] \arrow[d] & \omega(\mathbf{1}) \otimes R \arrow[d] \\
R \arrow[r, "\text{Id}"'] & R.
\end{tikzcd}
\end{equation}
\item[(3)] For all $X,Y \in \text{Ob}(\mathcal{T})$ and for every morphism $\alpha \in \mathop{\rm Hom}(X,Y)$, the following diagram commutes:
\begin{equation}
\begin{tikzcd}
\omega(X) \otimes R \arrow[r, "\lambda_X"] \arrow[d, "\omega(\alpha) \otimes \text{Id}"'] & \omega(X) \otimes R \arrow[d, "\omega(\alpha) \otimes \text{Id}"] \\
\omega(Y) \otimes R \arrow[r, "\lambda_Y"'] & \omega(Y) \otimes R.
\end{tikzcd}
\end{equation}
\end{itemize}
Deligne et al~\cite{Detal82} proved that all Tannakian categories are categories of finite-dimensional linear representations of a pro-algebraic group.
\begin{Theorem} \label{th: aut}
Let $\mathcal{T}$ be a Tannakian category over $\mathbb{K}$ with a fibre functor $\omega$.
\begin{itemize}\itemsep=0pt
\item[$(1)$] The functor $R \mapsto \underline{\mathop{\rm Aut}}^{\otimes}(\omega)(R)$ is representable by an affine group scheme over $\mathbb{K}$, which is denoted as $\underline{\mathop{\rm Aut}}^{\otimes}(\omega)$ or $G^{\omega}$, and is called the \textit{Tannaka group} of the pair $(\mathcal{T}, \omega)$.
\item[$(2)$] For every $X \in \text{Ob}(\mathcal{T})$, the group $\underline{\mathop{\rm Aut}}^{\otimes}(\omega)$ acts naturally on $\omega(X)$ and the functor
\begin{equation}
\begin{tikzcd}[row sep = -0.1em]
\mathcal{T} \arrow[r] & \mathop{\rm Rep}\nolimits_{\mathbb{K}}(G^{\omega}), \\[1ex]
X \arrow[r, mapsto] & \omega(X) \arrow[loop right, "G^{\omega}"]
\end{tikzcd}
\end{equation}
sending $X$ to the vector space $\omega(X)$ with this action of $\underline{\mathop{\rm Aut}}^{\otimes}(\omega)$, is an equivalence of~cate\-gories.
\end{itemize}
\end{Theorem}
Given a second fibre functor $\omega'$, we analogously define $\underline{\mathop{\rm Isom}}^{\otimes}(\omega,\omega')(R)$ to be the collection of families $(\tau_X)_{X \in \text{Ob}(\mathcal{T})}$ of $R$-linear isomorphisms
\begin{gather}
\tau_X\colon\ \omega(X) \otimes_{\mathbb{K}} R \longrightarrow \omega'(X) \otimes_{\mathbb{K}} R
\end{gather}
which are compatible with the tensor structure and functorial.
Deligne et al~\cite{Detal82} proved the following result.
\begin{Theorem}
Let $\mathcal{T}$ be a Tannakian category over $\mathbb{K}$ with two fibre functors $\omega$ and $\omega'$. The~fun\-c\-tor $R \mapsto \underline{\mathop{\rm Isom}}^{\otimes}(\omega,\omega')(R)$ is representable by an affine scheme over $\mathbb{K}$, which is denoted as $\underline{\mathop{\rm Isom}}^{\otimes}(\omega, \omega')$, and is a right torsor under $\underline{\mathop{\rm Aut}}^{\otimes}(\omega)$ and a left torsor under $\underline{\mathop{\rm Aut}}^{\otimes}(\omega')$.
\end{Theorem}
\begin{Remark}
In what follows, we denote $\mathop{\rm Aut}^{\otimes}(\omega) = \underline{\mathop{\rm Aut}}^{\otimes}(\omega)(\mathbb{C})$ the group of $\mathbb{C}$-linear auto\-morphisms of the fibre functor $\omega$, and analogously $\mathop{\rm Isom}\nolimits^{\otimes}(\omega, \omega') = \underline{\mathop{\rm Isom}}^{\otimes}(\omega, \omega')(\mathbb{C})$ the group of~$\mathbb{C}$-linear isomorphisms between the fibre functors $\omega$ and $\omega'$.
\end{Remark}

\subsection{Motivic Galois theory} \label{sec: mcategory}
Recall the notions of pure and mixed $H$-systems associated with algebraic varieties over $\mathbb{Q}$ given in Sections~\ref{sec: pure} and~\ref{sec: mixed}.
On the one hand, the algebraic de Rham and Betti cohomologies of a smooth projective $\mathbb{Q}$-variety are fundamentally described by a pure $H$-system.
On the other hand, applying the resolution of singularities by Hironaka~\cite{Hir64-I,Hir64-II}, the algebraic de Rham and Betti cohomologies of an arbitrary quasi-projective $\mathbb{Q}$-variety can be expressed in terms of the cohomologies of smooth projective varieties, and since pure $H$-systems of different weights get mixed in this expression, they are fundamentally described by a mixed system of realisations.
Because we specifically look at the application of the theory of motives to the theory of periods, it is sufficient to our purpose to work with the partial realisation of Grothendieck's notion of motives provided by $H$-systems\footnote{In the literature on motivic periods, mixed de Rham and Betti systems of realisations are sometimes called motives themselves.}. However, it turns out to be necessary and fruitful to enhance the na\"ive construction given in Section~\ref{sec: mot_alg} to its rigorous category-theoretic formulation.

Recall that $\mathbf{MHSy}(\mathbb{Q})$ is the category of mixed $H$-systems over $\mathbb{Q}$, and $\omega_{{\rm B}}$, $\omega_{\rm dR}$ are its two forgetful functors arising from the Betti and de Rham realisations, respectively.
All the defining properties of a Tannakian category, encoded in Definition~\ref{def: tannaka}, apply to $\mathbf{MHSy}(\mathbb{Q})$, thus justifying the use of the Tannakian machinery in the study of motivic periods.
\begin{Proposition}
$\mathbf{MHSy}(\mathbb{Q})$ is a Tannakian category over $\mathbb{Q}$ and the functors $\omega_{{\rm B}}$, $\omega_{\rm dR}$ are fibre functors.
\end{Proposition}
In what follows, we write $ \mathcal{H} = \mathbf{MHSy}(\mathbb{Q}) $.
The pro-algebraic group $\mathop{\rm Aut}^{\otimes}(\omega_{\rm dR})$ is denoted by $G^{\rm dR}$ and called the \textit{motivic Galois group}.
Observe that $G^{\rm dR}(H)$ is a group in ${\rm GL}(\omega_{\rm dR}(H))$ for every object $H \in \text{Ob}(\mathcal{H})$.
Following Theorem~\ref{th: aut}, the category of mixed $H$-systems is equivalent to the category of finite-dimensional $\mathbb{Q}$-linear representations of the motivic Galois group, that is
\begin{gather}
\mathcal{H} \simeq \mathop{\rm Rep}\nolimits_{\mathbb{Q}}\big(G^{\rm dR}\big).
\end{gather}
\begin{Remark}
We observe that the motivic Galois group can alternatively be realised via Betti cohomology as $G^{{\rm B}} = \mathop{\rm Aut}^{\otimes}(\omega_{{\rm B}})$ and the corresponding category of finite-dimensional $\mathbb{Q}$-linear representations is still the same category of mixed $H$-systems $\mathcal{H}$.
\end{Remark}

In Tannakian formalism, the space of motivic periods $\mathcal{P}^{\rm m}$ is expressed as
\begin{gather}
\mathcal{P}^{\rm m}= \mathbb{Q} \big\langle [H,\omega,\sigma]^{\rm m} \,|\, H \in \text{Ob}(\mathcal{H}), \; \omega \in \omega_{\rm dR}(H), \, \sigma \in \omega_{{\rm B}}(H)^{\vee} \big\rangle
\end{gather}
with implicit factorisation modulo bilinearity and functoriality.
Recall that
\begin{gather}
H = (H_{\rm dR}, \, H_{{\rm B}}, \, \mathop{\rm comp}\nolimits_H \colon H_{\rm dR} \otimes_{\mathbb{Q}} \mathbb{C} \xrightarrow{\sim} H_{{\rm B}} \otimes_{\mathbb{Q}} \mathbb{C}),
\end{gather}
where $H_{\rm dR} = \omega_{\rm dR}(H)$, $H_{{\rm B}} = \omega_{{\rm B}}(H)$ are finite-dimensional $\mathbb{Q}$-vector spaces and $\mathop{\rm comp} \in \mathop{\rm Isom}\nolimits^{\otimes}(\omega_{\rm dR},\omega_{{\rm B}})$ is a $\mathbb{C}$-linear isomorphism.
In this framework, an alternative but equivalent description of motivic periods follows.
\begin{Proposition}
$\mathcal{P}^{\rm m}$ is isomorphic to the space of regular functions on the affine $\mathbb{Q}$-scheme $\mathop{\rm Isom}\nolimits^{\otimes}(\omega_{\rm dR},\omega_{{\rm B}})$.
\end{Proposition}
We observe that such an isomorphism is explicitly written as
\begin{align}
\mathcal{P}^{\rm m} &\xlongrightarrow{\sim} \mathcal{O}\big(\mathop{\rm Isom}\nolimits^{\otimes}(\omega_{\rm dR},\omega_{{\rm B}})\big),
\\
 {}[H,\omega,\sigma ]^{\rm m} &\longmapsto \big[ (\lambda_X)_{X \in \text{Ob}(\mathcal{H})} \mapsto (\sigma \otimes_{\mathbb{Q}} \text{Id}_{\mathbb{C}}) \circ \lambda_H \circ (\omega \otimes 2 \pi {\rm i}) \big],
\end{align}
where we have
\begin{gather}
\begin{array}{rcccl}
\omega_{\rm dR}(H) \otimes_{\mathbb{Q}} \mathbb{C} &\xlongrightarrow{\lambda_H} &\omega_{{\rm B}}(H) \otimes_{\mathbb{Q}} \mathbb{C} &\xlongrightarrow{\sigma \otimes_{\mathbb{Q}} \text{Id}_{\mathbb{C}}} & \mathbb{C},
\\[1ex]
\omega &\longmapsto & \lambda_H(\omega) &\longmapsto & \sigma(\lambda_H(\omega)).
\end{array}
\end{gather}
Following Theorem~\ref{th: aut}, the motivic Galois group $G^{\rm dR}$ has a natural action on $\mathop{\rm Isom}\nolimits^{\otimes}(\omega_{\rm dR},\omega_{{\rm B}})$ which is written as
\begin{gather} \label{eq: action}
G^{\rm dR} \otimes \mathop{\rm Isom}\nolimits^{\otimes}(\omega_{\rm dR},\omega_{{\rm B}}) \longrightarrow \mathop{\rm Isom}\nolimits^{\otimes}(\omega_{\rm dR},\omega_{{\rm B}})
\end{gather}
and which induces a dual coaction on the corresponding space of regular functions $\mathcal{P}^{\rm m}$, that is
\begin{align} \label{eq: coaction}
\Delta\colon\ \mathcal{P}^{\rm m} &\longrightarrow \mathcal{O}\big(G^{\rm dR}\big) \otimes \mathcal{P}^{\rm m}, \\
 {}[H,\omega,\sigma]^{\rm m} &\longmapsto \sum_{i=1}^n \big[H,\omega,e_i^{\vee}\big]^{\rm dR} \otimes [H,e_i,\sigma]^{\rm m},
\end{align}
where $\{ e_i \}_{i = 1, \dots, n}$ is a basis of $\omega_{\rm dR}(H)$, and $\big\{ e_i^{\vee} \big\}_{i = 1, \dots, n}$ denotes the associated vector dual basis of~$\omega_{{\rm B}}(H)^{\vee}$, as introduced in Section~\ref{sec: periodmap}.
Here, $[H,e_i,\sigma]^{\rm m} \in \mathcal{P}^{\rm m}$ is called a \textit{Galois conjugate} of the motivic period $[H,\omega,\sigma]^{\rm m}$, while $\big[H,\omega,e_i^{\vee}\big]^{\rm dR} \in \mathcal{O}\big(G^{\rm dR}\big)$ is called a \textit{de Rham period}. We~denote by $\mathcal{P}^{\rm dR} = \mathcal{O}\big(G^{\rm dR}\big)$ the space of regular functions on the motivic Galois group, and we call it the \textit{space of de Rham periods}. The coaction $\Delta$ is known as the \textit{Galois coaction}.

\begin{Remark}
Note that the space of de Rham periods is naturally a Hopf algebra, while the space of motivic periods is not, thus making the coaction intrinsically asymmetric. Indeed, the Galois coaction turns the finite-dimensional $\mathbb{Q}$-vector space $\mathcal{P}^{\rm m}$ into a comodule over the Hopf algebra $\mathcal{P}^{\rm dR}$.
Moreover, there is a canonical single-valued map that associates a number to each de Rham period, thus representing the de Rham analogue of the period map. For a detailed discussion we refer to Brown~\cite{Bro17_2, Bro17_1}.
\end{Remark}

\begin{Example}
Consider the motivic logarithm $\log(z)^{\rm m}$ for $z \in \mathbb{Q} \backslash \{ 0,1 \}$. Following Section~\ref{sec: log}, we have
\begin{gather}
\log(z)^{\rm m} = \bigg[ H^1(\mathbb{G}_m, \{ 1, z \}), \bigg[ \frac{{\rm d}x}{x} \bigg], [ \gamma_1 ] \bigg]^{\rm m}.
\end{gather}
Let us denote $H=H^1(\mathbb{G}_m, \{ 1, z \})$ for simplicity. Adopting the canonical choice of bases, as in Example~\ref{ex: ex10}, the period matrix of $H$ is
\begin{gather}
\begin{pmatrix}
1 & \log(z) \\
0 & 2 \pi {\rm i}
\end{pmatrix}\!.
\end{gather}
Direct application of the prescription in~\eqref{eq: coaction} gives the explicit decomposition
\begin{gather}
\Delta\bigg[ H, \bigg[ \frac{{\rm d}x}{x} \bigg], [\gamma_1] \bigg]^{\rm m} = \bigg[ H, \bigg[ \frac{{\rm d}x}{x} \bigg], \bigg[ \frac{{\rm d}x}{z-1} \bigg]^{\vee} \bigg]^{\rm dR} \otimes \bigg[ H, \bigg[ \frac{{\rm d}x}{z-1} \bigg], [\gamma_1] \bigg]^{\rm m}
\\ \hphantom{\Delta\bigg[ H, \bigg[ \frac{{\rm d}x}{x} \bigg], [\gamma_1] \bigg]^{\rm m} =}
{} + \bigg[ H, \bigg[ \frac{{\rm d}x}{x} \bigg], \bigg[ \frac{{\rm d}x}{x} \bigg]^{\vee} \bigg]^{\rm dR} \otimes \bigg[ H, \bigg[ \frac{{\rm d}x}{x} \bigg], [\gamma_1] \bigg]^{\rm m}.
 \label{eq: coactLog00}
 \end{gather}
Observing that $\big[\frac{{\rm d}x}{z-1}\big]^{\vee} = [\gamma_1]$ and $\big[\frac{{\rm d}x}{x} \big]^{\vee} = [\gamma_2]$, and identifying the de Rham periods
\begin{gather}
\bigg[ H, \bigg[ \frac{{\rm d}x}{x} \bigg], \bigg[ \frac{{\rm d}x}{z-1} \bigg]^{\vee} \bigg]^{\rm dR} = \log(z)^{\rm dR} , \qquad
\bigg[ H, \bigg[ \frac{{\rm d}x}{x} \bigg], \bigg[ \frac{{\rm d}x}{x} \bigg]^{\vee} \bigg]^{\rm dR} =(2 \pi {\rm i})^{\rm dR},
\end{gather}
we find that the expression~\eqref{eq: coactLog00} is equivalent to
\begin{gather} \label{eq: coactLog}
\Delta \log(z)^{\rm m} = \log(z)^{\rm dR} \otimes 1^{\rm m} + (2 \pi {\rm i})^{\rm dR} \otimes \log(z)^{\rm m},
\end{gather}
where $1^{\rm m}$ and $\log(z)^{\rm m}$ are the Galois conjugates of $\log(z)^{\rm m}$.
\end{Example}

\begin{Example}
As for $\log(z)^{\rm m}$, the Galois coaction of the motivic multiple zeta value $\zeta(\mathbf{s})^{\rm m}$ can be computed explicitly.
For example, from the period matrix~\eqref{eq: matrixZ2}, we have that
\begin{gather}
\Delta \zeta(2)^{\rm m} = \zeta(2)^{\rm dR} \otimes 1^{\rm m} + \big((2 \pi {\rm i})^{\rm dR}\big)^2 \otimes \zeta(2)^{\rm m}.
\end{gather}
As powers of $(2 \pi {\rm i})^{\rm dR}$ naturally appear among de Rham conjugates, the Galois coaction is often intended with an implicit factorisation\footnote{The operation of factorisation of a $H$-system modulo $2 \pi {\rm i}$ is called a \textit{Tate twist}, and it can indeed be formally defined in terms of the Hodge--Tate systems introduced in Example~\ref{ex: hodgetate}.} of $\mathcal{P}^{\rm dR}$ modulo the ideal generated by $(2 \pi {\rm i})^{\rm dR}$. Let~us assume so.
For $n \ge 1$, we have that
\begin{gather} \label{eq: coactZ2}
\Delta \zeta(2)^{\rm m} = \zeta(2)^{\rm dR} \otimes 1^{\rm m} + 1^{\rm dR} \otimes \zeta(2)^{\rm m},
\\
\Delta \zeta(2n+1)^{\rm m} = \zeta(2n+1)^{\rm dR} \otimes 1^{\rm m} + 1^{\rm dR} \otimes \zeta(2n+1)^{\rm m},
\\
\Delta (\zeta(2)^{\rm m} \zeta(2n+1)^{\rm m}) = \zeta(2n+1)^{\rm dR} \otimes \zeta(2)^{\rm m} + 1^{\rm dR} \otimes \zeta(2)^{\rm m} \zeta(2n+1)^{\rm m}.
\end{gather}
\end{Example}

\subsection{Coaction conjecture}
Let us look at the example of scalar massless $\phi^4$ quantum field theory and consider the Galois coaction restricted to $\mathcal{P}_{\phi^4}^{\rm m}$. This is a priori valued in the whole space $\mathcal{P}^{\rm dR} \otimes \mathcal{P}^{\rm m}$.
However, after computing every known $\phi^4$-period with loop orders at most $7$ and several $\phi^4$-periods with higher loop orders, and explicitly verifying that in each case the Galois coaction preserves the space $\mathcal{P}_{\phi^4}^{\rm m}$, Panzer and Schnetz\footnote{Panzer and Schnetz~\cite{PS17} explicitly computed the first examples of $\phi^4$-amplitudes which are not MZVs. Such numbers are polylogarithms at 2nd and 6th roots of unity. The coaction conjecture is verified for them as well.}~\cite{PS17} proposed the following conjecture, known as the \textit{coaction conjecture}.
\begin{Conjecture} \label{conj: PanzerSchnetz}
The Galois coaction closes on $\phi^4$-periods. In other words, the Galois conjugates of a $\phi^4$-period are also $\phi^4$-periods, that is
\begin{gather}
\Delta\big(\mathcal{P}_{\phi^4}^{\rm m}\big)\subseteq\mathcal{P}^{\rm dR}\otimes\mathcal{P}_{\phi^4}^{\rm m}.
\end{gather}
\end{Conjecture}

Such a conjecture implies the existence of a fundamental hidden symmetry underlying the class of $\phi^4$-periods that we do not yet properly understand. Indeed, the unexpected observations by Panzer and Schnetz, and the resulting conjecture, have greatly stimulated research, moti\-va\-ting the search for a mathematical mechanism able to distinguish $\phi^4$-periods from periods of all graphs, and thus explain this surprising evidence.

Some advancements in this direction have already been made. Conjecture~\ref{conj: PanzerSchnetz} is the strongest among several reformulations of its statement obtained by suitably enlarging the space of ampli\-tu\-des under consideration. A~weaker version of the coaction conjecture has been proven by Brown~\cite{Bro17_2}.
To a scalar Feynman graph $G$, we associate the finite-dimensional $\mathbb{Q}$-vector space~$\mathcal{P}_{a}^{\rm m}(G)$ consisting of the motivic realisations of all affine integrals of globally-defined algebraic differential forms on the usual integration domain $\sigma$, called the \textit{affine periods} of $G$. They include convergent affine integrals of the form
\begin{gather}
\int_0^{\infty} \cdots \int_0^{\infty}\frac{q}{\Psi_G^k} \bigg|_{x_{n_G} = 1} {\rm d}^{n_G-1}x,
\end{gather}
where $k \ge 1$ is an integer, and $q$ is a polynomial in $\mathbb{Q}[x_1, \dots, x_{n_G -1}]$. However, the denominator of the integrand of an affine period of $G$ can also possibly involve linear factors of the form $\sum_{e \in E_{\gamma}} x_e$, where $\gamma$ is a subgraph of~$G$.
\begin{Theorem} \label{th: coactB}
The Galois group acts on $\mathcal{P}_{a}^{\rm m}(G)$, that is $\Delta(\mathcal{P}_{a}^{\rm m}(G)) \subseteq \mathcal{P}^{\rm dR} \otimes \mathcal{P}_{a}^{\rm m}(G)$.
\end{Theorem}

We write $\mathcal{P}_{a}^{\rm m}$ to denote the space spanned by all $\mathcal{P}_{a}^{\rm m}(G)$ for any\footnote{We observe that restricting to $\phi^4$-graphs does not change the resulting space $\mathcal{P}_{a}^{\rm m}$, which is sometimes also denoted by $\mathcal{P}_{\tilde{\phi}^4 }^{\rm m}$.} scalar Feynman graph~$G$. Since direct computations show that $\mathcal{P}_{a}^{\rm m} \simeq \mathcal{P}_{\phi^4}^{\rm m}$ at low loop orders, Theorem~\ref{th: coactB} directly supports Conjecture~\ref{conj: PanzerSchnetz}.

\begin{Remark}
The affine motive $H_{a}$ of a scalar Feynman graph $G$ is defined analogously to the projective $H$-system $H = H^{n_G-1}(P \backslash Y_G, B \backslash (B \cap Y_G))$, described in Section~\ref{sec: intGmot}, after replacing the projective space $P$ with an affine open subspace $A \subset P$ obtained by removing its hyperplanes with strictly positive coefficients. The inclusion $A \backslash Y_G \hookrightarrow P \backslash Y_G$ induces a morphism of objects $H \rightarrow H_a$ in $\mathcal{H}$, which gives an equivalence at the level of motivic periods. Every motivic period of $H$ is also a motivic period of $H_{a}$, but the contrary is not true. Many periods of $H_{a}$ are not periods of Feynman graphs in the ordinary sense.
\end{Remark}

Let $G$ be a scalar Feynman graph. We~define the \textit{generalised Feynman integrals} associated to $G$ as the projective integrals of parametric form\footnote{Note that the ordinary Feynman integral~\eqref{eq: I_G} of $G$ can be written in this form.}
\begin{gather}\label{eq: I_Gtilde}
\int_{\sigma} \frac{p \, \Omega}{\Psi_G^k \, \Xi_G^h(\{p_j, m_e\})},
\end{gather}
where $k,h \ge 1$ are integers, and $p$ is a homogeneous polynomial in $\mathbb{Q}[\{x_e\}]$ of degree $k l_G +h (l_G +1) - n_G$, so that the overall integrand is homogeneous of degree zero.
Although possibly divergent, generalised Feynman integrals are periods, and they can be promoted to their motivic realisations after being suitably regularised. We~define the finite-dimensional $\mathbb{Q}$-vector space $\mathcal{P}_{g}^{\rm m}$ to be the space of motivic realisations of regularised versions of generalised Feynman integrals.
An analogous statement to Conjecture~\ref{conj: PanzerSchnetz} is proposed by Brown~\cite{Bro17_2}.
\begin{Conjecture}
$\mathcal{P}_{g}^{\rm m}$ is stable under the Galois coaction, that is $\Delta(\mathcal{P}_{g}^{\rm m}) \subseteq \mathcal{P}^{\rm dR} \otimes \mathcal{P}_{g}^{\rm m}$.
\end{Conjecture}

\subsection{Weights and the small graph principle} \label{sec: weight}
Recall the notions of Hodge and weight filtrations introduced in Sections~\ref{sec: pure} and~\ref{sec: mixed}.
For $H \in \text{Ob}(\mathcal{H})$, the $\mathbb{Q}$-vector space $\omega_{\rm dR}(H)$ is equipped with a Hodge filtration $F^{\bullet}$ and a weight filtration $W^{\rm dR}_{\bullet}$, while the $\mathbb{Q}$-vector space $\omega_{{\rm B}}(H)$ is provided with a weight filtration $W^{{\rm B}}_{\bullet}$ only.
Mixed $H$-systems, contrary to pure ones, do not have a well-defined weight. However, their graded quotients with respect to the weight filtration do possess a pure structure of definite weight. These properties are used to define a notion of \textit{weight} for motivic periods.
\begin{Definition}
The weight filtration on $\omega_{\rm dR}(H)$ induces a weight filtration on the space of motivic periods by
\begin{gather}
W_{\bullet}^{\rm dR} \mathcal{P}^{\rm m} = \mathbb{Q} \big\langle [H,\omega,\sigma]^{\rm m} \,|\, \omega \in W_{\bullet}^{\rm dR} \big\rangle.
\end{gather}
We denote $W=W^{\rm dR}$ for simplicity. A~given motivic period $[H,\omega,\sigma]^{\rm m} $ is said to have weight at most $n$ if it belongs to $W_n \mathcal{P}^{\rm m}$, and to have weight $n$ if it is non-zero in the graded quotient ${\rm Gr}^W_n \mathcal{P}^{\rm m} = W_n \mathcal{P}^{\rm m} / W_{n-1} \mathcal{P}^{\rm m}$.
\end{Definition}
\begin{Remark}
We observe that the weight of motivic periods can alternatively, but equivalently, be defined from the Betti realisation via the weight filtration induced on $\mathcal{P}^{\rm m}$ by $W^{{\rm B}}$.
\end{Remark}

\begin{Example} \label{ex: weightLog}
Consider $H = H^1(\mathbb{G}_m, \{ 1, z \})$ for $z \in \mathbb{Q} \backslash \{ 0,1 \}$. Its weight filtration in the de Rham realisation is
\begin{gather}
W_{-1} = 0 \subseteq W_0 = W_1 = \mathbb{Q}(0) \subseteq W_2 = H^1(\mathbb{G}_m, \{ 1, z \}).
\end{gather}
Observing that $0,1 \in W_0$ and $2\pi {\rm i}, \log(z) \in W_2$, the weight of each entry of the period matrix of~$H$ is determined. Indeed, $0$, $1$ are periods of weight zero, while $2\pi {\rm i}$, $\log(z)$ have weight $2$.
\end{Example}
\begin{Example}
The weight filtration can be used to systematically study $\mathcal{P}_{\phi^4}^{\rm m}$ weight by weight.
For example, direct computation in low weight shows that $W_0 \mathcal{P}_{\phi^4}^{\rm m} =W_1 \mathcal{P}_{\phi^4}^{\rm m}=W_2\mathcal{P}_{\phi^4}^{\rm m} = \mathbb{Q}(0)$.
\end{Example}

More generally, the following proposition is due to Brown~\cite{Bro17_2}.
\begin{Proposition} \label{prop: Q02}
Let $G$ be a primitive log-divergent Feynman graph. Every Galois conjugate of its motivic Feynman integral $I_G^{\rm m}$ which has weight up to $2$ is a period of $\mathbb{Q}(0)$, that is a rational number.
\end{Proposition}
The Galois conjugates of Feynman periods are expected to satisfy the following conjecture by Brown~\cite{Bro17_2}, known as \textit{small graph principle}.\footnote{The small graph principle is also conjectured to hold for regularised versions of generalised Feynman integrals associated to arbitrary scalar Feynman graphs.}
\begin{Conjecture} \label{conj: sgp}
Let $G$ be a primitive log-divergent Feynman graph. Denote by $[ H_G, \omega_G, \sigma ]^{\rm m}$ the motivic realisation of its Feynman integral $I_G^{\rm m}$. The elements on the right-hand side of the coaction formula for $\Delta [ H_G, \omega_G, \sigma ]^{\rm m}$ can be expressed in the form
\begin{gather}
\prod_i [ H_{\gamma_i}, \omega_{\gamma_i}, \sigma ]^{\rm m},
\end{gather}
where the product runs over a subset $\{ \gamma_i \}$ of the set of subgraphs and quotient graphs of $G$.
\end{Conjecture}
The small graph principle implies that the Galois conjugates of weight at most $k$ of the motivic amplitude of a primitive log-divergent Feynman graph are associated to its sub-quotient graphs\footnote{Sub-quotient graphs are formally obtained by contracting and deleting edges of the original graph.} with at most $k+1$ edges.
In other words, when interested in Feynman periods of weight at most $k$, it suggests to look at graphs with up to $k+1$ edges. As a consequence, the topology of a given graph constrains the Galois theory of its amplitudes.

\begin{Example}
Consider the system of realisations $H = H^1(\mathbb{G}_m, \{ 1, p \})$ for $p \in \mathbb{Q} \backslash \{ 0,1 \}$ prime. Following Example~\ref{ex: weightLog}, $\log(p)^{\rm m}$ is a period of $H$ with weight 2. Then, the small graph principle suggests that any $\log(p)^{\rm m}$ appearing in the right-hand side of the coaction formula for a given $\phi^4$-periods comes from graphs with at most three edges. Proposition~\ref{prop: trivial} implies that all two-edge graphs are trivial, meaning that the associated graph motive $H_G$ is the Hodge--Tate system~$\mathbb{Q}(0)$, which does not have $\log(p)^{\rm m}$ in its period matrix. Writing down all possible graphs with three edges, we get the graphs shown in Fig.~\ref{fig: log} along with their first graph polynomials in~the Schwinger parameters.
\begin{figure}[htb!]
\centering
\subfloat[][{$x_1+x_2+x_3$}]
 {\includegraphics[scale=0.55]{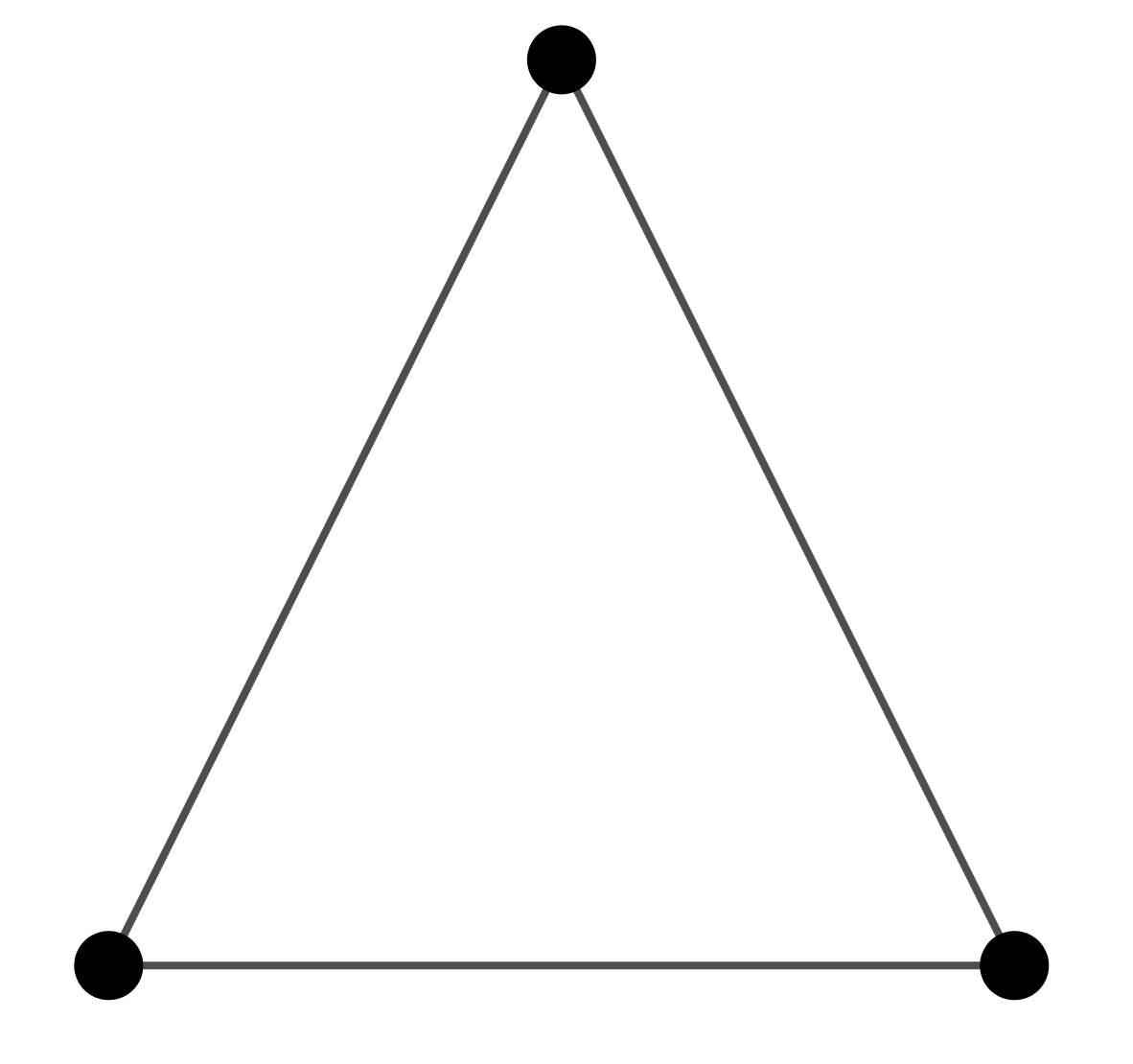}} \qquad
\subfloat[][{$x_1x_2+x_2x_3+x_3 x_1$}]
 {\includegraphics[scale=0.55]{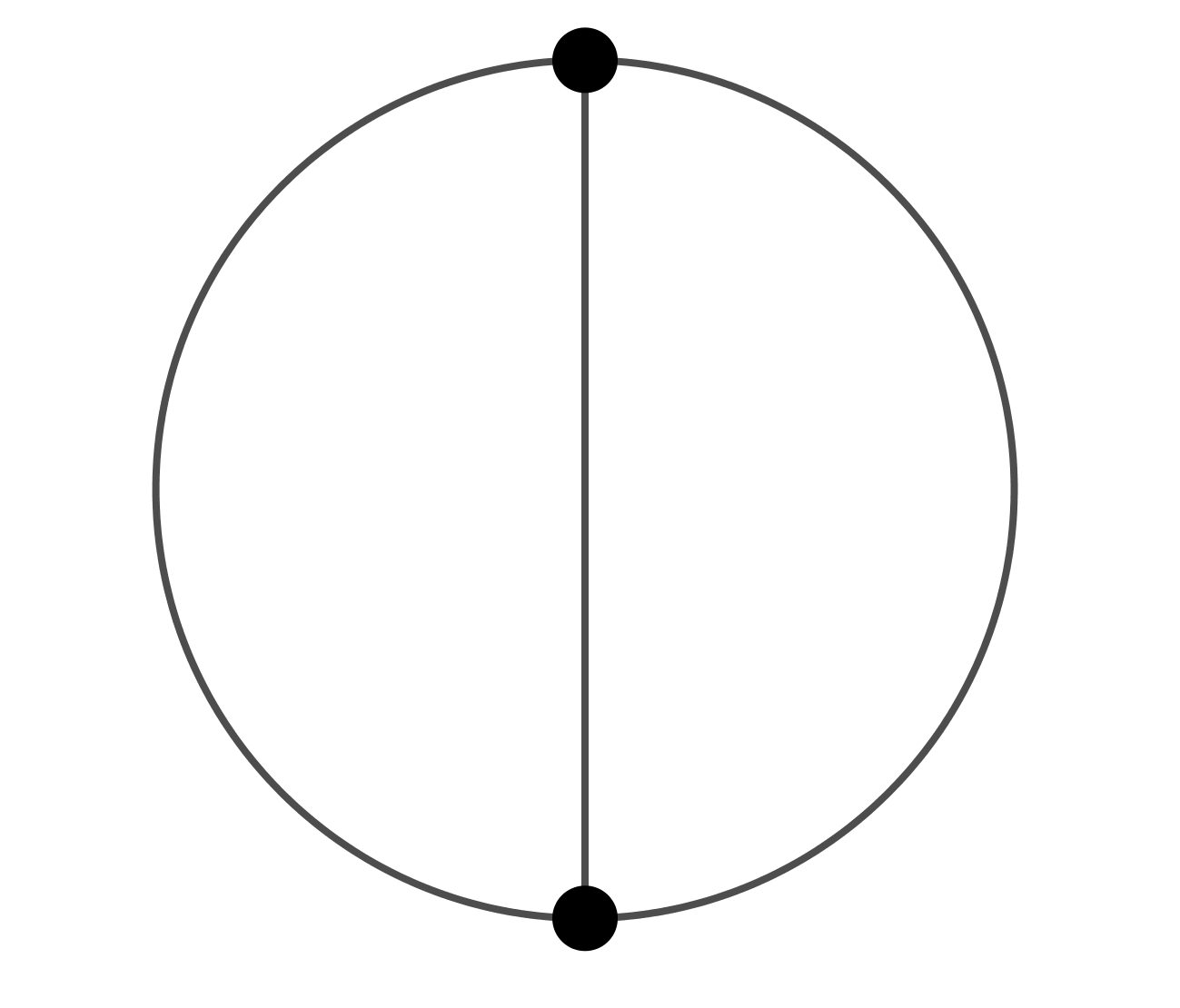}} \qquad
\subfloat[][{$x_1(x_2+x_3)$}]
 {\includegraphics[scale=0.55]{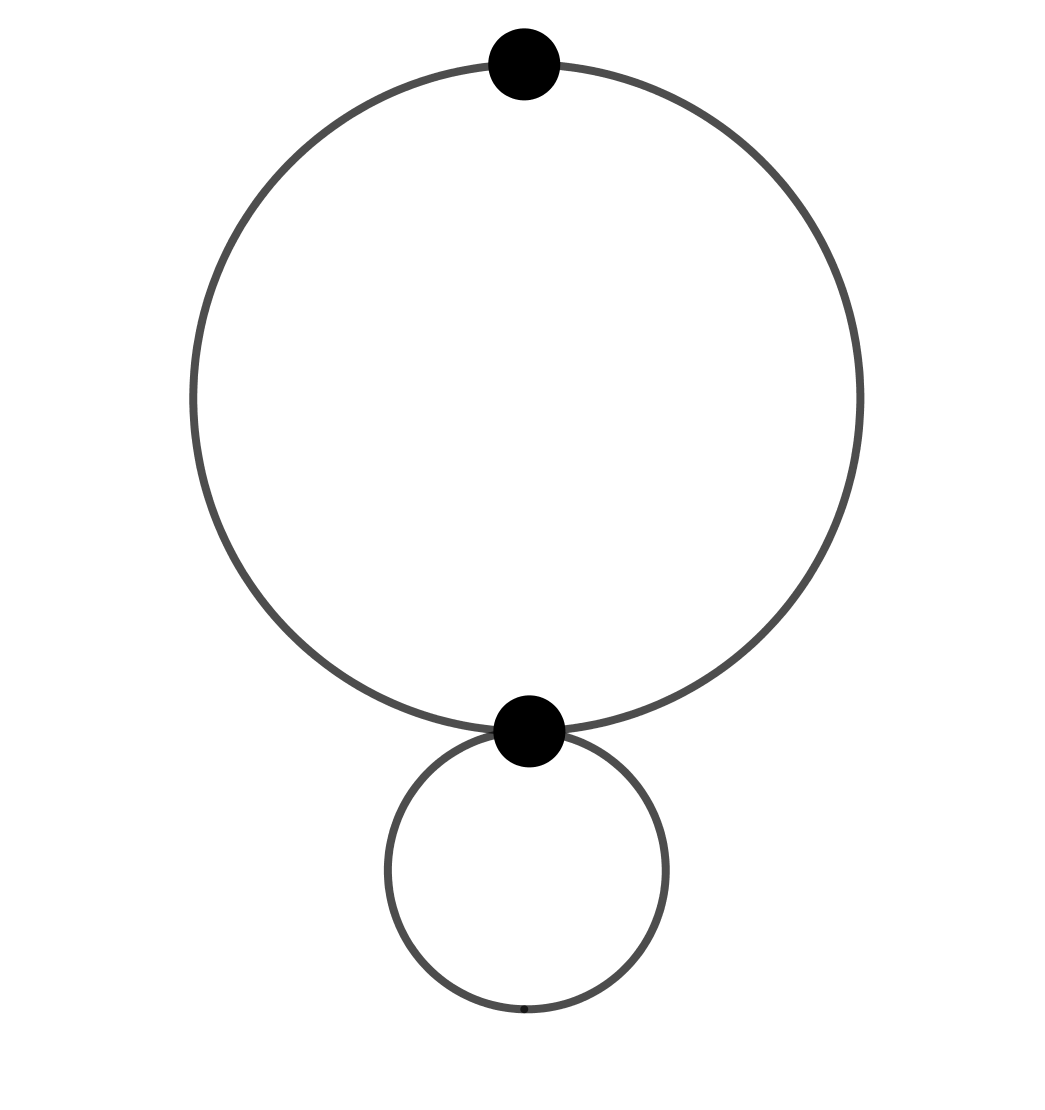}} \quad
\subfloat[][{$x_1x_2x_3$}]
 {\includegraphics[scale=0.55]{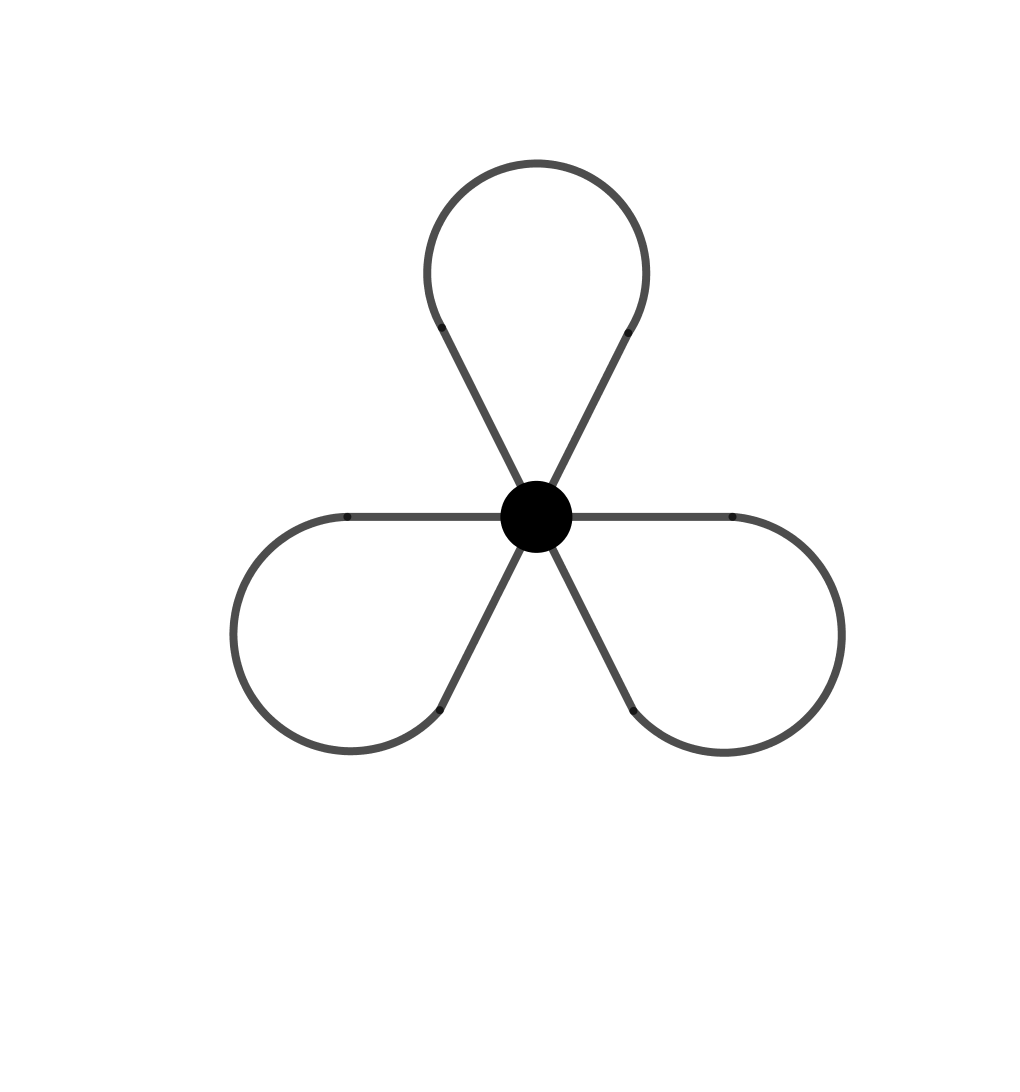}}
\caption{Feynman graphs with 3 edges and their first graph polynomials.}
\label{fig: log}
\end{figure}

The two outer graphs (a) and (d) are also trivial by Proposition~\ref{prop: trivial}, while the two middle graphs (b) and (c) satisfy $H_G = \mathbb{Q}(0) \oplus \mathbb{Q}(-1)$, which does not have $\log(p)^{\rm m}$ as a period. Indeed, $\log(p)$ cannot be obtained as an integral with a denominator equal to either of the first graph polynomials (b) or (c). We~conclude that $\log(p)^{\rm m}$ cannot be a Galois conjugate of a~$\phi^4$-period. By equation~\eqref{eq: coactLog}, we derive that $\log(p)^{\rm m} \notin \mathcal{P}_{\phi^4}^{\rm m}$. Note that this is consistent with Proposition~\ref{prop: Q02}.
\end{Example}

\begin{Example}
Direct computation by Panzer and Schnetz~\cite{PS17} shows that all $\phi^4$-periods of loop order up to 6 are $\mathbb{Q}$-linear combinations of multiple zeta values.
Following the small graph principle, we graphically order the set of MZVs by weight as
\begin{equation}
\begin{tikzcd}[row sep=-3.5, column sep=-2.5]
 1 & \zeta(2) & \zeta(3) & \zeta(2)^2 & \zeta(5) & \zeta(3)^2 & \zeta(7) & \zeta(3,5) & \cdots \\
 & & & & \zeta(2) \zeta(3) & \zeta(2)^3 & \zeta(2)\zeta(5) & \zeta(2)\zeta(3)^2 &\\
 & & & & & & \zeta(2)^2\zeta(3) & \vdots &
\end{tikzcd}
\end{equation}
As a consequence of the coaction conjecture and explicit expressions of the coaction formula, as~the ones in equation~\eqref{eq: coactZ2}, $\zeta(2)^{\rm m} \notin \mathcal{P}_{\phi^4}^{\rm m}$ implies that all MZVs which are linear in $\zeta(2)$ cannot be $\phi^4$-periods. Analogously, $ \big(\zeta(2)^2\big)^{\rm m} \notin \mathcal{P}_{\phi^4}^{\rm m}$ implies that all MZVs which are quadratic in $\zeta(2)$ are not $\phi^4$-periods.
The set of MZVs that can appear as $\phi^4$-periods would then be reduced to
\begin{equation}
\begin{tikzcd}[row sep=-3.5, column sep=-2.5]
 1 & & \zeta(3) & & \zeta(5) & \zeta(3)^2 & \zeta(7) & \zeta(3,5) & \cdots \\
 & & & & & \zeta(2)^3 & & \vdots &
\end{tikzcd}
\end{equation}
However, the statements $\zeta(2), \zeta(2)^2 \notin \mathcal{P}_{\phi^4}$ are only conjectural. Precisely, it is conjectured that $\zeta(2)^k \notin \mathcal{P}_{\phi^4}$ for $k \le 5$, while $\zeta(2)^k \in \mathcal{P}_{\phi^4}$ for $k \ge 6$. We~observe that such statements rely on~the control over weight drops, as conjectured by Panzer and Schnetz~\cite{PS17}, which excludes low weight MZVs coming from high loop order graphs.
\end{Example}

From similar considerations, other highly non-trivial constraints at all loop orders in perturbation theory can be derived using the Galois coaction and weight filtrations. For example, by~Conjecture~\ref{conj: PanzerSchnetz}, whenever it is shown that a given period is not a $\phi^4$-period, we conjecturally deduce that all periods that have the given one among their Galois conjugates cannot appear in~$\mathcal{P}_{\phi^4}$ either.

\begin{Remark}
Structures even more fundamental than those captured by the coaction conjecture and the small graph principle underly the space of motivic periods of Feynman graphs. Although not being sufficiently explored in the literature, the notion of \textit{operad} in the category of~motives imposes strong constraints on the admissible periods and it should be the object of~further investigation. The operad structure underlying the space of motivic Feynman integrals is interestingly the same structure governing the renormalisation group equation. Kaufmann and Ward~\cite{KW17} provide details on related notions in category theory.
\end{Remark}

\section{Conclusions}

Originally providing a framework for re-organising and re-interpreting much of the previous knowledge on Feynman integrals, the theory of motivic periods has revealed unexpected features, placing restrictions on the set of numbers which can occur as amplitudes and paving the way for a more comprehensive understanding of their general structure. Indeed, the coaction conjecture gives new constraints at each loop order, which in turn propagate to all higher loop orders because of the recursive structure inherent in perturbative quantum field theories. At the same time, the small graph principle makes finite computations at low-loop into all-order results.

Assume to deal with a Feynman integral of the form $\int_{\sigma} \omega$ in $\mathcal{P}$.
The general prescription for its~investigation via the theory of motivic periods can be summarised as follows:
\begin{itemize}\itemsep=0pt
\item[(1)] Associate a motivic representation $[H,\omega, \sigma]^{\rm m}$ to the integral $\int_{\sigma} \omega$, deriving explicitly the corresponding algebraic de Rham and Betti system of realisations, and cohomology and homology classes.
\item[(2)] Use all the known information about the mixed system of realisations $H$ to derive explicit filtrations.
\item[(3)] Write down the period matrix of $H$.
\item[(4)] Apply the Galois coaction and derive the Galois conjugates.
\item[(5)] Apply the theory of weights of mixed Hodge structures to reduce the calculation of the Galois conjugates to the study of motivic periods of small graphs.
\item[(6)] Analyse explicitly the few admissible small graphs and eliminate the excluded periods, sometimes called \textit{holes}.
\item[(7)] Possibly use other known symmetries of the specific example at hand to draw conclusions.
\end{itemize}
This picture is, however, extensively conjectural. The very first step of replacing periods with their motivic version requests the validity of the period conjecture. Moreover, even disregarding the conjectural status of current statements, the present state of understanding of motivic amplitudes is still far from building a theory. Although the given general prescription for the investigation of motivic Feynman integrals has been particularly fruitful for massless scalar $\phi^4$ quantum field theory, further advancements are needed to enlarge the reach of current results.

Speculating in full generality, consider the whole class of Feynman integrals in perturbative quantum field theory. We~expect them to have a natural motivic representation, and thus to generate a space $\mathcal{M}$ of motivic periods, a space $\mathcal{A}$ of de Rham periods, and a corresponding coaction $\Delta\colon\mathcal{M} \longrightarrow \mathcal{P}^{\rm dR} \otimes \mathcal{P}^{\rm m}$. A~potential coaction principle would then state that $\Delta(\mathcal{M}) \subseteq \mathcal{A} \otimes \mathcal{M}$. Being $\mathcal{A}$ a Hopf algebra, we could canonically introduce the group $C$ of homomorphisms from $\mathcal{A}$ to any commutative ring. It~would follow that the coaction principle can be recast in~terms of the group action $C \times \mathcal{M} \longrightarrow \mathcal{M}$, that is, the space of amplitudes is stable under the action of the group $C$, often referred to as \textit{cosmic Galois group}. This speculative construction, that broadly reproduces the general prescription summarised above, motivates a programme of research leading towards a systematic study of scattering amplitudes via the representation theory of groups.

Although practically harder than the $\phi^4$-case, like-minded attempts are already on the way to gather information about the numbers that come from evaluating other classes of Feynman integrals.
\begin{itemize}\itemsep=0pt
\item[$(a)$] Towards a general motivic description of scalar quantum field theories, Abreu et al.~\cite{Aetal17, Aetal18, Aetal20} give evidence suggesting that scalar Feynman integrals of small graphs with non-trivial masses and momenta satisfy similar properties to $\phi^4$-periods. A~diagrammatic coaction for specific families of integrals appearing in the evaluation of scalar Feynman dia\-g\-rams, such as multiple polylogarithms and generalised hypergeometric functions, is proposed and a connection between this diagrammatic coaction and graphical operations on Feynman diagrams is conjectured. At one-loop order, a fully explicit and very compact representation of the coaction in terms of one-loop integrals and their cuts is found.
Moreover, Brown and Dupont~\cite{BD19} investigate a rigorous theory of motives associated to certain hypergeometric integrals.
\item[$(b)$] A subsequent generalisation arises transitioning from scalar quantum field theories to gauge theories.
The problem of dealing with much more involved parametric integrands which are not explicitly expressed in terms of the Symanzik polynomials of the associated Feynman graphs has only recently been tackled. A~combinatoric and graph-theoretic approach to Schwinger parametric Feynman integrals in quantum electrodynamics by Golz~\cite{Gol18} has revealed that the parametric integrands can be explicitly written in terms of new types of~graph polynomials related to specific subgraphs. The tensor structure of quantum electrodynamics is given a diagrammatic interpretation. The resulting significant simplification of the integrands paves the way for a systematic motivic description of gauge theories.
\item[$(c)$] In the same research direction, a high-precision computation of the 4-loop contribution to the electron anomalous magnetic moment $g-2$ by Laporta~\cite{Lap17} shows the presence of polylogarithmic parts with fourth and sixth roots of unity. This result is conjecturally recast into the motivic $f$-alphabet by Schnetz~\cite{Sch18}, giving a more compact expression which explicitly reveals a Galois structure. In this work, the $\mathbb{Q}$-vector spaces of Galois conjugates of the $g-2$ are conjectured up to weight four.
\end{itemize}

As a final remark, we mention that scattering amplitudes do not appear exclusively in perturbative quantum field theory. Among other settings, there are string perturbation theory and $\mathcal{N} = 4$ super Yang--Mills theory.
In each of these theories, after suitably defining the space of integrals or amplitudes\footnote{In various modern approaches to $\mathcal{N} = 4~\text{SYM}$, including the bootstrap method, on-shell techniques, and the amplituhedron, the amplitude is constructed independently of the Feynman graphs. In these settings, the coaction principle operates on the entire amplitude, contrary to the case of perturbative quantum field theory, where it operates graph by graph.} under consideration, a version of the coaction principle is expected to hold and some promising preliminary results have already been found. We~refer to the work of Schlotterer, Stieberger and Taylor~\cite{SS13, ST14} and subsequent developments for superstring perturbation theory, and to the work of Caron-Huot et al.~\cite{CDetal20, CDetal19} for the planar limit of $\mathcal{N} = 4$ super Yang--Mills theory.

\subsection*{Acknowledgements}
I thank Francis Brown and Lionel Mason for useful discussions.
I thank the three anonymous referees for their detailed reports that have provided a valuable guide in improving the paper.
Fina\-lly, I thank Evgeny Mukhin and the rest of the organisers of the Conference on Representation Theory and Integrable Systems (ETH Z\"urich, 2019) for the opportunity to speak and to contribute to the special issue.
This work is partially supported by the Italian Department of Edu\-ca\-tion, Research and University (Torno Subito 13474/19.09.2018 POR-Lazio-FSE/2014-2020) and the Swiss National Centre of Competence in Research SwissMAP (NCCR 51NF40-141869 The Mathematics of Physics).

\LastPageEnding


\addcontentsline{toc}{section}{References}
\begin{thebibliography}{99}
\footnotesize\itemsep=0pt

\bibitem{Aetal17}
Abreu S., Britto R., Duhr C., Gardi E., Diagrammatic {H}opf algebra of cut
 {F}eynman integrals: the one-loop case, \href{https://doi.org/10.1007/jhep12(2017)090}{\textit{J.~High Energy Phys.}}
 \textbf{2017} (2017), no.~12, 090, 73~pages, \href{https://arxiv.org/abs/1704.07931}{arXiv:1704.07931}.

\bibitem{Aetal18}
Abreu S., Britto R., Duhr C., Gardi E., Matthew J., Coaction for {F}eynman
 integrals and diagrams, \textit{PoS Proc. Sci.} (2018), PoS(LL2018), 047,
 14~pages, \href{https://arxiv.org/abs/1808.00069}{arXiv:1808.00069}.

\bibitem{Aetal20}
Abreu S., Britto R., Duhr C., Gardi E., Matthew J., From positive geometries to
 a coaction on hypergeometric functions, \href{https://doi.org/10.1007/jhep02(2020)122}{\textit{J.~High Energy Phys.}}
 \textbf{2020} (2020), no.~2, 122, 44~pages, \href{https://arxiv.org/abs/1910.08358}{arXiv:1910.08358}.

\bibitem{And04}
Andr\'e Y., Une introduction aux motifs (motifs purs, motifs mixtes,
 p\'eriodes), \textit{Panoramas et Synth\`eses}, Vol.~17, Soci\'et\'e
 Math\'ematique de France, Paris, 2004.

\bibitem{BEK06}
Bloch S., Esnault H., Kreimer D., On motives associated to graph polynomials,
 \href{https://doi.org/10.1007/s00220-006-0040-2}{\textit{Comm. Math. Phys.}} \textbf{267} (2006), 181--225,
 \href{https://arxiv.org/abs/math.AG/0510011}{arXiv:math.AG/0510011}.

\bibitem{BBV10}
Bl\"umlein J., Broadhurst D.J., Vermaseren J.A.M., The multiple zeta value data
 mine, \href{https://doi.org/10.1016/j.cpc.2009.11.007}{\textit{Comput. Phys. Comm.}} \textbf{181} (2010), 582--625,
 \href{https://arxiv.org/abs/0907.2557}{arXiv:0907.2557}.

\bibitem{BWdiv}
Bogner C., Weinzierl S., Periods and {F}eynman integrals, \href{https://doi.org/10.1063/1.3106041}{\textit{J.~Math.
 Phys.}} \textbf{50} (2009), 042302, 16~pages, \href{https://arxiv.org/abs/0711.4863}{arXiv:0711.4863}.

\bibitem{BW10}
Bogner C., Weinzierl S., Feynman graph polynomials, \href{https://doi.org/10.1142/S0217751X10049438}{\textit{Internat.~J. Modern
 Phys.~A}} \textbf{25} (2010), 2585--2618, \href{https://arxiv.org/abs/1002.3458}{arXiv:1002.3458}.

\bibitem{BT82}
Bott R., Tu L.W., Differential forms in algebraic topology, \textit{Graduate
 Texts in Mathematics}, Vol.~82, \href{https://doi.org/10.1007/978-1-4757-3951-0}{Springer-Verlag}, New York~-- Berlin, 1982.

\bibitem{Bro13}
Broadhurst D., Multiple zeta values and modular forms in quantum field theory,
 in Computer Algebra in Quantum Field Theory, \textit{Texts Monogr. Symbol. Comput.},
 \href{https://doi.org/10.1007/978-3-7091-1616-6_2}{Springer}, Vienna, 2013, 33--73.

\bibitem{BK95}
Broadhurst D.J., Kreimer D., Knots and numbers in {$\phi^4$} theory to {$7$}
 loops and beyond, \href{https://doi.org/10.1142/S012918319500037X}{\textit{Internat.~J. Modern Phys.~C}} \textbf{6} (1995),
 519--524, \href{https://arxiv.org/abs/hep-ph/9504352}{arXiv:hep-ph/9504352}.

\bibitem{BK97}
Broadhurst D.J., Kreimer D., Association of multiple zeta values with positive
 knots via {F}eynman diagrams up to {$9$} loops, \href{https://doi.org/10.1016/S0370-2693(96)01623-1}{\textit{Phys. Lett.~B}}
 \textbf{393} (1997), 403--412, \href{https://arxiv.org/abs/hep-th/9609128}{arXiv:hep-th/9609128}.

\bibitem{Brown2012}
Brown F., On the decomposition of motivic multiple zeta values, in
 Galois--{T}eichm\"uller theory and arithmetic geometry, \href{https://doi.org/10.2969/aspm/06310031}{\textit{Adv. Stud.
 Pure Math.}}, Vol.~63, Math. Soc. Japan, Tokyo, 2012, 31--58,
 \href{https://arxiv.org/abs/1102.1310}{arXiv:1102.1310}.

\bibitem{Bro17_2}
Brown F., Feynman amplitudes, coaction principle, and cosmic {G}alois group,
 \href{https://doi.org/10.4310/CNTP.2017.v11.n3.a1}{\textit{Commun. Number Theory Phys.}} \textbf{11} (2017), 453--556,
 \href{https://arxiv.org/abs/1512.06409}{arXiv:1512.06409}.

\bibitem{Bro17_1}
Brown F., Notes on motivic periods, \href{https://doi.org/10.4310/CNTP.2017.v11.n3.a2}{\textit{Commun. Number Theory Phys.}}
 \textbf{11} (2017), 557--655, \href{https://arxiv.org/abs/1512.06410}{arXiv:1512.06410}.

\bibitem{BD13}
Brown F., Doryn D., Framings for graph hypersurfaces, \href{https://arxiv.org/abs/1301.3056}{arXiv:1301.3056}.

\bibitem{BD19}
Brown F., Dupont C., Lauricella hypergeometric functions, unipotent fundamental
 groups of the punctured Riemann sphere, and their motivic coactions,
 \href{https://arxiv.org/abs/1907.06603}{arXiv:1907.06603}.

\bibitem{BK13}
Brown F., Kreimer D., Angles, scales and parametric renormalization,
 \href{https://doi.org/10.1007/s11005-013-0625-6}{\textit{Lett. Math. Phys.}} \textbf{103} (2013), 933--1007,
 \href{https://arxiv.org/abs/1112.1180}{arXiv:1112.1180}.

\bibitem{BS2012}
Brown F., Schnetz O., A {K}3 in {$\phi^4$}, \href{https://doi.org/10.1215/00127094-1644201}{\textit{Duke Math.~J.}} \textbf{161}
 (2012), 1817--1862, \href{https://arxiv.org/abs/1006.4064}{arXiv:1006.4064}.

\bibitem{BS12}
Brown F., Schnetz O., Single-valued multiple polylogarithms and a proof of the
 zig-zag conjecture, \href{https://doi.org/10.1016/j.jnt.2014.09.007}{\textit{J.~Number Theory}} \textbf{148} (2015), 478--506,
 \href{https://arxiv.org/abs/1208.1890}{arXiv:1208.1890}.

\bibitem{CDetal20}
Caron-Huot S., Dixon L.J., Drummond J.M., Dulat F., Foster J., G\"urdo\u{g}an
 O., von Hippel M., McLeod A.J., Papathanasiou G., The {S}teinmann cluster
 bootstrap for ${\mathcal N}=4$ super {Y}ang--{M}ills amplitudes, \textit{PoS
 Proc. Sci.} (2020), PoS(CORFU2019), 003, 37~pages, \href{https://arxiv.org/abs/2005.06735}{arXiv:2005.06735}.

\bibitem{CDetal19}
Caron-Huot S., Dixon L.J., Dulat F., von Hippel M., McLeod A.J., Papathanasiou
 G., The cosmic {G}alois group and extended {S}teinmann relations for planar
 {${\mathcal N}=4$} {SYM} amplitudes, \href{https://doi.org/10.1007/jhep09(2019)061}{\textit{J.~High Energy Phys.}}
 \textbf{2019} (2019), no.~9, 061, 65~pages, \href{https://arxiv.org/abs/1906.07116}{arXiv:1906.07116}.

\bibitem{CS01}
Colmez P., Serre J.P. (Editors), Correspondance {G}rothendieck--{S}erre,
 \textit{Documents Math\'ematiques (Paris)}, Vol.~2, Soci\'et\'e
 Math\'ematique de France, Paris, 2001.

\bibitem{Der31}
De~Rham G., Sur l'analysis situs des vari\'et\'es \`a {$n$} dimensions, 1931,
 available ar \url{http://www.numdam.org/item?id=THESE_1931__129__1_0}.

\bibitem{Del71_1}
Deligne P., Th\'eorie de {H}odge. {I}, in Actes du {C}ongr\`es {I}nternational
 des {M}ath\'ematiciens ({N}ice, 1970), {T}ome~1, 1971, 425--430.

\bibitem{Del71_2}
Deligne P., Th\'eorie de {H}odge.~{II}, \href{https://doi.org/10.1007/BF02684692}{\textit{Inst. Hautes \'Etudes Sci.
 Publ. Math.}} \textbf{40} (1971), 5--57.

\bibitem{Del74}
Deligne P., Th\'eorie de {H}odge.~{III}, \href{https://doi.org/10.1007/BF02685881}{\textit{Inst. Hautes \'Etudes Sci.
 Publ. Math.}} \textbf{44} (1974), 5--77.

\bibitem{DG05}
Deligne P., Goncharov A.B., Groupes fondamentaux motiviques de {T}ate mixte,
 \href{https://doi.org/10.1016/j.ansens.2004.11.001}{\textit{Ann. Sci. \'Ecole Norm. Sup.~(4)}} \textbf{38} (2005), 1--56,
 \href{https://arxiv.org/abs/math.NT/0302267}{arXiv:math.NT/0302267}.

\bibitem{Detal82}
Deligne P., Milne J.S., Ogus A., Shih K., Hodge cycles, motives, and {S}himura
 varieties, \textit{Lecture Notes in Math.}, Vol.~900, \href{https://doi.org/10.1007/978-3-540-38955-2}{Springer-Verlag},
 Berlin~-- New York, 1982.

\bibitem{Dem71}
Demazure M., Motifs des vari\'et\'es alg\'ebriques, in S{\'e}minaire Bourbaki,
 Vol.~1991/92, Exp.\ No.~364--381, \textit{Lecture Notes in Math.}, Vol.~180,
 \href{https://doi.org/10.1007/BFb0058822}{Springer-Verlag}, Berlin~-- Heidelberg, 1971, 19--38.

\bibitem{Dys49}
Dyson F.J., The radiation theories of {T}omonaga, {S}chwinger, and {F}eynman,
 \href{https://doi.org/10.1103/PhysRev.75.486}{\textit{Phys. Rev.}} \textbf{75} (1949), 486--502.

\bibitem{Dys52}
Dyson F.J., Divergence of perturbation theory in quantum electrodynamics,
 \href{https://doi.org/10.1103/PhysRev.85.631}{\textit{Phys. Rev.}} \textbf{85} (1952), 631--632.

\bibitem{EH13}
Elvang H., Huang Y., Scattering amplitudes, \href{https://arxiv.org/abs/1308.1697}{arXiv:1308.1697}.

\bibitem{Fey49}
Feynman R.P., Space-time approach to quantum electrodynamics, \href{https://doi.org/10.1103/PhysRev.76.769}{\textit{Phys.
 Rev.}} \textbf{76} (1949), 769--789.

\bibitem{FG}
Gil J.I.B., Fres\'an J., Multiple zeta values: from numbers to motives,
 {a}vailable at \url{https://javier.fresan.perso.math.cnrs.fr/mzv.pdf}.

\bibitem{Gol18}
Golz M., Parametric quantum electrodynamics, Ph.D.~Thesis, {H}umboldt
 University, Berlin, 2018.

\bibitem{Gro66}
Grothendieck A., On the de {R}ham cohomology of algebraic varieties,
 \href{https://doi.org/10.1007/BF02684807}{\textit{Inst. Hautes \'Etudes Sci. Publ. Math.}} \textbf{29} (1966), 95--103.

\bibitem{Har77}
Hartshorne R., Algebraic geometry, \textit{Graduate Texts in Mathematics}, Vol.~52, \href{https://doi.org/10.1007/978-1-4757-3849-0}{Springer-Verlag}, New York~-- Heidelberg, 1977.

\bibitem{Henn15}
Henn J.M., Lectures on differential equations for {F}eynman integrals,
 \href{https://doi.org/10.1088/1751-8113/48/15/153001}{\textit{J.~Phys.~A: Math. Theor.}} \textbf{48} (2015), 153001, 35~pages,
 \href{https://arxiv.org/abs/1412.2296}{arXiv:1412.2296}.

\bibitem{Hir64-I}
Hironaka H., Resolution of singularities of an algebraic variety over a field
 of characteristic zero.~{I}, \href{https://doi.org/10.2307/1970486}{\textit{Ann. of Math.}} \textbf{79} (1964),
 109--203.

\bibitem{Hir64-II}
Hironaka H., Resolution of singularities of an algebraic variety over a field
 of characteristic zero.~{II}, \href{https://doi.org/10.2307/1970547}{\textit{Ann. of Math.}} \textbf{79} (1964),
 205--326.

\bibitem{Hod41}
Hodge W.V.D., The theory and applications of harmonic integrals, Cambridge
 University Press, Cambridge, Macmillan Company, New York, 1941.

\bibitem{HM17}
Huber A., M\"uller-Stach S., Periods and {N}ori motives, \textit{Ergebnisse der
 Mathematik und ihrer Grenzgebiete. 3.~Folge. A~Series of Modern Surveys in
 Mathematics}, Vol.~65, \href{https://doi.org/10.1007/978-3-319-50926-6}{Springer}, Cham, 2017.

\bibitem{KS06}
Kashiwara M., Schapira P., Categories and sheaves, \textit{Grundlehren der
 Mathematischen Wissenschaften}, Vol.~332, \href{https://doi.org/10.1007/3-540-27950-4}{Springer-Verlag}, Berlin, 2006.

\bibitem{KW17}
Kaufmann R.M., Ward B.C., Feynman categories, \textit{Ast\'erisque}
 \textbf{387} (2017), vii+161~pages, \href{https://arxiv.org/abs/1312.1269}{arXiv:1312.1269}.

\bibitem{Kle72}
Kleiman S.L., Motives, in Algebraic Geometry, {O}slo 1970 ({P}roc. {F}ifth
 {N}ordic {S}ummer-{S}chool in {M}ath., {O}slo, 1970), 1972, 53--82.

\bibitem{KZ01}
Kontsevich M., Zagier D., Periods, in Mathematics Unlimited~-- 2001 and Beyond,
 \href{https://doi.org/10.1007/978-3-642-56478-9_39}{Springer}, Berlin, 2001, 771--808.

\bibitem{Kre97}
Kreimer D., Renormalization and knot theory, \href{https://doi.org/10.1142/S0218216597000315}{\textit{J.~Knot Theory
 Ramifications}} \textbf{6} (1997), 479--581, \href{https://arxiv.org/abs/hep-th/9412045}{arXiv:hep-th/9412045}.

\bibitem{Lap17}
Laporta S., High-precision calculation of the 4-loop contribution to the
 electron $g-2$ in QED, \href{https://doi.org/10.1016/j.physletb.2017.06.056}{\textit{Phys. Lett.~B}} \textbf{772} (2017), 232--238,
 \href{https://arxiv.org/abs/1704.06996}{arXiv:1704.06996}.

\bibitem{LR96}
Laporta S., Remiddi E., The Analytical value of the electron $g-2$ at order
 $\alpha^3$ in QED, \href{https://doi.org/10.1016/0370-2693(96)00439-X}{\textit{Phys. Lett.~B}} \textbf{379} (1996), 283--291,
 \href{https://arxiv.org/abs/hep-ph/9602417}{arXiv:hep-ph/9602417}.

\bibitem{MNP13}
Murre J.P., Nagel J., Peters C.A.M., Lectures on the theory of pure motives,
 \textit{University Lecture Series}, Vol.~61, \href{https://doi.org/10.1090/ulect/061}{Amer. Math. Soc.}, Providence,
 RI, 2013.

\bibitem{PS17}
Panzer E., Schnetz O., The {G}alois coaction on {$\phi^4$} periods,
 \href{https://doi.org/10.4310/CNTP.2017.v11.n3.a3}{\textit{Commun. Number Theory Phys.}} \textbf{11} (2017), 657--705,
 \href{https://arxiv.org/abs/1603.04289}{arXiv:1603.04289}.

\bibitem{Pet57}
Petermann A., Fourth order magnetic moment of the electron, \textit{Helv. Phys.
 Acta} \textbf{30} (1957), 407--408.

\bibitem{Saa72}
Saavedra~Rivano N., Cat\'egories {T}annakiennes, \textit{Lecture Notes in
 Math.}, Vol.~265, \href{https://doi.org/10.1007/BFb0059108}{Springer-Verlag}, Berlin~-- New York, 1972.

\bibitem{Sal51_2}
Salam A., Divergent integrals in renormalizable field theories, \href{https://doi.org/10.1103/PhysRev.84.426}{\textit{Phys.
 Rev.}} \textbf{84} (1951), 426--431.

\bibitem{Sal51_1}
Salam A., Overlapping divergences and the {$S$}-matrix, \href{https://doi.org/10.1103/PhysRev.82.217}{\textit{Phys. Rev.}}
 \textbf{82} (1951), 217--227.

\bibitem{SS13}
Schlotterer O., Stieberger S., Motivic multiple zeta values and superstring
 amplitudes, \href{https://doi.org/10.1088/1751-8113/46/47/475401}{\textit{J.~Phys.~A: Math. Theor.}} \textbf{46} (2013), 475401,
 37~pages, \href{https://arxiv.org/abs/1205.1516}{arXiv:1205.1516}.

\bibitem{Sch10}
Schnetz O., Quantum periods: a census of {$\phi^4$}-transcendentals,
 \href{https://doi.org/10.4310/CNTP.2010.v4.n1.a1}{\textit{Commun. Number Theory Phys.}} \textbf{4} (2010), 1--47,
 \href{https://arxiv.org/abs/0801.2856}{arXiv:0801.2856}.

\bibitem{Sch18}
Schnetz O., The {G}alois coaction on the electron anomalous magnetic moment,
 \href{https://doi.org/10.4310/cntp.2018.v12.n2.a4}{\textit{Commun. Number Theory Phys.}} \textbf{12} (2018), 335--354,
 \href{https://arxiv.org/abs/1711.05118}{arXiv:1711.05118}.

\bibitem{Sch16}
Schnetz O., Numbers and functions in quantum field theory, \href{https://doi.org/10.1103/physrevd.97.085018}{\textit{Phys.
 Rev.~D}} \textbf{97} (2018), 085018, 20~pages, \href{https://arxiv.org/abs/1606.08598}{arXiv:1606.08598}.

\bibitem{SSV17}
Serone M., Spada G., Villadoro G., The power of perturbation theory,
 \href{https://doi.org/10.1007/JHEP05(2017)056}{\textit{J.~High Energy Phys.}} \textbf{2017} (2017), no.~5, 056, 41~pages,
 \href{https://arxiv.org/abs/1702.04148}{arXiv:1702.04148}.

\bibitem{GAGA}
Serre J.P., G\'eom\'etrie alg\'ebrique et g\'eom\'etrie analytique,
 \href{https://doi.org/10.5802/aif.59}{\textit{Ann. Inst. Fourier (Grenoble)}} \textbf{6} (1956), 1--42.

\bibitem{Sha82}
Shanks D., Dihedral quartic approximations and series for~{$\pi $},
 \href{https://doi.org/10.1016/0022-314X(82)90075-0}{\textit{J.~Number Theory}} \textbf{14} (1982), 397--423.

\bibitem{Smi06}
Smirnov V.A., Feynman integral calculus, \href{https://doi.org/10.1007/3-540-30611-0}{Springer-Verlag}, Berlin, 2006.

\bibitem{Sre10}
Srednicki M., Quantum field theory, Cambridge University Press, Cambridge,
 2010.

\bibitem{ST14}
Stieberger S., Taylor T.R., Closed string amplitudes as single-valued open
 string amplitudes, \href{https://doi.org/10.1016/j.nuclphysb.2014.02.005}{\textit{Nuclear Phys.~B}} \textbf{881} (2014), 269--287,
 \href{https://arxiv.org/abs/1401.1218}{arXiv:1401.1218}.

\bibitem{tHooftVeltman}
't~Hooft G., Veltman M., Regularization and renormalization of gauge fields,
 \href{https://doi.org/10.1016/0550-3213(72)90279-9}{\textit{Nuclear Phys.~B}} \textbf{44} (1972), 189--213.

\bibitem{Voe00}
Voevodsky V., Triangulated categories of motives over a field, in Cycles,
 Transfers, and Motivic Homology Theories, \textit{Ann. of Math. Stud.}, Vol.~143, Princeton University Press, Princeton, NJ, 2000, 188--238.

\bibitem{Voi02}
Voisin C., Hodge theory and complex algebraic geometry.~{I}, \textit{Cambridge
 Studies in Advanced Mathematics}, Vol.~76, \href{https://doi.org/10.1017/CBO9780511615344}{Cambridge University Press},
 Cambridge, 2002.

\bibitem{Wal06}
Waldschmidt M., Transcendence of periods: the state of the art, \href{https://doi.org/10.4310/PAMQ.2006.v2.n2.a3}{\textit{Pure
 Appl. Math.~Q.}} \textbf{2} (2006), 435--463.

\bibitem{Wei94}
Weibel C.A., An introduction to homological algebra, \textit{Cambridge Studies
 in Advanced Mathematics}, Vol.~38, \href{https://doi.org/10.1017/CBO9781139644136}{Cambridge University Press}, Cambridge,
 1994.

\bibitem{Wei60}
Weinberg S., High-energy behavior in quantum field-theory, \href{https://doi.org/10.1103/PhysRev.118.838}{\textit{Phys. Rev.}}
 \textbf{118} (1960), 838--849.

\bibitem{Zee03}
Zee A., Quantum field theory in a nutshell, Princeton University Press,
 Princeton, NJ, 2003.

\end{thebibliography}
\end{document}